\pdfoutput=1
\documentclass[letterpaper,12pt]{article}

\usepackage{graphicx}
\usepackage{amsmath,amsthm,amssymb,mathrsfs,mathtools}
\usepackage[OT1]{fontenc} 
\usepackage{etoc}
\usepackage{float}
\usepackage{fancyhdr}

\def\subsectiontitle{}
\def\subsubsectiontitle{}
\usepackage{tikz}
\usetikzlibrary{automata, positioning}

\usepackage[noamsmath]{kpfonts}
\usepackage{zi4}

\usepackage{enumitem}
\usepackage[margin=1in]{geometry}

\usepackage[colorlinks,breaklinks=true]{hyperref}
\usepackage{breakcites}

\usepackage{xcolor}
\AtBeginDocument{%
	\hypersetup{%
		linkcolor={red!50!black},%
		citecolor={blue!50!black},%
		urlcolor={blue!80!black}%
	}%
}%

\usepackage[nameinlink,noabbrev,sort,capitalise]{cleveref}
\makeatletter
\usepackage{etoolbox}
\patchcmd{\pprintMaketitle}
{\fi\hrule}%
{\fi\ifvoid\extrainfobox\else\unvbox\extrainfobox\par\vskip10pt\fi\hrule}
{}{}

\newsavebox\extrainfobox

\usepackage{comment}
\usepackage{multirow}
\usepackage{multicol}
\usepackage{subcaption}

\allowdisplaybreaks

\let\oldfootnote\footnote
\renewcommand\footnote[1]{\oldfootnote{\hspace{.4mm}#1}}

\makeatletter
\renewenvironment{proof}[1][\proofname] {\par\pushQED{\qed}\normalfont\topsep6\p@\@plus6\p@\relax\trivlist\item[\hskip\labelsep\bfseries#1\@addpunct{.}]\ignorespaces}{\popQED\endtrivlist\@endpefalse}
\makeatother

\let\oldFootnote\footnote
\newcommand\nextToken\relax

\renewcommand\footnote[1]{%
	\oldFootnote{#1}\futurelet\nextToken\isFootnote}

\newcommand\isFootnote{%
	\ifx\footnote\nextToken\textsuperscript{,}\fi}

\usepackage{blkarray}
\usepackage{pgfpages,epsfig,lscape}
\usepackage{epic}
\usepackage{setspace}
\usetikzlibrary{arrows,decorations,decorations.pathreplacing,calc,matrix}
\usepackage{natbib}
\DeclareFontFamily{U}{mathb}{\hyphenchar\font45}
\DeclareFontShape{U}{mathb}{m}{n}{
	<-6> mathb5 <6-7> mathb6 <7-8> mathb7
	<8-9> mathb8 <9-10> mathb9
	<10-12> mathb10 <12-> mathb12
}{}
\DeclareSymbolFont{mathb}{U}{mathb}{m}{n}
\DeclareMathSymbol{\llcurly}{\mathrel}{mathb}{"CE}
\DeclareMathSymbol{\ggcurly}{\mathrel}{mathb}{"CF}

\hypersetup{colorlinks = true}
\hypersetup{allcolors=blue}

\theoremstyle{plain}

\newtheorem{theorem}{Theorem}
\newtheorem*{theorem*}{Theorem}
\newtheorem{proposition}{Proposition}
\newtheorem{lemma}{Lemma}
\newtheorem{corollary}{Corollary}

\newtheorem{claim}{Claim}
\newenvironment{claimproof}[1][\proofname]{\begin{proof}[#1]}{\end{proof}}

\theoremstyle{definition}
\newtheorem{definition}{Definition}
\newtheorem{example}{Example}
\newtheorem{remark}{Remark}
\theoremstyle{plain}

\def\citeapos#1{\citeauthor{#1}'s (\citeyear{#1})}

\def\w{\omega}

\def\O{\mathcal{O}}
\def\I{\mathcal{I}}

\begin{document}

	\title{\Large Consistency and the exclusion core under co-ownership\thanks{I acknowledge financial support from the National Natural Science Foundation of China (72394391, 72122009) and the Wu Jiapei Foundation of the China Information Economics Society (E21103567).%
	}}

	\author{Jun Zhang\thanks{Institute for Social and Economic Research, Nanjing Audit University. Email: zhangjun404@gmail.com}
	}

	\date{\today}

	\maketitle

	\begin{abstract}

		In discrete allocation with co-ownership, the exclusion core of \citet{balbuzanov2019endowments} grants owners exclusion rights and secures nonemptiness by keeping unaffected agents out of blocking coalitions. This restriction can defeat legitimate ownership claims and make the exclusion core's prediction hinge on agents who have nothing at stake. Consistency diagnoses this failure and dictates the repair: an unaffected group that could depart with its assignments may join a blocking coalition, but only to complete the coalition's joint exclusion right. The resulting \textit{consistent exclusion core} is the largest consistent refinement of the exclusion core and remains nonempty. Consistency has a price: no strategy-proof mechanism selects from any nonempty, consistent, Pareto efficient, and individually rational solution, including the consistent exclusion core. The conflict vanishes when every object is privately or publicly owned: the consistent exclusion core then coincides with the exclusion core and admits a strategy-proof selection.

	\end{abstract}



	\thispagestyle{empty}
	\setcounter{page}{0}
	\newpage

\section{Introduction}

The core is a foundational solution concept in economic theory: it selects the outcomes that no coalition can improve upon using only what its members control. In assignments of indivisible objects without transfers, control comes from ownership, and the canonical models rest on simple ownership structures in which each object is either privately owned by a single agent or publicly owned by the entire population: the housing market \citep{shapley1974cores} has only privately owned objects, house allocation \citep{HZ1979,abdulkadirouglu1998random} only publicly owned ones, and house allocation with existing tenants \citep{AbduSonmez1999} both. Ownership in practice is rarely so simple: laboratory space is shared among research groups, offices are controlled jointly by several university departments, and housing units are held in common by their residents. \cite{balbuzanov2019endowments} (BK hereafter) introduce a model that places such \emph{co-ownership} at its center and contains the three canonical models as special cases. Its cost is that nonemptiness and Pareto efficiency, which conventional core concepts deliver together in the canonical models, come apart.

Underlying this difficulty is a basic question: what rights does ownership confer? The conventional cores answer with \emph{exchange rights}: a blocking coalition may only reallocate its own endowments among its members. They differ in whether an \emph{unaffected} agent, one whose assignment does not change, may join a blocking coalition. The \emph{strong core} admits such agents and is Pareto efficient but may be empty; the \emph{weak core} keeps them out and is nonempty but may be Pareto inefficient.

BK answer instead with \emph{exclusion rights}, which are widely regarded in property law as the defining incident of ownership. A coalition owns an object when it contains all of the object's owners, and it may then exclude others from the object; since an agent who occupies such an object can be pressed, under threat of exclusion, to act on the coalition's behalf, the coalition also controls whatever it owns together with her, and so on. Their \emph{exclusion core} rejects an allocation when some coalition can make all of its members strictly better off while every outsider made worse off loses an object the coalition controls. Like the weak core, it keeps unaffected agents out of blocking coalitions, which secures its nonemptiness; unlike the weak core, it retains Pareto efficiency, since a Pareto improvement harms no outsider. The restriction, however, makes the exclusion core fragile.

The restriction can defeat an entirely legitimate ownership claim. In Example~\ref{Example:co-ownership}, an object that two agents co-own and both want is assigned to a third agent who owns nothing, leaving both owners empty-handed. The two owners together fail to exclude the third agent, because only one of them can receive the object; the other stands to gain nothing and so is kept out of the blocking coalition, which then does not own the object. The exclusion core thus retains an implausible allocation that the strong core readily rejects.

The restriction can even disable every exclusion right at once, making the exclusion core's prediction hinge on an agent who has nothing at stake. Add to any economy an artificial agent who co-owns every object with its original owners but wants none of them; she cannot be made strictly better off, so she joins no blocking coalition, and no coalition without her controls anything. The exclusion core of the enlarged economy therefore consists of all Pareto efficient allocations, exactly what it would select if every object were publicly owned. The effect is starkest in the housing market of \cite{shapley1974cores}: the exclusion core selects only the top trading cycles (TTC) allocation, yet in the enlarged market it contains every Pareto efficient allocation. The artificial agent dilutes the original owners' control, but she would lose nothing by consenting to a block; the strong core, which rests on the same joint ownership but lets her join as an unaffected member, is unchanged (Proposition~\ref{prop:fragility}). It is thus the restriction that keeps the agent out, not the dilution, that makes the exclusion core disregard the original owners' claims.

This fragility is a failure of \emph{consistency}, a standard criterion for evaluating solutions across economic theory, which requires that when some agents depart with their assignments under a recommended allocation, a solution continue to recommend what that allocation prescribes for those who remain. In both failures above, the exclusion core retains an allocation that it rejects once an agent receiving nothing departs. In Example~\ref{Example:co-ownership}, the departure of one co-owner leaves the other co-owner as the object's private owner, who can then exclude the third agent; in the enlarged economy, the departure of the artificial agent restores the original economy, whose exclusion core rejects the allocations it rejected before she was added. The exclusion core thus violates consistency (Corollary~\ref{prop:exclusioncore:consistency}), whereas the strong core is consistent in the same environment (Proposition~\ref{prop:strongcore:consistency}); the failure thus lies in the exclusion core's design, not in the environment.

Consistency in this paper allows departures only by \emph{self-feasible} groups, which leave with objects drawn from their own endowments, as is standard in the literature when agents hold heterogeneous endowments. A departure with an object that a remaining agent partly or even privately owns would extinguish her claim, and permitting such departures would rule out every nonempty, Pareto efficient, and individually rational solution (Remark~\ref{remark:strongerconsistency}). A self-feasible departure instead leaves every remaining claim intact: an object the departing group co-owns with agents who remain passes to those agents alone, and an object whose owners have all left becomes publicly owned.

Consistency also guides the repair. Suppose that a self-feasible group departs with its assignments and some coalition then blocks, by the exclusion core's own standard, what an allocation prescribes for the remaining agents. Any refinement of the exclusion core that satisfies consistency must reject the allocation, and lifting the block back to the original economy reinstates the departing group as unaffected members of the coalition, who complete its \emph{joint} exclusion right without exercising one of their own. This blocking test defines our \emph{consistent exclusion core} (Definition~\ref{definition:C1core}): unaffected agents may join a blocking coalition, but they must form a self-feasible group, and every harmed outsider must lose an object that the coalition controls but that this group does not control on its own. Theorem~\ref{theorem:C1core:char} shows that this test, which refers to no other economy, characterizes precisely the largest consistent refinement of the exclusion core: the concept removes an allocation only when the exclusion core itself rejects what the allocation prescribes after some self-feasible departure. Unlike the exclusion core, it is not fragile: adding the artificial agent leaves it unchanged, as it leaves the strong core.

The consistent exclusion core inherits Pareto efficiency from the exclusion core and is nonempty. To find its elements, we generalize the ``you request my house--I get your turn'' (YRMH-IGYT) algorithm of \cite{AbduSonmez1999} to arbitrary co-ownership: when a cycle is removed, remaining agents with a claim on an object in the cycle inherit the departing agents' claims on the remaining objects, and every object points to its own owners before any agent whose claim is merely acquired. Every outcome of the algorithm belongs to the consistent exclusion core (Proposition~\ref{prop:YRMH-IGYT}).

The generalized algorithm is manipulable, and the difficulty lies not in the algorithm but in what is asked of a solution. By Theorem~\ref{theorem:general:strategyproof}, no strategy-proof mechanism selects from any nonempty solution that is consistent, Pareto efficient, and individually rational, although individual rationality protects only an agent's claim to what she privately owns. The three requirements are weak enough that the theorem binds every nonempty consistent refinement of the exclusion core (Corollary~\ref{cor:exclusion:strategyproof}), even though the exclusion core itself admits a strategy-proof selection. What strategy-proofness cannot accommodate is therefore consistency, which protects latent ownership claims: a co-owner becomes the private owner of an object once its other owners depart.

These difficulties arise only under intricate co-ownership. When every object is privately or publicly owned, as in the canonical models, the exclusion core itself is consistent and thus coincides with the consistent exclusion core. It consists exactly of the outcomes of the generalized YRMH-IGYT, and for any fixed initial order of agents, the algorithm is strategy-proof, indeed group strategy-proof (Proposition~\ref{prop:specialsetting}).

We also compare our concept with the exchange-rights cores of \cite{sun2020core} for the same model. Their \emph{effective core} admits a strategy-proof selection and is therefore not consistent (Corollary~\ref{cor:effective:inconsistency}); their sharper \emph{rectified core} is consistent and therefore admits no strategy-proof selection (Corollary~\ref{cor:sun:strategyproof}). Neither concept stands in an inclusion relation to the consistent exclusion core, because exclusion rights and exchange rights authorize different blocks. Finally, consistency removes only what it must: the consistent exclusion core may retain an allocation that a coalition owning every object involved cannot block, because an unaffected member is not self-feasible on her own (Example~\ref{Example:combinedfeature}). Online Appendix~\ref{appendix:refinedEC:definition} proposes a refinement of the consistent exclusion core that admits such a member whenever the coalition owns her assignment; it rules out this allocation, remains nonempty and consistent, and turns out to lie inside both cores of \cite{sun2020core}. Under simple ownership, BK's exclusion core coincides with that refinement and hence predicts at least as sharply as its exchange-rights counterparts (Corollary~\ref{cor:special:rectified}).

The remainder of the paper proceeds as follows. Section~\ref{section:preliminary} presents the model, the consistency principle, and the cores we examine. Section~\ref{section:exclusion} establishes the exclusion core's fragility and its failure of consistency. Section~\ref{section:consistentEC} develops the consistent exclusion core and its characterization, Section~\ref{section:YRMH} proves its nonemptiness, and Section~\ref{section:incentives} derives the impossibility theorem. Section~\ref{section:special} studies simple ownership, Section~\ref{section:Sunetal} compares the two views of ownership, and Section~\ref{section:discussion} concludes. Proofs are collected in the appendix; the online appendix contains the remaining proofs, examples, and further results.

\subsection*{Related literature}

This paper joins a growing literature on cores for discrete allocation with complex property rights. \cite{balbuzanov2019endowments} introduce the co-ownership model and the exclusion core, and further embed both into a more general priority-based setting, within which BK propose two refinements. In the co-ownership model, neither refinement attains what ours does: their strong exclusion core may be empty, while their weak exclusion core is nonempty but not consistent (Remark~\ref{remark:BKrefinements}). \cite{balbuzanov2019property} extend their approach to economies with production networks. \cite{ishida2025group} study a special case of BK's model in which agents are partitioned into groups, each collectively owning a number of objects equal to its size, and construct group strategy-proof mechanisms whose outcomes lie in both the strong core and the exclusion core; footnote~\ref{fn:ishidapark} in Section~\ref{section:special} discusses their relation to our results.

\cite{sun2020core} propose the effective core and the rectified core, both generalizing the strong core under the exchange-rights view, and construct a strategy-proof mechanism that selects from the former. Their concepts and ours take parallel positions within the two views of ownership: starting from the strong core, which admits every unaffected agent into blocking coalitions, theirs limit how a self-feasible group of unaffected members may contribute its redundant endowments to a blocking coalition; starting from the exclusion core, which admits no unaffected agents, ours admits such a group but never lets the block rest on the group's own exclusion right. Theirs are calibrated to restore nonemptiness, whereas ours is dictated by consistency; we show that the rectified core is consistent but is not the largest consistent refinement of the effective core (Proposition~\ref{prop:rectifiedcore:consistency}).

\cite{tan2026freeriding}, in work subsequent to an earlier version of this paper,\footnote{They position their concepts relative to the \emph{refined exclusion core} in that earlier version \citep{zhang2020cores}. The concepts in this paper have since changed substantially and now differ from those in \cite{zhang2020cores}.} propose the \emph{bilateral core}. It is built on a veto-type exclusion right weaker than BK's, yet it lies between two exchange-rights concepts: it contains the strong core and refines \citeapos{sun2020core} rectified core. Online Appendix~\ref{appendix:bilateral} shows that their concept is neither consistent nor comparable to BK's exclusion core or the consistent exclusion core; it nonetheless contains our refinement in Online Appendix~\ref{appendix:refinedEC:definition}. They also identify a conflict among strategy-proofness, Pareto efficiency, and a fairness requirement in their paper. Our impossibility instead requires individual rationality and imposes consistency on the solution from which a mechanism selects rather than on the mechanism itself.

\cite{cheng2024proper} adjust a different lever: in a priority-based setting with strict priorities, they treat the system of exclusion rights as the object to be chosen rather than as a primitive, and their \emph{proper exclusion right system} yields a core containing exactly the TTC outcome. Where they calibrate the rights and let the core follow, we hold BK's rights fixed and let consistency discipline the solution.

Methodologically, we build on the consistency principle, surveyed by \cite{driessen1991survey} and \cite{thomson1990consistency,thomson2011consistency,thomsonconsistency}, which has long served as a standard for selecting solutions in many fields, including bargaining \citep{lensberg1987stability}, bankruptcy problems \citep{aumann1985game}, cost allocation \citep{moulin1985separability}, fair allocation \citep{thomson1988study,tadenuma1991no}, rationing problems \citep{moulin1999rationing}, and cooperative games without side payments \citep{peleg1985axiomatization}. Market design has used two versions of the principle. The stronger permits any group of agents to depart and has characterized the core of the marriage model \citep{sasaki1992consistency} and consistent rules in house allocation \citep{ergin2000consistency,ergin2002efficient,ehlers2007consistent}; the weaker restricts departures to groups that leave a well-defined reduced economy and has characterized the TTC mechanism and its variants \citep{sonmez2010house,ehlers2014top,karakaya2019top,dougan2022robust}. Our model requires the weaker version, since an arbitrary departure can extinguish the ownership claims of those who remain.

	\section{Preliminaries}\label{section:preliminary}

	\subsection{The co-ownership model}\label{section:model}
	Let $ \I $ be the grand set of agents and $ \O $ the grand set of indivisible and heterogeneous objects, both infinite. An economy is represented by a tuple $ \Gamma=(I,O,\succ_I,\{C_o\}_{o\in O}) $, where $ I\subseteq \I $ is a finite nonempty set of agents, $ O\subseteq \O $ is a finite, possibly empty, set of objects, $ \succ_I=(\succ_i)_{i\in I} $ is a preference profile of agents, and for each $ o\in O $, $ C_o\subseteq I$ is a nonempty set of agents who hold $ o $ jointly. Let $ o^*\notin \O $ denote a null object, available in unlimited copies in every economy, with the convention $ C_{o^*}=\emptyset $.

	Every nonempty $ C\subseteq I $ is called a \textit{coalition}, and a coalition $ C' $ is a \textit{subcoalition} of $ C $ if $ C'\subseteq C $. An agent \textit{owns} an object $ o\in O $ if she belongs to $ C_o $, but ownership is exercised by the owners only as a group: a coalition \textit{owns} $ o $ if it contains $ C_o $. Hence $ C_o $ is the smallest coalition that owns $ o $, and a coalition that owns $ o $ may include agents who do not own $ o $. For convenience, we record coalitional ownership in an \textit{endowment} function $ \w: 2^I \rightarrow 2^O$ such that, for every $ C\subseteq I $, $ \w(C)=\{o\in O:C_o\subseteq C\} $; we treat $ \w(C) $ as the set of endowments of $ C $. In BK's description of the model, $ \w $ is the primitive and each $ C_o $ is defined as the minimal coalition that owns $ o $. If $C_o=\{i\}$ for some agent $i$, we say $o$ is privately owned by $i$. If $C_o=I$, we say $o$ is publicly owned, and whenever $|C_o|\ge 2$, we say $o$ is co-owned.

	Each agent $ i\in I $ demands one object and has a \textit{strict} preference relation $ \succ_i $ over $ O\cup \{o^*\} $. Object $ o\in O $ is \textit{acceptable} to $ i $ if $ o\succ_i o^* $; otherwise, $ o $ is \textit{unacceptable} to $ i $. For any two objects $ o$ and $o' $, we write $ o\succsim_i o' $ if $ o=o' $ or $ o\succ_i o' $.   	

	An \textit{allocation} is a function $ \mu:I\rightarrow O\cup \{o^*\} $ such that, for all distinct $ i,j\in I $, $ \mu(i)=\mu(j) $ implies $ \mu(i)=o^* $; that is, every agent receives exactly one object and every object in $ O $ is assigned to at most one agent. An agent who receives $ o^* $ receives nothing. For every $ C\subseteq I $, write $ \mu(C)=\{\mu(i):i\in C\} $; $ C $ is \textit{self-feasible} under an allocation $ \mu $ if $ \mu(C)\subseteq \w(C)\cup \{o^*\} $.  For any $C\subseteq I$ and any two allocations $ \mu $ and $ \sigma $, define $ C_{\sigma >\mu}=\{i\in C: \sigma(i)\succ_i \mu(i)\} $ and $ C_{\sigma= \mu}=\{i\in C: \sigma(i)=\mu(i)\} $. We call $ C_{\sigma= \mu} $ the set of \textit{unaffected agents}  within $ C $ from $ \mu $ to $ \sigma $.

	An allocation $ \mu $ is Pareto dominated by another allocation $ \sigma $ if $ \sigma(i)\succsim_i \mu(i) $ for all $ i\in I$ and $ \sigma(j)\succ_j \mu(j) $ for some $ j\in I$. An allocation is \textit{Pareto efficient} if it is not Pareto dominated by any other allocation; otherwise, it is \textit{Pareto inefficient}.

	An allocation $ \mu $ is \textit{individually rational} if $\mu(i)\succsim_i o$ for all $i\in I$ and all $o\in\w(\{i\})\cup\{o^*\}$. An agent acting alone can secure only her privately owned objects or the null object, so individual rationality asks no more than that her assignment be at least as good as each of them; her claims to co-owned objects are left unprotected. The requirement is therefore weak, and the impossibility results that invoke it are correspondingly strong.

	An economy has \textit{simple ownership} if every object is privately or publicly owned; otherwise some object is co-owned by a proper subset of the agents, and we say that the economy has \textit{intricate co-ownership}. 

	Let $ \mathcal{E} $ denote the set of all economies, which we call the \emph{full domain}: given $ I $ and $ O $, every ownership structure $ \{C_o\}_{o\in O} $ and every profile of strict preferences over $ O\cup\{o^*\} $ form an economy. In each $ \Gamma\in \mathcal{E} $, let $ \mathcal{A}(\Gamma) $ denote the set of allocations.

	A \textit{solution} is a correspondence $ f:\mathcal{E}\rightarrow \bigcup_{\Gamma\in \mathcal{E}}2^{\mathcal{A}(\Gamma)}$ such that, for every $ \Gamma\in \mathcal{E} $, $ f(\Gamma)\in 2^{\mathcal{A}(\Gamma)} $; note that $ f(\Gamma) $ may be empty. A solution $ f $ is \textit{Pareto efficient} if, for every $ \Gamma\in \mathcal{E} $, every allocation in $ f(\Gamma) $ is Pareto efficient in $ \Gamma $; likewise, $ f $ is \textit{individually rational} if every allocation in $ f(\Gamma) $ is individually rational in $ \Gamma $. It is \textit{nonempty} if $ f(\Gamma)\neq\emptyset $ for every $ \Gamma \in \mathcal{E} $.

	For any two solutions $f$ and $g$, we write $g\subseteq f$ if, for every $\Gamma \in \mathcal{E}$, $g(\Gamma)\subseteq f(\Gamma)$; we then call $g$ a \textit{refinement} of $f$.

	A \textit{mechanism} is a single-valued nonempty solution $ \varphi $; write $ \varphi(\Gamma) $ for the allocation selected by $ \varphi $ for $\Gamma$. A mechanism $ \varphi $ \textit{selects from} a nonempty solution $ f $ if $ \varphi(\Gamma)\in f(\Gamma) $ for every $ \Gamma\in \mathcal{E} $. A mechanism $ \varphi $ is \textit{strategy-proof} if, for every $ \Gamma=(I,O,\succ_I,\{C_o\}_{o\in O}) $, every agent $ i $, and every alternative report $ \succ'_i $, $\varphi(\Gamma)(i)\succsim_i \varphi(\Gamma')(i)$, where 
	$\Gamma'=(I,O,(\succ'_i,\succ_{-i}),\{C_o\}_{o\in O})$. Thus, no agent obtains a strictly better assignment by misreporting her preference. A mechanism $ \varphi $ is \textit{group strategy-proof} if, for every $ \Gamma $, no coalition $ C $ has alternative reports $ \succ'_C $ such that, with $ \Gamma'=(I,O,(\succ'_C,\succ_{-C}),\{C_o\}_{o\in O}) $, $ \varphi(\Gamma')(i)\succsim_i \varphi(\Gamma)(i) $ for every $ i\in C $ and $ \varphi(\Gamma')(j)\succ_j \varphi(\Gamma)(j) $ for some $ j\in C $. Group strategy-proofness implies strategy-proofness.

	Three canonical models are special cases of the model here. Let all objects be acceptable to all agents. If $ |I|=|O| $ and each object is privately owned by a distinct agent, we obtain the \textit{housing market model}. If $ |I|=|O| $ and all objects are publicly owned, we obtain the \textit{house allocation model}. If there exist a nonempty $ I'\subsetneq I $ and a nonempty $ O'\subsetneq O$ with $ |I'|=|O'|$ such that each $ o\in O' $ is privately owned by a distinct agent in $ I' $, and the other objects are publicly owned, we obtain the \textit{house allocation with existing tenants model}. All three models have simple ownership.

\subsection{The consistency principle}\label{section:consistency:definition}

Consistency is a principle of coherence across populations: if a solution recommends an allocation and some agents then depart with their assignments, the recommendation should continue to hold for those who remain; violation of it would make what the remaining agents receive depend on who else happens to be present. The departures are hypothetical: consistency asks what a solution recommends if a group leaves, not whether the group would wish to leave. To make this precise, we must specify how the remaining objects are owned once a group departs, in a way compatible with the original economy. We study the following reduction rule: a remaining object stays with whichever of its original owners remain, and if all of its owners have left, it becomes publicly owned. The rule treats ownership as a claim held jointly by the owners: the claim survives as long as some owner remains, and an object that no remaining agent claims reverts to the common pool, as an office whose co-owning departments have all left reverts to the university. Only then is the smaller economy genuinely a reduced version of the larger one.

The principle's force depends on which departures we allow. If any group may leave with its assignments, a departure can extinguish the claims of those who remain: a departing agent may take an object partly or even privately owned by a remaining agent, who is thereby stripped of a right she still holds. We therefore require the departing group to be \emph{self-feasible}, so that no remaining agent's ownership is disturbed. This restriction is common in the literature when agents hold heterogeneous ownership. Indeed, the stronger requirement that permits any departure is untenable in our model (Remark~\ref{remark:strongerconsistency}).

Formally, in any $ \Gamma=(I,O,\succ_I,\{C_o\}_{o\in O}) $, given an allocation $ \mu $ and a nonempty $ I'\subsetneq I $ that is self-feasible under $\mu$, suppose the agents in $I'$ leave with their assignments. We denote the reduced economy by $ \Gamma(\mu,I\backslash I')=(I\backslash I',O\backslash \mu(I'),\succ'_{I\backslash I'},\{C'_o\}_{o\in O\backslash\mu(I')}) $ such that:
\begin{itemize}
 \item For each $ i\in I\backslash I' $, $ \succ'_i $ is the restriction of $ \succ_i $ to $ (O\backslash\mu(I'))\cup\{o^*\} $;
 \item For each $ o\in O\backslash\mu(I') $, if $ C_o\cap (I\backslash I')\neq \emptyset $, then $ C'_o= C_o\cap (I\backslash I') $; otherwise $ C'_o=I\backslash I' $.
\end{itemize}
For any coalition $ C $, let $ \mu_{C} $ denote the restriction of $ \mu $ to $ C $: for each $ i\in  C$, $ \mu_{C}(i)=\mu(i) $. Then, $ \mu_{I\backslash I'} $ is a well-defined allocation in $ \Gamma(\mu,I\backslash I') $. We also write $ \Gamma(\mu,I)=\Gamma $ for the case in which no agent departs.

\begin{definition}
	A solution $ f $ is \textbf{consistent} if, for every $ \Gamma=(I,O,\succ_I,\{C_o\}_{o\in O}) $ with $ f(\Gamma)\neq \emptyset $, for every $ \mu\in f(\Gamma) $ and every nonempty self-feasible $ I'\subsetneq I $ under $\mu$, $ \mu_{I\backslash I'} \in f(\Gamma(\mu,I\backslash I')) $.
\end{definition}

\begin{remark}[Stronger consistency]\label{remark:strongerconsistency}
	Consider the stronger requirement that permits every nonempty group $ I'\subsetneq I $ to depart with its assignments. No nonempty, Pareto efficient, and individually rational solution satisfies it (Example~\ref{Example:stronger:consistency} in Online Appendix~\ref{appendix:extraexample}).
\end{remark}

We use consistency in two ways. First, it serves as a criterion for evaluating solutions; we apply it to existing cores in Section~\ref{section:exclusion}. Second, and more importantly, it plays a constructive role, which is captured by a general operator $\Phi$.

For any solution $ f $, define the solution $ \Phi(f) $ by
\[
\Phi(f)(\Gamma)=\bigl\{\mu\in f(\Gamma):\ \mu_{I\backslash I'}\in f\bigl(\Gamma(\mu,I\backslash I')\bigr)\ \text{for every nonempty self-feasible } I'\subsetneq I \text{ under } \mu \bigr\}.
\]
Thus $ \Phi(f) $ retains precisely those allocations of $ f $ whose restrictions remain in $ f $ after every admissible departure.

\begin{lemma}[Largest consistent refinement]\label{lemma:maxrefinement}
	For every solution $ f $, $ \Phi(f) $ is the largest consistent refinement of $ f $: $ \Phi(f)\subseteq f $, $ \Phi(f) $ is consistent, and every consistent solution contained in $ f $ is contained in $ \Phi(f) $.%
\end{lemma}

The operator $ \Phi $ therefore extracts from a solution its largest consistent part, and a single application suffices because admissible departures compose: a self-feasible departure followed by another from the reduced economy amounts to a single self-feasible departure from the original economy (Appendix~\ref{appendix:preliminary}). A solution $f$ itself is consistent if and only if it is a fixed point of $\Phi$: $f=\Phi(f)$.

Two remarks follow. First, given a nonempty solution $ f $, the operator $ \Phi $ does not ensure that $ \Phi(f) $ is nonempty; nonemptiness must therefore be established separately.

Second, Lemma~\ref{lemma:maxrefinement} describes $ \Phi(f) $ only through reduced economies: whether an allocation $ \mu $ belongs to $ \Phi(f)(\Gamma) $ depends on what $ f $ selects after every admissible departure from $ \Gamma $, not on what it selects in $ \Gamma $ alone. These reduced economies are not simply copies of $ \Gamma $ with some agents deleted: an object loses the owners who depart and becomes publicly owned once all of them have gone, so a departure can confront the remaining agents with a different ownership structure. The definition of $ \Phi(f) $ thus says which allocations consistency removes, but not which coalition in $ \Gamma $ could object to them; it gives $ \Phi(f) $ no cooperative foundation in $ \Gamma $ of the kind that defines a core. In general, there is no reason why $ \Phi(f) $ should admit such a foundation.

	\subsection{Core concepts}\label{section:standardcore}

	Conventionally, ownership confers the right to \emph{exchange}: a coalition blocks an allocation if its members can benefit from reallocating their endowments among themselves. Depending on whether all members must strictly benefit, the core concept has two variants. 

	\begin{definition}\label{definition:weakcore}
		(a) In an economy, an allocation $ \mu $ is \emph{weakly blocked} by a coalition $ C $ via another allocation $ \sigma $ if (1) $ \forall i\in C $, $ \sigma(i) \succsim_i \mu(i) $ and $ \exists j\in C $, $ \sigma(j)\succ_j\mu(j) $; (2) $ \sigma(C)\subseteq \w(C)\cup \{o^*\}$. 		
	The \textbf{strong core} consists of allocations that are not weakly blocked.

		(b) In an economy, an allocation $ \mu $ is \emph{strongly blocked} by a coalition $ C $ via another allocation $ \sigma $ if (1) $ \forall i\in C $, $ \sigma(i) \succ_i \mu(i) $; (2) $ \sigma(C)\subseteq \w(C)\cup \{o^*\}$. 	 	
		 The \textbf{weak core} consists of allocations that are not strongly blocked.
	\end{definition}

Taking $ C $ to be a singleton in part (b) recovers individual rationality: an allocation is individually rational exactly when no agent alone can strongly block it. The strong core is contained in the weak core. In the housing market model, the strong core is a singleton, containing only the TTC outcome.\footnote{In the housing market model, TTC works as follows: in each step, let each agent point to the agent who owns her most preferred object; then, clear cycles by letting involved agents exchange their objects.} In the house allocation model, the strong core coincides with the set of Pareto efficient allocations. In the general co-ownership model, however, the strong core is Pareto efficient yet may be empty; the weak core is nonempty yet may be Pareto inefficient and may retain implausible allocations.

\cite{balbuzanov2019endowments} propose instead that ownership confers the right to \emph{exclude}: a coalition may exclude an agent from any object it controls, directly or indirectly. In an economy $\Gamma$, a coalition $ C $ directly controls its endowments $\w(C)$ and may indirectly control additional objects by leveraging its exclusion right along the chains of occupation.

Formally, under an allocation $\mu$ in $\Gamma$, the set of objects controlled by $ C $ is defined as
 \[
 \Omega(C|\w,\mu)=\w(\cup_{k=0}^\infty C^k),
 \]
 where $ C^0=C $ and $ C^k=C^{k-1}\cup \{i\in I\backslash C^{k-1}: \mu(i)\in \w(C^{k-1}) \} $ for every $ k\ge 1 $. Since $ I $ is finite, the recursion stabilizes after finitely many steps.

 The recursion traces the chains of occupation. An agent outside $C$ who occupies an object in $\w(C)$ can be excluded from it by $C$, so she depends on $C$ and can be pressed to exercise her ownership on $C$'s behalf; adding her to the group brings the enlarged group's endowments within its reach, and whoever occupies those objects can be pressed in turn. Each step of the recursion follows every such chain one link further. Thus $ \Omega(C|\w,\mu) $ is the endowment of the group the chains reach.

 The exclusion core consists of allocations in which no coalition can make all of its members strictly better off by excluding others from the objects it controls.

 \begin{definition}[\cite{balbuzanov2019endowments}]\label{Definition:exclusion:core}
   In an economy, an allocation $ \mu $ is \emph{exclusion blocked} by a coalition $ C $ via another allocation $ \sigma $ if (1) $ \forall i\in C $, $ \sigma(i) \succ_i \mu(i) $; (2) $ \forall j\in I\backslash C $, $\mu(j)\succ_j \sigma(j)\implies \mu(j)\in  \Omega(C|\w,\mu)$.  	
 	The \textbf{exclusion core} consists of allocations that are not exclusion blocked.
 \end{definition}

The exclusion core differs from the conventional cores along two dimensions. The first is what a blocking coalition may do. Under conventional cores it must reallocate the objects it owns; an exclusion blocking coalition need not be self-feasible, and may instead exclude outsiders from any object it controls. The second is who may belong to the coalition. The exclusion core follows the weak core, both keeping unaffected agents out of blocking coalitions, whereas the strong core lets them join. The following example shows why the exclusion core keeps them out: admitting them would leave it empty.

\begin{example}[Keeping unaffected agents out ensures the exclusion core's nonemptiness]\label{Example:Kingdom}
		Consider the following economy in which all objects are privately owned by one agent; the other two agents own nothing. Here and in later examples, we omit $o^*$ from a preference table when it ranks below all real objects.

		\begin{center}
			\begin{tabular}{rcc}
				& $ a $ & $ b $  \\ \hline
				$ C_o $: & $ 1 $ & $ 1 $ \\ \hline
				$ \mu $: & $ 1 $ &  \\
				$ \sigma $: & $ 1 $ & $ 2 $\\
				$ \delta $: & $ 1 $ & $ 3 $ 
			\end{tabular}
			\quad \quad
			\begin{tabular}{ccc}
				$ \succ_1 $	& $ \succ_2 $ & $ \succ_3 $  \\ \hline
				$ a $ & $ a $ & $ a $ \\
				$ b $ & $ b $ & $ b $ \\
				& \\
				&
			\end{tabular}
		\end{center}

		The exclusion core equals $\{\sigma,\delta\}$: $ \mu $ is exclusion blocked by  $ \{2\} $ via $ \sigma $, or by $ \{3\} $ via $ \delta $, since in both cases, no agent is worse off; $\sigma$ and $\delta$ are not exclusion blocked, because the only agent who can be improved, $2$ or $3$, does not control any object.

		However, the exclusion core would be empty if unaffected agents could join blocking coalitions: $ \{1,3\} $ could exclusion block $ \sigma $ via $ \delta $, and $ \{1,2\} $ could exclusion block $ \delta $ via $ \sigma $.
\end{example}

In Example \ref{Example:Kingdom}, the weak core equals $ \{\mu,\sigma, \delta\}$, in which $\mu$ is not Pareto efficient, yet the strong core is empty. The exclusion core would be empty if unaffected agents could join blocking coalitions, for precisely the same reason the strong core is empty: agent $ 1 $, unaffected in every block since she keeps object $ a $, alternately partners with $ 2 $ and $ 3 $ to form blocking coalitions. It therefore keeps them out to ensure nonemptiness. However, Section \ref{section:exclusion} shows that this restriction is also the source of its weakness.

\section{Fragility of the exclusion core}\label{section:exclusion}

The restriction that keeps unaffected agents out of blocking coalitions secures the exclusion core's nonemptiness without sacrificing Pareto efficiency. This section shows what it costs: the restriction makes the exclusion core fragile, since what it selects can hinge on agents who have nothing at stake. A co-owner who cannot gain from a block must be left out of it, and her absence strips the coalition of control over the object she co-owns. Example \ref{Example:co-ownership}, due to BK, shows this in the smallest economy; Proposition \ref{prop:fragility} shows that adding a single artificial agent, who co-owns every object but wants none of them, triggers the same force in every economy.

	\begin{example}[Exclusion core admits implausible allocations]\label{Example:co-ownership}
		Consider the following economy in which the only object is co-owned by two agents, while a third agent owns nothing.
		\begin{center}
			\begin{tabular}{rc}
				& $ a $   \\ \hline
				$ C_o $: & $ 1,2 $ \\ \hline
				$ \mu $: & $ 1 $  \\
				$ \sigma $: & $ 2 $  \\
				$ \delta $: &  $ 3 $ 
			\end{tabular}
			\quad \quad
			\begin{tabular}{ccc}
				$ \succ_1 $	& $ \succ_2 $ & $ \succ_3 $  \\ \hline
				$ a $ & $ a $ & $ a $ \\
				& \\
				&\\
				&
			\end{tabular}
		\end{center}

		The exclusion core equals $ \{\mu,\sigma,\delta\} $, whereas the strong core equals $ \{\mu,\sigma\} $. As BK argue, the allocation $ \delta $ is implausible: the only object is assigned to the agent who owns nothing, while its two owners prefer the object but receive nothing. The allocation $ \delta $ is weakly blocked by $ \{1,2\} $ via $ \mu $, but it cannot be exclusion blocked, because $1$ and $2$ cannot be strictly improved simultaneously and neither of them alone controls the object.
	\end{example}

Scaled up, the force at work in Example~\ref{Example:co-ownership} can disable the exclusion right over every object in every economy. Take any economy $ \Gamma=(I,O,\succ_I,\{C_o\}_{o\in O}) $ and construct an economy $ \Gamma^*=(I\cup \{i^*\},O,\succ_{I\cup \{i^*\}},\{C^*_o\}_{o\in O}) $ by adding an artificial agent $ i^*\in \I\backslash I $ who co-owns every object with its original owners but wants none of them: for each $ o\in O $, $ C^*_o=C_o\cup\{i^*\} $, and $ \succ_{i^*} $ ranks $ o^* $ first. We call $ \Gamma^* $ the \textit{augmentation} of $ \Gamma $. A publicly owned object of $ \Gamma $ remains publicly owned in $ \Gamma^* $, and every other object acquires $ i^* $ as an additional owner.

Since $ i^* $ most prefers the null object, a natural heuristic in $ \Gamma^* $ is to assign $ i^* $ the null object, then treat $ i^* $ as absent and solve the problem for the remaining agents. For every allocation $ \mu $ in $ \Gamma $, let $ \mu^* $ denote the allocation in $ \Gamma^* $ with $ \mu^*(i^*)=o^* $ and $ \mu^*(i)=\mu(i) $ for all $ i\in I $; we call $ \mu^* $ the augmentation of $ \mu $. The heuristic thus recommends for $ \Gamma^* $ exactly the augmentations of what it recommends for $ \Gamma $.

The strong core satisfies the heuristic. The exclusion core, however, violates it wherever ownership matters: since $ i^* $ co-owns every object, no coalition without her owns anything, and since she cannot be strictly improved, she cannot join any exclusion blocking coalition. No coalition can therefore exclude anyone from anything, and the exclusion core of $ \Gamma^* $ consists of all Pareto efficient allocations. This is exactly what the exclusion core selects when every object is publicly owned.

\begin{proposition}\label{prop:fragility}
	For every economy $ \Gamma $, the strong core of $ \Gamma^* $ consists of the augmentations of the strong core allocations of $ \Gamma $, whereas the exclusion core of $ \Gamma^* $ consists of the augmentations of all Pareto efficient allocations of $ \Gamma $; this set equals the exclusion core that $ \Gamma^* $ would have if every object were publicly owned.
\end{proposition}

Because the exclusion core is Pareto efficient, the augmentation never shrinks it; whenever the exclusion core of $ \Gamma $ is a proper subset of its Pareto efficient allocations, the augmentation strictly enlarges it and makes its prediction independent of the original ownership structure. The contrast is sharpest in the canonical housing market model, in which the exclusion core coincides with the strong core, both selecting only the TTC allocation. In the augmentation of a housing market, which we call an \textit{augmented housing market}, the strong core remains the singleton containing the augmented TTC allocation, but the exclusion core expands to all Pareto efficient allocations---the predictions it makes in the house allocation model, where every object is publicly owned. Thus, adding a single agent who consumes nothing makes the exclusion core disregard every private endowment in the housing market. The added agent does dilute the original owners' control, but the dilution alone is harmless: the strong core also requires a blocking coalition to own what it reallocates, yet its prediction is unchanged, because the artificial agent can join a blocking coalition as an unaffected member.

This fragility is a failure of consistency. Every $ \Gamma $ is a reduced economy of its augmentation: in every Pareto efficient allocation of $ \Gamma^* $, the artificial agent receives the null object, so she is self-feasible, and her departure restores $ \Gamma $. Proposition \ref{prop:fragility} therefore implies that the exclusion core violates consistency in the augmentation of every $\Gamma$ in which it is a proper subset of Pareto efficient allocations: for each Pareto efficient allocation of $ \Gamma $ outside its exclusion core, its augmentation is selected in $ \Gamma^* $, although it is rejected in $ \Gamma $. Note that this violation requires only the departure of a single agent with the null object.

\begin{corollary}\label{prop:exclusioncore:consistency}
	The exclusion core is not consistent.
\end{corollary}

In contrast, the strong core is consistent, though the weak core is not.

\begin{proposition}\label{prop:strongcore:consistency}
	The strong core is consistent, whereas the weak core is not.
\end{proposition}

Corollary~\ref{prop:exclusioncore:consistency} and Proposition~\ref{prop:strongcore:consistency} together locate the source of the problem: since the strong core is consistent in the same environment, the exclusion core's fragility lies in its own design---specifically, in the restriction that keeps unaffected agents out of blocking coalitions---rather than in the difficulty of the environment.

\begin{remark}[BK's other solution concepts]\label{remark:BKrefinements}
	BK anticipate the weakness of the exclusion core illustrated in Example \ref{Example:co-ownership} and extend the concept to a setting in which agents' rights over objects are described by a priority structure rather than by owner sets. From the priorities they derive three nested solutions, and the co-ownership model embeds into that setting so that their \emph{unconditional} exclusion core coincides with the exclusion core; the other two, their \emph{strong} and \emph{weak} exclusion cores, are therefore refinements of it. Neither refinement attains both nonemptiness and consistency: the strong exclusion core may be empty, and the weak exclusion core, though always nonempty, is not consistent. BK also define an auxiliary \emph{direct} exclusion core, which requires that every agent made worse off lose an object the coalition directly owns; it contains the exclusion core and is not consistent either. Online Appendix \ref{appendix:strongEC} gives the details.
\end{remark}

\section{The consistent exclusion core}\label{section:consistentEC}

\subsection{Definition}\label{sub:CEC:definition}

Under co-ownership, the right to exclude is held jointly: a coalition controls an object directly only if it contains all of the object's owners, and indirectly only if every owner outside the coalition can be pressed, along a chain of occupation, to act on its behalf. An exclusion block, however, requires every member of the coalition to gain strictly (Definition \ref{Definition:exclusion:core}). When a block benefits only some owners of an object, the rest must be left out, and the coalition is then stripped of control over the object it owns---the source of the failure in Example \ref{Example:co-ownership}. Yet admitting unaffected agents without restriction is not the answer either: keeping them out is what secures nonemptiness of the exclusion core (Example \ref{Example:Kingdom}).

The two examples differ in what an unaffected agent's participation would contribute to the blocking coalition. In Example \ref{Example:co-ownership}, the unaffected co-owner would complete a right the coalition holds \emph{jointly}, restoring the very control that her absence denies it. In Example \ref{Example:Kingdom}, agent $ 1 $ controls every object outright, so her participation would exercise an independent exclusion right rather than complete a joint one, and it is exactly this independent right that would threaten nonemptiness. The two examples thus suggest a principle: an unaffected agent may join a coalition when her participation completes the coalition's joint exclusion right, but not when it exercises an exclusion right of her own.

This principle is intuitive but not yet precise, since it does not say which unaffected agents may join or what must be withheld from them. Consistency supplies both answers. Suppose a self-feasible group $ I' $ departs with its assignments under $ \mu $, and in the resulting reduced economy the restriction $ \mu_{I\backslash I'} $ is exclusion blocked by a coalition $ C $ via $ \sigma $. Any consistent refinement of the exclusion core must then reject $ \mu $ in the original economy, since consistency would otherwise require selecting $ \mu_{I\backslash I'} $ in an economy where the exclusion core's own standard rejects it. Lifting this block back to the original economy reinstates the departing group $ I' $ as unaffected participants in the coalition. Consistency thus identifies which groups of unaffected agents may join: those that are self-feasible under the allocation. It also identifies what must be withheld from them: the block being lifted is formed by $ C $ via $ \sigma $ in the reduced economy $ \Gamma(\mu,I\backslash I') $, from whose population $ I' $ is entirely absent, so it cannot rely on any exclusion right belonging to $ I' $. An outsider harmed by the block must therefore lose an object controlled by the coalition as a whole, but not one that the unaffected agents can control on their own.

The logic of co-ownership and the requirement of consistency thus converge on the same rule. To formalize it, we distinguish a coalition's joint exclusion right from a subcoalition's own exclusion right.

\begin{definition}\label{defn:ownexclusionright}
	In any economy, given any allocation $ \mu $, for any coalition $ C $ and any $ C' \subsetneq C$, possibly empty, the set of objects that are controlled by $ C' $ in the absence of $ C\backslash C' $ is
	\[
	\Omega(C'|C,\w,\mu)=\w(\cup_{k=0}^\infty C^k),
	\]
	where $ C^0=C' $,
	and $ C^k=C^{k-1}\cup \{i\in I\backslash (C \cup C^{k-1}): \mu(i)\in \w(C^{k-1}) \} $ for every $ k\ge 1 $.
\end{definition}

Unlike $ \Omega(C'|\w,\mu) $, the set $ \Omega(C'|C,\w,\mu) $ is computed without the members of $ C\backslash C' $: the recursion never absorbs them, so it ignores every chain of occupation that passes through them. Example \ref{example:defineOmega} at the end of this subsection illustrates the difference and its role.

With this distinction in hand, we can state the rule. It has two parts, both bearing on a blocking coalition $ C $ that contains unaffected agents. The first concerns membership: the group $ C_{\sigma=\mu} $ must be self-feasible under $ \mu $. The second concerns the rights the block may use: every outsider harmed by the block must lose an object that $ C $ controls but that $ C_{\sigma=\mu} $ does not control on its own, so that the block executes the exclusion right of $ C $ but not that of $ C_{\sigma=\mu} $. We now define our refinement of BK's exclusion core.

\begin{definition}\label{definition:C1core}
	In an economy, an allocation $ \mu $ is \emph{consistently exclusion blocked} by a coalition $ C $ via another allocation $ \sigma $ if
	\begin{enumerate}
		\item $ \forall i\in C $, $ \sigma(i) \succsim_i \mu(i) $ and $ \exists j\in C $, $ \sigma(j)\succ_j\mu(j) $;
		\item $ \mu(C_{\sigma=\mu})\subseteq \w(C_{\sigma=\mu})\cup \{o^*\} $;
		\item $ \forall j\in I\backslash C $, $\mu(j)\succ_j \sigma(j)\implies \mu(j)\in  \Omega(C|\w,\mu)$ but $ \mu(j)\notin \Omega(C_{\sigma=\mu}|C,\w,\mu)$.
	\end{enumerate}
	The \textbf{consistent exclusion core} consists of allocations not consistently exclusion blocked.
\end{definition}

Every exclusion block defined by BK is a consistent exclusion block, because $ C_{\sigma=\mu}=\emptyset $. Hence the consistent exclusion core refines the exclusion core.

\begin{example}[The role of $ \Omega(C'|C,\w,\mu) $]\label{example:defineOmega}
	Consider the following economy.
	\begin{center}
		\begin{tabular}{rccc}
			& $ a $ & $ b $ & $ c $ \\ \hline
			$ C_o $: & $ 1,2$ & $ 1 $ & $ 1 $\\ \hline
			$ \mu $: & $ 3 $ & $ 1 $ & $ 2 $\\
			$ \sigma $: & $ 2 $ & $ 1 $ & $ 3 $
		\end{tabular}
		\quad \quad
		\begin{tabular}{ccc}
			$ \succ_1 $	& $ \succ_2 $ & $ \succ_3 $  \\ \hline
			$ b $ & $ a $ & $ a $  \\
			$ c $ & $ b $ & $ b $  \\
			$ a $& $ c $ & $ c $
		\end{tabular}
	\end{center}

	Take $ C=\{1,2\} $ and $ C'=\{1\} $. Since object $ a $ is co-owned, $ \w(C')=\{b,c\} $. The recursion defining $ \Omega(C'|C,\w,\mu) $ cannot reach agent $ 3 $, since $ \mu(3)=a\notin \w(C') $. Hence $ \Omega(C'|C,\w,\mu)=\{b,c\} $. The recursion defining $ \Omega(C'|\w,\mu) $ instead absorbs agent $ 2 $, since $ \mu(2)=c\in \w(C') $; object $ a $ then lies in $ \w(\{1,2\}) $, which absorbs agent $ 3 $, so $ \Omega(C'|\w,\mu)=\{a,b,c\}=\Omega(C|\w,\mu) $. The chain that brings object $ a $ within agent $ 1 $'s reach therefore runs through agent $ 2 $, a member of $ C $ itself, and it is exactly this chain that $ \Omega(C'|C,\w,\mu) $ suppresses.

	This difference decides whether $ \mu $ is consistently exclusion blocked. According to Definition \ref{definition:C1core}, $ C $ consistently exclusion blocks $ \mu $ via $ \sigma $ by excluding agent $ 3 $ from $ a $: agent $ 2 $ strictly gains, agent $ 1 $ is unaffected, so $ C_{\sigma=\mu}=\{1\} $ is self-feasible, and the only harmed agent, $ 3 $, loses $ a $, which lies in $ \Omega(C|\w,\mu) $ but not in $ \Omega(C_{\sigma=\mu}|C,\w,\mu)=\{b,c\} $. Had condition (3) used $ \Omega(C_{\sigma=\mu}|\w,\mu)=\{a,b,c\} $ instead, the block would fail, although excluding agent $ 3 $ from $ a $ exercises $ \{1,2\} $'s joint ownership rather than any exclusion right that agent $ 1 $ holds on her own. This block is exactly what consistency requires: once agent $ 1 $ departs with $ b $, agent $ 2 $ becomes the private owner of $ a $ and excludes agent $ 3 $ in the reduced economy.
\end{example}

\subsection{Characterization}\label{sub:CEC:characterization}

Condition (3) of Definition~\ref{definition:C1core} constrains the block within the economy at hand, whereas consistency looks to the economy that remains after a self-feasible group departs with its assignments. Lemma~\ref{lemma:firstcondition} expresses condition (3) as a requirement in the latter economy, which gives the condition an economic interpretation; Theorem~\ref{theorem:C1core:char} then builds on this lemma to provide a characterization of the consistent exclusion core.

Specifically, fix allocations $ \mu $ and $ \sigma $ and a coalition $ C $ satisfying conditions (1) and (2) of Definition \ref{definition:C1core}. Let 
\begin{equation}\label{eq:cdiamond}\tag{$\diamond$}
C^\diamond=\cup_{k=0}^\infty C^k,
\end{equation}
where $ C^0=C_{\sigma=\mu} $ and $ C^k=C^{k-1}\cup\{i\in I\backslash (C\cup C^{k-1}):\mu(i)\in\w(C^{k-1})\} $ for every $ k\ge 1 $, as in Definition \ref{defn:ownexclusionright}, so that $ \w(C^\diamond)= \Omega(C_{\sigma=\mu}|C,\w,\mu) $: $ C^\diamond $ collects the unaffected members of $ C $ together with the outsiders whose assignments $ C_{\sigma=\mu} $ reaches on its own. Every agent of $ C^\diamond $ holds an object owned by $ C^\diamond $ or the null object, so $ C^\diamond $ is self-feasible under $ \mu $ and can depart with its assignments. Since $C^\diamond\cap C= C_{\sigma=\mu}$, the nonempty $C_{\sigma>\mu}$ remains after the departure. Lemma \ref{lemma:firstcondition} shows that condition (3) of Definition \ref{definition:C1core} holds if and only if every outsider harmed by the block loses an object that $ C_{\sigma>\mu} $ controls in the reduced economy $ \Gamma(\mu,I\backslash C^\diamond) $. Condition (3) thus requires the block to rest on the exclusion right of $ C_{\sigma>\mu} $ alone, once the unaffected members $ C_{\sigma=\mu} $ have left with all they control on their own.

\begin{lemma}\label{lemma:firstcondition}
Let $ \mu $ and $ \sigma $ be allocations in an economy $ \Gamma $, and let $ C $ be a coalition satisfying conditions (1) and (2) of Definition \ref{definition:C1core}. Then, for every $  j\in I\backslash C $, 
\begin{center}
	$ \mu(j)\in  \Omega(C|\w,\mu)$ but $ \mu(j)\notin \Omega(C_{\sigma=\mu}|C,\w,\mu)$ $\Longleftrightarrow$ $ \mu(j)\in   \Omega(C_{\sigma>\mu}|\w',\mu_{I\backslash C^\diamond})$,
\end{center} where $ C^\diamond $ is given by \eqref{eq:cdiamond} and $ \w' $ denotes the endowment function in $ \Gamma(\mu,I\backslash C^\diamond) $.
\end{lemma}

Example~\ref{example:defineOmega} illustrates the equivalence. There $ C^\diamond=\{1\} $, since agent $ 1 $ reaches no outsider on her own. Agent $ 3 $'s loss of object $ a $ satisfies the left-hand side, as computed in the example, and the right-hand side as well: once agent $ 1 $ departs, agent $ 2 $ privately owns $ a $ and can exclude agent $ 3 $ on her own.

This equivalence drives our main result. A consistent exclusion block by $ C $ in $ \Gamma $ becomes an exclusion block by $ C_{\sigma>\mu} $ in $ \Gamma(\mu,I\backslash C^\diamond) $, and since $ C^\diamond $ can depart with its assignments, an allocation that admits such a block in $\Gamma$ leaves behind a restriction in $\Gamma(\mu,I\backslash C^\diamond)$ that the exclusion core itself rejects; every consistent refinement of the exclusion core must therefore reject the allocation. This yields the following characterization.

\begin{theorem}\label{theorem:C1core:char}
	The consistent exclusion core is the largest consistent refinement of the exclusion core.
\end{theorem}

Write $ \mathcal{EC} $ and $ \mathcal{CEC} $ for the exclusion core and the consistent exclusion core. Lemma \ref{lemma:maxrefinement} already identifies $ \Phi(\mathcal{EC}) $ as the largest consistent refinement of the exclusion core, so the content of Theorem \ref{theorem:C1core:char} is the identity 
\[
\mathcal{CEC}=\Phi(\mathcal{EC}).
\]

The two sides are defined differently: membership in $ \Phi(\mathcal{EC})(\Gamma) $ is checked in every economy that a self-feasible departure leaves behind, and membership in $ \mathcal{CEC}(\Gamma) $ by a blocking test in $ \Gamma $ alone. Viewed from the side of $ \mathcal{CEC} $, the identity gives the largest consistent refinement the cooperative foundation that $ \Phi $ does not supply: an allocation of the exclusion core is removed exactly when some coalition in $ \Gamma $ can block it, with a self-feasible group of unaffected members only completing the coalition's joint exclusion right. Viewed from the side of $ \Phi $, it justifies each removal by consistency alone: the consistent exclusion core deletes an allocation exactly when some self-feasible group can depart with its assignments and leave behind a restriction that the exclusion core itself rejects, so every deletion appeals to no standard beyond BK's own. Theorem \ref{theorem:C1core:char} is silent on nonemptiness; Section \ref{section:YRMH} establishes it constructively by an algorithm that produces an element of the refinement in every economy. The refinement therefore achieves consistency without forfeiting the nonemptiness that gives the exclusion core its advantage over the strong core.

The consistent exclusion core repairs both failures identified in Section \ref{section:exclusion}. In Example \ref{Example:co-ownership}, $ \delta $ is consistently exclusion blocked by $ \{1,2\} $ via $ \mu $: agent $ 1 $ strictly gains, the unaffected agent $ 2 $ holds the null object and so is self-feasible, and the only harmed agent, $ 3 $, loses object $ a $, which $ \{1,2\} $ owns but agent $ 2 $ cannot control on her own; by symmetry, $ \delta $ is also consistently exclusion blocked via $ \sigma $. Likewise, in the augmentation of any economy $ \Gamma $, whenever a Pareto efficient allocation $ \nu $ of $ \Gamma $ is exclusion blocked by a coalition $ C $ via $ \sigma $, the coalition $ C\cup\{i^*\} $ consistently exclusion blocks $ \nu^* $ via $ \sigma^* $: the members of $ C $ strictly gain, the unaffected artificial agent holds the null object and so is self-feasible, and every harmed agent loses an object that $ C\cup\{i^*\} $ controls in $ \Gamma^* $, as $ C $ does in $ \Gamma $, but that $ i^* $ cannot control on her own. In fact, the consistent exclusion core of $ \Gamma^* $ consists of the augmentations of the consistent exclusion core of $ \Gamma $, just as the strong core does (Proposition~\ref{prop:fragility}), by a parallel argument. Therefore, unlike the exclusion core, the consistent exclusion core is unaffected by the augmentation.

\section{Finding elements of the consistent exclusion core}\label{section:YRMH}

To show that the consistent exclusion core is nonempty, we generalize the YRMH-IGYT algorithm of \cite{AbduSonmez1999}, originally designed for the special case of house allocation with existing tenants, and show that every outcome of the algorithm belongs to the consistent exclusion core.

The original YRMH-IGYT proceeds as follows. Fix a linear order of agents and let each private endowment initially point to its owner. In each step, the first agent in the order points to her favorite object. If that object is publicly owned or its owner has already been removed, the agent obtains that object and is removed. Otherwise, the owner of that object is moved to the top of the order and the step is repeated. If a cycle forms in some step, the agents in the cycle exchange their private endowments and are removed.

Our generalization differs from the original algorithm in two respects. First, agents may acquire shared ownership when a cycle is removed, so that a remaining agent with a claim on an object in the cycle does not lose it with nothing in return.\footnote{This ownership-sharing feature follows \cite{sun2020core}, but unlike their algorithm, we distinguish between agents who are initial owners of an object and those who acquire shared ownership. The outcomes of their algorithm may not belong to the consistent exclusion core. In Example \ref{Example:YRMH-IGYT} below, it can produce the allocation $ \mu $: endow $ a $ to agent $ 3 $, and $ b $ and $ c $ to agent $ 4 $; agent $ 4 $ takes $ c $, after which $ b $, having lost its only owner, passes to the remaining agents and is endowed to agent $ 2 $; agents $ 2 $ and $ 3 $ then exchange $ a $ and $ b $, and agent $ 1 $ receives nothing. Yet $ \mu $ is consistently exclusion blocked by $ \{1,3,4\} $ via $ \sigma $: agent $ 1 $ strictly gains; the unaffected group $ \{3,4\} $ is self-feasible, since agent $ 4 $ owns $ b $ and $ c $; and the only harmed agent, $ 2 $, loses $ a $, which $ \{1,3,4\} $ owns but $ \{3,4\} $ cannot control on its own.} Specifically, after a cycle is removed in a step, the departing agents' rights over remaining objects, whether held as owners or through acquired shared ownership, are shared with every remaining agent who owns, or has acquired shared ownership of, an object in the cycle. Second, in each step, each remaining object points to one of its remaining owners if any exist; otherwise it points to one of the remaining agents who have acquired shared ownership of it, if any exist; ties in both cases are broken by an exogenous order $ \rhd $. If neither type of agent exists, the object points to no one. We continue to call it YRMH-IGYT.

\begin{center}
	\textbf{YRMH-IGYT}
\end{center}

\begin{quote}
	\textbf{Notation}:
	After each step $ t $, let $ I(t) $ denote the set of remaining agents and $ O(t) $ the set of remaining objects. For each $ o\in O(t) $, let $ C_o(t) $ denote the set of remaining owners and $ S_o(t) $ the set of remaining agents who have acquired shared ownership of $ o$.

	\textbf{Initialization}: Let
	$ I(0)=I $, $ O(0)=O $, and for each $ o\in O $, $ C_o(0)=C_o $ and $ S_o(0)=\emptyset $. Fix a linear order of agents $ \rhd $. Let each $ o\in O $ initially point to the $ \rhd $-highest agent in $ C_o $.

	\textbf{Step} $ t\ge 1 $: If any remaining object still points to a remaining agent from the previous step, maintain every such pointing. If the agent pointed to by any $ o\in O(t-1) $ is removed in the previous step, then: if $ C_o(t-1)\neq \emptyset $, let $ o $ point to the $ \rhd $-highest agent in $ C_o(t-1) $;\footnote{Although the order of agents may change across steps in the algorithm, when we refer to the $ \rhd $-highest agent in a set, $\rhd$ denotes the initial order of agents chosen at the beginning of the algorithm.} if $ C_o(t-1)= \emptyset $ and $ S_o(t-1)\neq \emptyset $, let $ o $ point to the $ \rhd $-highest agent in $ S_o(t-1) $; otherwise, let $ o $ not point to any agent.

	Let the highest-ranked agent in the current order (which is $ \rhd $ if $ t=1 $), say $ i $, point to her favorite remaining object among $O(t-1)\cup \{o^*\}$, say $ o $.
	\begin{itemize}
		\item If $ o $ does not point to any agent, or $ o=o^* $, let $ i $ obtain $ o $ and then depart.

		\item If $ o $  points to an agent and a cycle forms, let every agent in the cycle obtain the object she points to and then depart.

		\item If $ o $  points to an agent but a cycle does not form, move the agent pointed to by $ o $ to the top of the current order of agents.
	\end{itemize}

   	\textbf{Updating ownership}: For each $ j\in I(t) $ and $ a\in O(t) $, if $ j\in C_a(t-1) $, let $ j\in C_a(t) $; if $ j\in S_a(t-1) $, let $ j\in S_a(t) $. Moreover, if a cycle is removed in step $ t $, for every $ j $ and every $ a $ involved in the cycle and for every $ b\in O(t) $ and $ j'\in I(t) $ satisfying $ j\in C_b(t-1)\cup S_b(t-1) $, $ j'\in C_a(t-1)\cup S_a(t-1) $, but $ j'\notin C_b(t-1)\cup S_b(t-1) $, let $ j'\in S_b(t) $. Go to the next step.
\end{quote}

	In each step, either at least one agent is removed or some agent is moved to the top of the order. The latter cannot recur indefinitely: each such step follows an object's pointing to an agent not yet at the top, and since there are finitely many agents, the chain must return to an agent already in it, at which point a cycle forms and its members are removed. Hence the algorithm terminates after finitely many steps. Different choices of the initial order $ \rhd $ may yield different allocations. We call them the \textit{outcomes} of YRMH-IGYT.

\begin{proposition}\label{prop:YRMH-IGYT}
	Every outcome of YRMH-IGYT belongs to the consistent exclusion core. Consequently, the consistent exclusion core is nonempty.
\end{proposition}

The converse fails: YRMH-IGYT may not find all elements of the consistent exclusion core. The following example shows this and illustrates how ownership is shared.

\begin{example}[Shared ownership in YRMH-IGYT]\label{Example:combinedfeature}
	Consider the following economy, in which agents $ 1 $ and $ 2 $ co-own object $ a $, agent $ 1 $ privately owns object $ b $, and agent $ 3 $ owns nothing.
	\begin{center}
		\begin{tabular}[t]{rcc}
			& $ a $ & $ b $  \\ \hline
			$ C_o $: & $ {1,2} $ & $ 1 $ \\ \hline
			$ \mu $: & $ 1 $ & $ 2 $ \\
			$ \sigma $: & $ 2 $ & $ 1 $\\
			$ \delta $: & $ 1 $ & $ 3 $
		\end{tabular}
		\quad \quad
		\begin{tabular}[t]{ccc}
			$ \succ_1 $	& $ \succ_2 $ & $ \succ_3 $  \\ \hline
			$ a $ & $ a $ & $ a $ \\
			$ b $ & $ b $ & $ b $
		\end{tabular}
	\end{center}

	With the initial order $ 1\rhd 3\rhd 2 $, the outcome of YRMH-IGYT is $ \mu $. In Step 1, $ 1 $ points to $ a $ and $ a $ points to $ 1 $. So, $ 1 $ obtains $ a $ and is removed; since $ 1 $ owns the remaining object $ b $ and $ 2 $ is a remaining owner of $ a $, agent $ 2 $ acquires shared ownership of $ b $. In Step 2, $ 3 $ points to $ b $ and $ b $ points to $ 2 $. So, $ 2 $ is moved to the top of the order. In Step 3, $ 2 $ points to $ b $ and $ b $ points to $ 2 $. So, $ 2 $ obtains $ b $ and is removed, and $ 3 $ then obtains $ o^* $. Without shared ownership, $ b $ would point to no agent in Step 2, and $ 3 $ would obtain it, which yields $ \delta $.

	In fact, no initial order leads to $ \delta $. If $ 2 $ precedes $ 1 $, then $ a $ points to $ 2 $ as long as she remains, and she does not leave empty-handed while $ a $ remains; so $ 1 $ never obtains $ a $. If $ 1 $ precedes $ 2 $, then $ 1 $ can obtain $ a $ only in a cycle while $ 2 $ and $ b $ remain; $ 2 $ then acquires shared ownership of $ b $, and any request by $ 3 $ for $ b $ moves $ 2 $ to the top of the order, where she obtains $ b $. Yet $ \delta $ belongs to the consistent exclusion core. Agent $ 1 $ cannot gain, and no block can take $ a $ from her, since only a coalition containing her controls $ a $. Agent $ 2 $ can gain $ b $ only through a block that harms $ 3 $ and so must control $ b $; such a block needs agent $ 1 $ as an unaffected member, but she is not self-feasible, since $ a $ is co-owned with agent $ 2 $.
\end{example}

In house allocation with existing tenants, for every initial order $\rhd$, YRMH-IGYT is a strategy-proof mechanism. In our general model, however, it is not strategy-proof, as the next example shows.

\begin{example}[YRMH-IGYT is not strategy-proof]\label{Example:YRMH-IGYT}
	Consider the following economy.
	\begin{center}
		\begin{tabular}{rccc}
			& $ a $ & $ b $ & $ c $ \\ \hline
			$ C_o $: & $ 1,3 $ & $ 4$ & $ 4 $\\ \hline
			$ \mu $: & $ 2 $ & $ 3 $ & $ 4 $\\
			$ \sigma $: & $ 1 $ & $ 3 $ & $ 4 $
		\end{tabular}
		\quad \quad
		\begin{tabular}{cccc}
			$ \succ_1 $	& $ \succ_2 $ & $ \succ_3 $ & $ \succ_4 $  \\ \hline
			$ b $ & $ a $ & $ b $ & $ c $ \\
			$ a $ & $ b $ & $ a $ & $ b $\\
			$ c $ & $ c $ & $ c $ & $ a $
		\end{tabular}
	\end{center}

With the initial order $ 2\rhd 4 \rhd 3 \rhd 1  $, the outcome of YRMH-IGYT is $ \sigma $. In Step 1, $ 2 $ points to $ a $ and $ a $ points to $ 3 $. So, $ 3 $ is moved to the top of the order. In Step 2, $ 3 $ points to $ b $ and $ b $ points to $ 4 $. So, $ 4 $ is moved to the top of the order. In Step 3, $ 4 $ points to $ c $ and $ c $ points to $ 4 $. So, $ 4 $ obtains $ c $ and is removed. In Step 4, $ 3 $ points to $ b $ and $ b $ does not point to any agent. So, $ 3 $ obtains $ b $ and is removed. In Step 5, $ 2 $ points to $ a $ and $ a $ points to $ 1 $. So, $ 1 $ is moved to the top of the order. In Step 6, $ 1 $ points to $ a $ and $ a $ points to $ 1 $. So, $ 1 $ obtains $ a $ and is removed.

Agent $ 2 $ thus receives nothing. If she instead reports $ b\succ'_2 a \succ'_2 c $, then she points to $ b $ in Step 1, which moves $ 4 $ to the top of the order; after $ 4 $ obtains $ c $, object $ b $ points to no agent, so $ 2 $ obtains it, and then $ 3 $ obtains $ a $ and $ 1 $ receives nothing. Hence $ 2 $ benefits from misreporting.
\end{example}

The manipulation in Example \ref{Example:YRMH-IGYT} works because object $ a $ has two owners. When $ 2 $ requests object $ a $ in the algorithm, both owners of object $ a $ get $ 2 $'s turn and thereby crowd out $ 2 $'s future opportunities. In contrast, in house allocation with existing tenants, every object that is not publicly owned has a single owner. If agent $ i $ requests agent $ j $'s private endowment and $ j $ then gets $ i $'s turn to receive an assignment without being involved in a cycle, $ i $ must receive $ j $'s remaining endowment in a subsequent step. 
Section \ref{section:incentives} shows that the failure is not specific to YRMH-IGYT.

\section{An impossibility result on strategy-proof selection}\label{section:incentives}

Section \ref{section:YRMH} shows that YRMH-IGYT is manipulable. The difficulty is in fact unavoidable: no strategy-proof mechanism can select from a nonempty solution that is consistent, individually rational, and Pareto efficient. All three are properties of the solution, defined in Section \ref{section:preliminary}; the mechanism that selects from it need not itself be consistent. Individual rationality, in particular, protects only an agent's claim to what she privately owns.

\begin{theorem}\label{theorem:general:strategyproof}
	No strategy-proof mechanism selects from a nonempty solution that is consistent, individually rational, and Pareto efficient.
\end{theorem}

Every nonempty consistent refinement of the exclusion core satisfies the theorem's hypotheses, since Pareto efficiency and individual rationality are both inherited from the exclusion core. We thus obtain the following corollary.

\begin{corollary}\label{cor:exclusion:strategyproof}
	No strategy-proof mechanism selects from any nonempty consistent refinement of the exclusion core.
\end{corollary}

The exclusion core itself, by contrast, admits a strategy-proof selection \citep{balbuzanov2019endowments}. It is therefore consistency, not the right to exclude, that strategy-proofness cannot accommodate. Theorem~\ref{theorem:general:strategyproof} can be restated to show what consistency adds. Let $ \mathcal{PI} $ select, in every economy, all Pareto efficient and individually rational allocations. Since Pareto efficiency is itself consistent, $ \Phi(\mathcal{PI}) $ consists of the Pareto efficient allocations that are individually rational and remain so after every self-feasible departure. By Lemma~\ref{lemma:maxrefinement}, every consistent, individually rational, and Pareto efficient solution, the consistent exclusion core included, is contained in $ \Phi(\mathcal{PI}) $, so Theorem~\ref{theorem:general:strategyproof} is equivalent to the statement that no strategy-proof mechanism selects from $ \Phi(\mathcal{PI}) $. What consistency adds to individual rationality is thus the protection of latent ownership claims: a co-owner becomes the private owner of an object once its other owners depart.

\begin{remark}[Independence of the requirements]\label{remark:independence}
	The three requirements on the solution in Theorem \ref{theorem:general:strategyproof} are independent. \emph{Consistency} cannot be dropped: BK's exclusion core is nonempty, individually rational, and Pareto efficient, and admits a strategy-proof selection; it is not consistent (Corollary \ref{prop:exclusioncore:consistency}). \emph{Individual rationality} cannot be dropped either: for any fixed order of agents, serial dictatorship is a consistent and Pareto efficient solution that is itself strategy-proof, and it fails only individual rationality, since it disregards ownership. \emph{Pareto efficiency} is also indispensable: a stable-matching correspondence with priorities favoring owners is consistent and individually rational, and admits a strategy-proof deferred-acceptance selection, yet it can contain Pareto inefficient allocations. Online Appendix \ref{appendix:independence} presents the details for the last two claims.
\end{remark}

The full domain is also essential: Section~\ref{section:special} shows that the conflict disappears once intricate co-ownership is absent. When every object is privately or publicly owned, YRMH-IGYT is strategy-proof and its outcomes form exactly the consistent exclusion core.

\section{Economies with simple ownership}\label{section:special}

This section examines economies with simple ownership, in which every object is either privately or publicly owned. The class of these economies is closed under self-feasible departures, since an object whose owners all leave becomes publicly owned, so consistency can be assessed within it. We show that the exclusion core is consistent since it coincides with the consistent exclusion core, and both consist exactly of YRMH-IGYT outcomes. Moreover, for any fixed initial order, YRMH-IGYT is strategy-proof, and in fact group strategy-proof.\footnote{\label{fn:ishidapark}\cite{ishida2025group} also obtain group strategy-proof mechanisms, on a different domain: agents are partitioned into groups, each collectively owning as many objects as it has members, and there is no outside option. Their domain admits intricate co-ownership and overlaps ours only in the housing market and house allocation. On their domain, a self-feasible departure from a Pareto efficient allocation removes whole groups with exactly their own objects, so the exclusion core is consistent there and their mechanisms select from the consistent exclusion core. Simple ownership is thus sufficient, but not necessary, for the positive results of this section.}

\begin{proposition}\label{prop:specialsetting}
	When every object is either privately owned or publicly owned:
	\begin{enumerate}
		\item The consistent exclusion core coincides with the exclusion core, and both coincide with the set of YRMH-IGYT outcomes;
		\item For any fixed initial order, YRMH-IGYT is group strategy-proof. \label{prop:YRMH:strategyproof:special}
	\end{enumerate}
\end{proposition}

The first statement adds consistency to BK's results for this domain. BK show that the exclusion core equals the set of outcomes of their generalized TTC algorithm and, in house allocation with existing tenants, the set of YRMH-IGYT outcomes. Proposition~\ref{prop:specialsetting} shows that the exclusion core is moreover consistent, and that YRMH-IGYT, which may miss elements of the consistent exclusion core under intricate co-ownership (Example~\ref{Example:combinedfeature}), finds all of them throughout this domain.

Together with Theorem~\ref{theorem:general:strategyproof}, Proposition~\ref{prop:specialsetting} shows that the impossibility on the full domain stems from intricate co-ownership, not from public ownership or from an agent privately owning several objects:

\begin{corollary}\label{cor:special:strategyproof}
	When every object is either privately owned or publicly owned, the exclusion core is consistent and the consistent exclusion core admits a strategy-proof selection.
\end{corollary}

To prove Proposition \ref{prop:specialsetting}, we classify the economies of this section into three classes, each generalizing a canonical model: an economy is called (1) \textit{public-ownership} if $C_o=I$ for each $o\in O$, generalizing house allocation; (2) \textit{private-ownership} if $ |C_o|=1 $ for each $ o\in O $, generalizing the housing market; (3) \textit{mixed-ownership}, generalizing house allocation with existing tenants, if there is a nonempty $ O' \subsetneq O$ such that $ |C_o|=1 $ for each $ o\in O' $ and $ C_o=I $ for each $ o\in O\backslash O' $. 

In public-ownership economies, YRMH-IGYT becomes a serial dictatorship algorithm, so it is group strategy-proof, and by varying the initial order, it can find all Pareto efficient allocations. Since both the exclusion core and the consistent exclusion core are Pareto efficient and contain all YRMH-IGYT outcomes, the two cores must coincide and equal the set of YRMH-IGYT outcomes. 

In private-ownership economies, for every allocation in the exclusion core, our proof explicitly constructs an initial order and shows that the YRMH-IGYT procedure must find that allocation. Since the exclusion core contains all YRMH-IGYT outcomes, the exclusion core must equal the set of YRMH-IGYT outcomes. Since the consistent exclusion core lies between them, it must therefore coincide with them. YRMH-IGYT is group strategy-proof because its outcome coincides with that of a TTC algorithm with fixed object priorities, which is group strategy-proof.

Our proof for mixed-ownership economies rests on an invariance property of the three solutions after reformulating mixed ownership as private ownership, which might be of independent interest: for all three solutions, a publicly owned object behaves as if it were privately owned by an agent who wants nothing.

Specifically, for each mixed-ownership economy $ \Gamma=(I,O,\succ_I,\{C_o\}_{o\in O}) $, let $ O^P $ denote the set of publicly owned objects. We construct a private-ownership economy $ \Gamma^P=(I\cup \{i^P\},O,\succ_{I\cup \{i^P\}},\{C^P_o\}_{o\in O}) $ by adding an artificial agent $ i^P\in \I\backslash I $ who most prefers $ o^* $ and privately owns every object in $ O^P $: $ C^P_o=\{i^P\} $ for each $ o\in O^P $, and $ C^P_o=C_o $ for each $ o\in O\backslash O^P $. For every allocation $ \mu $ in $ \Gamma $, let $ \mu^P $ denote its extension to $ \Gamma^P $, which assigns $ o^* $ to $ i^P $ and agrees with $ \mu $ on $ I $. 

	\begin{lemma}\label{lemma:consistent}
	For every mixed-ownership economy $ \Gamma $, the exclusion core of $ \Gamma^P $ consists of the extensions of the exclusion core allocations of $ \Gamma $. The same holds for the consistent exclusion core and for the set of YRMH-IGYT outcomes.
\end{lemma} 

The intuition is as follows. In a Pareto efficient allocation of any mixed-ownership economy $ \Gamma $, an agent holding a public endowment cannot be excluded by any proper subcoalition of $ I $. Treating public endowments as privately owned by $ i^P $, who can never gain strictly, therefore leaves the exclusion core unchanged, and also the consistent exclusion core, since condition (3) of Definition \ref{definition:C1core} would prevent $ i^P $ from exercising her own exclusion right.\footnote{Unlike the artificial agent of Proposition \ref{prop:fragility}, who is interposed between the original owners and the objects they own, $ i^P $ takes over only objects from which no proper subcoalition could exclude anyone in the first place, and so leaves all three solutions unchanged.} For YRMH-IGYT, the outcomes in $ \Gamma $ and $ \Gamma^P $ agree on $ I $ whenever the initial orders rank $ I $ identically, although the procedures may differ: once $ i^P $ has left, a public endowment points to no agent in $ \Gamma^P $ but to the $ \rhd $-highest remaining agent in $ \Gamma $, so a chain of promotions that closes into a single cycle in $ \Gamma $ is resolved one agent at a time in $ \Gamma^P $.

Lemma \ref{lemma:consistent} implies that the coincidence among the three solutions extends from private-ownership economies to mixed-ownership economies, and YRMH-IGYT with any initial order is group strategy-proof.

The equivalence in Lemma \ref{lemma:consistent} is specific to exclusion rights. Under the strong core, $ i^P $ can join a blocking coalition as an unaffected member and lend it her ownership of $ O^P $, so the strong core of $ \Gamma^P $ can be empty even when that of $ \Gamma $ is not.\footnote{Let $ I=\{1,2\} $ and $ O=\{a,b\} $, where agent $ 1 $ privately owns $ a $ and $ b $ is publicly owned, and let both agents rank $ b\succ a\succ o^* $. The Pareto efficient allocations are the two that assign both objects, and both belong to the strong core of $ \Gamma $. In $ \Gamma^P $, where $ i^P $ privately owns $ b $, the coalition $ \{1,i^P\} $ weakly blocks the extension of the allocation that gives $ b $ to agent $ 2 $, and $ \{2,i^P\} $ weakly blocks the extension of the other, so the strong core of $ \Gamma^P $ is empty.} Since $ \Gamma^P $ has simple ownership, the strong core may be empty even on this domain.

\section{Comparison with the exchange-rights view of ownership}\label{section:Sunetal}

The previous sections have developed the consistent exclusion core within BK's \emph{exclusion-rights} view of ownership. This section compares it with the two concepts that \cite{sun2020core} propose for the same model under the competing \emph{exchange-rights} view, in which a blocking coalition may only reallocate the objects it owns. The two views share a common structure: in each, the base core is not consistent, and a sharper consistent concept arises by permitting blocks that consistency alone does not require. Proofs and the remaining formal definitions are in Online Appendices~\ref{appendix:rectified} and~\ref{appendix:refinedEC:definition}.

\paragraph{\citeauthor{sun2020core}'s concepts.} Following BK, \cite{sun2020core} study the same co-ownership model but pursue a different route: within the exchange-rights view, they restrict how self-feasible unaffected agents' redundant endowments may support a blocking coalition and define the following concept.

\begin{definition}[\cite{sun2020core}]\label{definition:effectivecore}
	In an economy, an allocation $ \mu $ is \emph{effectively blocked} by a coalition $ C $ via another allocation $ \sigma $ if (1) $ \forall i\in C $, $ \sigma(i) \succsim_i \mu(i) $ and $ \exists j\in C $, $ \sigma(j)\succ_j\mu(j) $; (2) $ \sigma(C)\subseteq \w(C)\cup \{o^*\}$ and, in addition, 
	\begin{itemize}
	\item[(3)] for every self-feasible $ C'\subseteq C_{\sigma=\mu} $ and every $ o\in \w(C')\backslash \sigma(C') $, if $ C\subsetneq I $, then $ o\notin \sigma(C) $.
	\end{itemize}
	The \textbf{effective core} consists of allocations that are not effectively blocked.
\end{definition}

The effective core therefore contains the strong core, and it is contained in the weak core. \cite{sun2020core} show that it is nonempty and Pareto efficient, and construct a strategy-proof mechanism that selects from it. It is plainly individually rational, so Theorem \ref{theorem:general:strategyproof} implies that it is not consistent.

\begin{corollary}\label{cor:effective:inconsistency}
	The effective core of \cite{sun2020core} is not consistent.
\end{corollary}

The effective core's violation of consistency can be demonstrated via Example \ref{Example:YRMH-IGYT}, in which the effective core admits the allocation $\mu$, yet it is hard to defend: object $ a $, co-owned by $ \{1,3\} $, is consumed by agent $ 2 $, who owns nothing, while the co-owner $ 1 $ receives nothing and wants it. The natural block exists---$ \{1,3,4\} $ weakly blocks $ \mu $ via $ \sigma $---but the effective core censors it: condition (3) of Definition \ref{definition:effectivecore} forbids the coalition from allocating $ b $, the object that the unaffected agent $ 4 $ owns but does not consume. That protection lasts only while agent $ 4 $ is present, and she is among the agents entitled to leave: once $ \{3,4\} $ departs with its assignments, agent $ 1 $ privately owns $ a $ and agent $ 2 $ owns nothing, yet $ \mu_{\{1,2\}} $ violates even individual rationality in the reduced economy.

BK's exclusion core and \citeapos{sun2020core} effective core thus share the same defect: neither is coherent across populations. By Lemma \ref{lemma:maxrefinement}, each nonetheless has a largest consistent refinement. For the exclusion core, Theorem \ref{theorem:C1core:char} identifies it as the consistent exclusion core. For the effective core, the largest consistent refinement contains \citeapos{sun2020core} second concept, to which we now turn.

Their second concept, the \textbf{rectified core}, replaces condition (3) of Definition \ref{definition:effectivecore} with
\begin{itemize}
\item[(3')] for every self-feasible $ C' \subseteq C_{\sigma=\mu} $ and every $ o\in \w(C')\backslash \sigma(C') $, if $ o\in \mu(I\backslash C) $, then $ o\notin \sigma(C) $.
\end{itemize}
That is, a redundant object of a self-feasible unaffected subcoalition $C'$ cannot be reassigned within the blocking coalition $C$ when it is currently assigned to an outsider. This condition is more permissive than that under the effective core, which does not permit such a reassignment of the object as long as there exists an agent outside the coalition. 

The rectified core is consistent, but it goes strictly beyond what consistency requires.

\begin{proposition}\label{prop:rectifiedcore:consistency}
	The rectified core of \cite{sun2020core} is consistent, but it is not the largest consistent refinement of the effective core.
\end{proposition}

The proof of Proposition \ref{prop:rectifiedcore:consistency} is in Online Appendix \ref{appendix:rectifiedcore}. \cite{sun2020core} have shown that the rectified core is nonempty. Since it is Pareto efficient, individually rational, and consistent, Theorem \ref{theorem:general:strategyproof} yields the following.

\begin{corollary}\label{cor:sun:strategyproof}
	The rectified core admits no strategy-proof selection.
\end{corollary}

\paragraph{Comparison.} The consistent exclusion core does not have a set-inclusion relationship with the two cores of \cite{sun2020core}, because exclusion rights and exchange rights authorize different blocks. In Example \ref{Example:HET} of Online Appendix \ref{appendix:comparisons}, the allocation $ \mu $ belongs to the rectified core and hence to the effective core, but not to the exclusion core and hence not to the consistent exclusion core. Conversely, in Example \ref{Example:combinedfeature}, the allocation $ \delta $ belongs to the consistent exclusion core but not to the effective core, and therefore not to the rectified core either.

The rectified core sharpens the effective core by permitting blocks that consistency alone does not require. The exclusion-rights view admits a similar sharpening, and Example \ref{Example:combinedfeature} illustrates the motivation: the allocation $ \delta $ gives $ b $ to agent $ 3 $, who owns nothing, and nothing to agent $ 2 $, who co-owns $ a $; the coalition $ \{1,2\} $ owns both objects but cannot consistently exclusion block $ \delta $, because the unaffected agent $ 1 $ is not self-feasible and thus kept out of a blocking coalition by condition (2) of Definition \ref{definition:C1core}.

Online Appendix \ref{appendix:refinedEC:definition} introduces the refinement of the consistent exclusion core that we find most natural, which permits an unaffected agent to join a blocking coalition $ C $ exactly when her assignment is owned by $ C $, so that the coalition can support its unaffected members out of its own endowment. We show that this condition means that every unaffected agent outside every self-feasible subcoalition of $ C_{\sigma=\mu}  $ must rely, directly or indirectly, on the ownership of $ C_{\sigma>\mu}  $ to keep her assignment (Lemma \ref{lemma:secondcondition}), which may justify her participation. The resulting concept, the \emph{autonomous exclusion core}, is consistent, contains every outcome of YRMH-IGYT, and is contained in the consistent exclusion core (Theorem \ref{thm:refined EC}); although it is built on exclusion rights, it is also contained in the two cores of \cite{sun2020core} (Proposition \ref{prop:subset:rectified core}). These results, together with Proposition \ref{prop:specialsetting}, imply that BK's original exclusion core already predicts at least as sharply as its exchange-rights counterparts when intricate co-ownership is absent.\footnote{By Theorem \ref{thm:refined EC}, the autonomous exclusion core contains the set of YRMH-IGYT outcomes and is contained in the exclusion core. When every object is privately or publicly owned, these two sets coincide by Proposition \ref{prop:specialsetting}, so the exclusion core coincides with the autonomous exclusion core, which Proposition \ref{prop:subset:rectified core} places in the rectified core.}

\begin{corollary}\label{cor:special:rectified}
	When every object is either privately owned or publicly owned, the exclusion core is contained in the rectified core and hence in the effective core.
\end{corollary}

Table~\ref{table:comparison} summarizes the properties of the core concepts discussed in this section.

\begin{table}[H]
	\centering
	\footnotesize
	\setlength{\tabcolsep}{4pt}
	\begin{tabular}{l|cccc|ccc}
		\hline
		& \multicolumn{4}{c|}{Exchange rights} & \multicolumn{3}{c}{Exclusion rights} \\
		Core & Strong & Weak & Effective & Rectified & Exclusion & Consistent & Autonomous \\ \hline
		Nonempty & $\times$ & $\checkmark$ & $\checkmark$ & $\checkmark$ & $\checkmark$ & $\checkmark$ & $\checkmark$ \\
		Pareto efficient & $\checkmark$ & $\times$ & $\checkmark$ & $\checkmark$ & $\checkmark$ & $\checkmark$ & $\checkmark$ \\
		Consistent & $\checkmark$ & $\times$ & $\times$ & $\checkmark$ & $\times$ & $\checkmark$ & $\checkmark$ \\
		Strategy-proof selection & -- & $\checkmark$ & $\checkmark$ & $\times$ & $\checkmark$ & $\times$ & $\times$ \\ \hline
	\end{tabular}
	\caption{--: not applicable, since the strong core may be empty. YRMH-IGYT outcomes $\subseteq$ autonomous $\subseteq$ consistent $\subseteq$ exclusion core; strong $\subseteq$ rectified $\subseteq$ effective $\subseteq$ weak core; autonomous $\subseteq$ rectified core. When every object is privately or publicly owned, the first chain holds with equality.}\label{table:comparison}
\end{table}

\section{Conclusion}\label{section:discussion}

Under co-ownership, a core concept must decide what rights ownership confers, and consistency asks whether its interpretation of ownership survives when some agents leave with their assignments. For BK's exclusion core, the exclusion-rights interpretation survives but its implementation does not: the rule that keeps every unaffected agent out of a blocking coalition makes the exclusion core's prediction hinge on agents who have nothing at stake (Proposition~\ref{prop:fragility}). Consistency, usually a test that a solution passes or fails, here also shows how to repair this rule while leaving BK's exclusion rights unchanged: a self-feasible unaffected group may join a blocking coalition only to complete the coalition's joint exclusion right, never to exercise its own. The resulting consistent exclusion core is the largest consistent refinement of the exclusion core (Theorem \ref{theorem:C1core:char}).

Our results also indicate where the difficulty of co-ownership arises. Once intricate co-ownership is admitted, consistency and immunity to manipulation are incompatible: no strategy-proof mechanism selects from a nonempty solution that is consistent, individually rational, and Pareto efficient (Theorem \ref{theorem:general:strategyproof}). When every object is privately or publicly owned, the conflict disappears: the exclusion core is already consistent, coincides with the set of YRMH-IGYT outcomes, and YRMH-IGYT is strategy-proof (Proposition \ref{prop:specialsetting}). Simple ownership also reconciles the two views of ownership: with intricate co-ownership, exclusion rights and exchange rights authorize different blocks, and the consistent exclusion core stands in no inclusion relation to \citeapos{sun2020core} concepts in general; with simple ownership, BK's exclusion core already predicts at least as sharply as both (Corollary \ref{cor:special:rectified}).

Several questions remain open. Theorem~\ref{theorem:general:strategyproof} concerns the full domain, and simple ownership is sufficient but not necessary for strategy-proof selection from the consistent exclusion core; characterizing the ownership structures on which YRMH-IGYT, or another selection, is (group) strategy-proof would locate exactly where the conflict between consistency and incentives begins. Under simple ownership, the outcomes of YRMH-IGYT are exactly the elements of the consistent exclusion core, but under intricate co-ownership they need not be; whether some family of procedures generates exactly the consistent exclusion core in general is open. Finally, Theorem~\ref{theorem:C1core:char} describes the consistent exclusion core relative to BK's concept; an axiomatization from primitive properties, in the spirit of \cite{peleg1985axiomatization} for cooperative games and \cite{sasaki1992consistency} for the marriage model, would identify it without reference to the exclusion core.

		\appendix	

\section{Appendix: Proofs}\label{appendix:proofs}

	\subsection{Proofs of results in Section \ref{section:preliminary}}\label{appendix:preliminary}

	\begin{proof}[Proof of Lemma \ref{lemma:maxrefinement}]
		We first prove a composition property of reductions. Let $ \mu $ be an allocation in an economy $ \Gamma $. Let $ I_1\subsetneq I $ be nonempty and self-feasible under $ \mu $ in $\Gamma$, and let $ I_2\subsetneq I\backslash I_1 $ be nonempty and  self-feasible under $ \mu_{I\backslash I_1} $ in $ \Gamma_1=\Gamma(\mu,I\backslash I_1) $. We then show that $ I_1\cup I_2 \subsetneq I$ is self-feasible under $ \mu $ in $ \Gamma $, and $\Gamma_1\bigl(\mu_{I\backslash I_1},\,I\backslash(I_1\cup I_2)\bigr)=\Gamma\bigl(\mu,\,I\backslash(I_1\cup I_2)\bigr)$.

		Write $ R_1=I\backslash I_1 $ and $ R=I\backslash(I_1\cup I_2) $, and let $ \{C'_o\} $ denote the ownership structure of $ \Gamma_1 $. Since $ I_2\subsetneq R_1 $, $ R $ is nonempty, so $ I_1\cup I_2 $ is a nonempty proper subset of $ I $. For $ i\in I_1 $, self-feasibility of $ I_1 $ gives $ \mu(i)\in\w(I_1)\cup\{o^*\}\subseteq\w(I_1\cup I_2)\cup\{o^*\} $. For $ i\in I_2 $ with $ \mu(i)=o\neq o^* $, self-feasibility of $ I_2 $ in $ \Gamma_1 $ gives $ C'_o\subseteq I_2 $. If $ C_o\cap R_1\neq\emptyset $, then $ C'_o=C_o\cap R_1\subseteq I_2 $, so $ C_o\subseteq I_1\cup I_2 $ and $ o\in\w(I_1\cup I_2) $. The case $ C_o\cap R_1=\emptyset $ is impossible: the object $ o $ would be publicly owned in $ \Gamma_1 $, so $ C'_o=R_1\subseteq I_2 $, contradicting $ I_2\subsetneq R_1 $. Hence $ I_1\cup I_2 $ is self-feasible under $ \mu $.

		Consider the ownership of any remaining object $ o\in O\backslash\mu(I_1\cup I_2) $. In $\Gamma\bigl(\mu,\,I\backslash(I_1\cup I_2)\bigr)$, the set of owners of $o$ is $ C_o\cap R $ if this set is nonempty, and $ R $ otherwise. In $\Gamma_1\bigl(\mu_{I\backslash I_1},\,I\backslash(I_1\cup I_2)\bigr)$, the set of owners of $o$ is $ C''_o=C'_o\cap R $ if this set is nonempty, and $ R $ otherwise. If $ C_o\cap R\neq\emptyset $, then $ C_o\cap R_1\supseteq C_o\cap R\neq\emptyset $, so $ C'_o=C_o\cap R_1 $ and $ C''_o=C_o\cap R_1\cap R=C_o\cap R $. If $ C_o\cap R=\emptyset $ and $ C_o\cap R_1\neq\emptyset $, then $ C'_o=C_o\cap R_1 $ and $ C'_o\cap R\subseteq C_o\cap R=\emptyset $, so $ C''_o=R $. If $ C_o\cap R_1=\emptyset $, then $ C'_o=R_1 $ and $ C'_o\cap R=R\neq\emptyset $, so $ C''_o=R $. In every case $ C''_o $ coincides with the set of owners of $o$ in $\Gamma\bigl(\mu,\,I\backslash(I_1\cup I_2)\bigr)$. So, $\Gamma_1\bigl(\mu_{I\backslash I_1},\,I\backslash(I_1\cup I_2)\bigr)=\Gamma\bigl(\mu,\,I\backslash(I_1\cup I_2)\bigr)$.

		We then prove the lemma. The inclusion $ \Phi(f)\subseteq f $ holds by definition. For maximality, let $ g $ be a consistent solution with $ g\subseteq f $, and let $ \mu\in g(\Gamma) $. For every nonempty self-feasible $ I'\subsetneq I $ under $ \mu $, consistency of $ g $ gives $ \mu_{I\backslash I'}\in g(\Gamma(\mu,I\backslash I'))\subseteq f(\Gamma(\mu,I\backslash I')) $. Hence $ \mu\in\Phi(f)(\Gamma) $, and thus $ g\subseteq\Phi(f) $.

		It remains to prove that $ \Phi(f) $ is consistent. Let $ \mu\in\Phi(f)(\Gamma) $ and let a nonempty $ I_1\subsetneq I $ be self-feasible under $ \mu $; write $ \Gamma_1=\Gamma(\mu,I\backslash I_1) $ and $ \mu_1=\mu_{I\backslash I_1} $. We show $ \mu_1\in\Phi(f)(\Gamma_1) $. First, $ \mu_1\in f(\Gamma_1) $, since $ I_1 $ is one of the departures tested in the definition of $ \Phi(f)(\Gamma) $. Second, let a nonempty $ I_2\subsetneq I\backslash I_1 $ be self-feasible under $ \mu_1 $ in $ \Gamma_1 $. By the composition property just established, $ I_1\cup I_2 $ is a nonempty proper self-feasible group under $ \mu $ in $ \Gamma $, and $ \Gamma_1(\mu_1,I\backslash(I_1\cup I_2))=\Gamma(\mu,I\backslash(I_1\cup I_2)) $. Since $ \mu\in\Phi(f)(\Gamma) $, the restriction of $ \mu_1 $ to $ I\backslash(I_1\cup I_2) $, which equals the restriction of $ \mu $, belongs to $ f(\Gamma(\mu,I\backslash(I_1\cup I_2))) $. Hence $ \mu_1\in\Phi(f)(\Gamma_1) $.
	\end{proof}

	\subsection{Proofs of results in Section \ref{section:exclusion}}\label{appendix:Section3}

	\begin{proof}[Proof of Proposition \ref{prop:fragility}]
		We first record two facts used throughout. In every Pareto efficient allocation of $ \Gamma^* $, the artificial agent $ i^* $ receives $ o^* $. Every Pareto efficient allocation of $ \Gamma^* $ is therefore the augmentation of an allocation of $ \Gamma $, and $ \nu^* $ is Pareto efficient in $ \Gamma^* $ if and only if $ \nu $ is Pareto efficient in $ \Gamma $. Moreover, let $ \w^* $ denote the endowment function of $ \Gamma^* $. Since $ C^*_o=C_o\cup\{i^*\} $ for every $ o\in O $, every $ C\subseteq I $ satisfies $ \w^*(C)=\emptyset $ and $ \w^*(C\cup\{i^*\})=\w(C) $.

		\textbf{The strong core.} For every allocation $ \nu $ of $ \Gamma $, its augmentation $ \nu^* $ is weakly blocked in $ \Gamma^* $ if and only if $ \nu $ is weakly blocked in $ \Gamma $. Indeed, if $ C $ weakly blocks $ \nu $ via $ \sigma $, then $ C\cup\{i^*\} $ weakly blocks $ \nu^* $ via the augmentation of $ \sigma $, since $ i^* $ is unaffected and $ \w^*(C\cup\{i^*\})=\w(C) $. Conversely, suppose that $ C $ weakly blocks $ \nu^* $ via $ \sigma $. If $ i^*\notin C $, then $ \w^*(C)=\emptyset $, so $ \sigma(C)\subseteq\{o^*\} $, and $ C $ weakly blocks $ \nu $ in $ \Gamma $ via $ \sigma_I $. If $ i^*\in C $, then $ \sigma(i^*)=o^* $, since $ \succ_{i^*} $ ranks $ o^* $ first, and $ C\backslash\{i^*\} $, which contains the strictly better-off member of $ C $, weakly blocks $ \nu $ via $ \sigma_I $, because $ \w^*(C)=\w(C\backslash\{i^*\}) $. Hence the augmentation of an allocation of $ \Gamma $ belongs to the strong core of $ \Gamma^* $ if and only if that allocation belongs to the strong core of $ \Gamma $.

		\textbf{The exclusion core.} We prove that it equals the set of Pareto efficient allocations in $ \Gamma^* $; by the first fact, these are the augmentations of the Pareto efficient allocations of $ \Gamma $. The exclusion core is Pareto efficient, which gives one inclusion. For the other, let $ \nu $ be Pareto efficient and suppose that $ \nu $ is exclusion blocked by a coalition $ C $ via another $ \sigma $. Since $ \nu $ is Pareto efficient, some agent outside $ C $ is worse off, and that agent must lose an object in $ \Omega(C|\w^*,\nu) $. If $ i^*\notin C $, then $ \w^*(C)=\emptyset $ and $ \Omega(C|\w^*,\nu)=\emptyset $; hence $ i^*\in C $. But $ \nu(i^*)=o^* $ by the first fact above, and $ o^* $ is the most preferred outcome of $ i^* $, so $ i^* $ cannot be made strictly better off and cannot belong to $ C $, a contradiction. Finally, if every object of $ \Gamma^* $ were publicly owned, no coalition other than the grand coalition would own anything, so the exclusion core would again be the set of Pareto efficient allocations of $ \Gamma^* $.
	\end{proof}

	\begin{proof}[Proof of Proposition \ref{prop:strongcore:consistency}]
		\textbf{The strong core.} Consider any economy $ \Gamma $. Suppose that the strong core is nonempty in $ \Gamma $ and $ \mu $ is any allocation within it. Let a nonempty $ I'\subsetneq I $ be self-feasible under $ \mu $. We want to prove that $ \mu_{I\backslash I'} $ belongs to the strong core of $ \Gamma(\mu,I\backslash I') $. 

		Suppose that in $ \Gamma(\mu,I\backslash I') $, $ \mu_{I\backslash I'} $ is weakly blocked by a coalition $ C\subseteq  I\backslash I'$ via another $ \sigma' $. Since $ \mu $ is Pareto efficient in $ \Gamma $, $ \mu_{I\backslash I'} $ must be Pareto efficient in $ \Gamma(\mu,I\backslash I') $. So, $ C\subsetneq I\backslash I' $. We then prove that in $ \Gamma $, $ \mu $ is weakly blocked by $ C\cup I' $ via an allocation $ \sigma $ that agrees with $ \sigma' $ on $ I\backslash I' $ and with $ \mu $ on $ I' $, a contradiction.

		We verify the two blocking conditions. First, every member of $ C\cup I' $ is weakly better off under $ \sigma $ than under $ \mu $, and at least one member of $ C $ is strictly better off.
		Second, since $ \mu_{I\backslash I'} $ is weakly blocked by $ C$ via $ \sigma' $ in $ \Gamma(\mu,I\backslash I') $, $ \sigma'(C) \subseteq \w'(C) \cup \{o^*\}$, where $ \w' $ denotes the endowment function in $ \Gamma(\mu,I\backslash I') $. We claim that $ \w'(C)\subseteq \w(C\cup I') $. Let $ o\in\w'(C) $, that is, $ C'_o\subseteq C $. If $ C_o\cap(I\backslash I')=\emptyset $, then $ o $ is publicly owned in the reduced economy, so $ C'_o=I\backslash I'\subseteq C $, contradicting $ C\subsetneq I\backslash I' $. Hence $ C_o\cap(I\backslash I')\neq\emptyset $, so $ C'_o=C_o\cap(I\backslash I')\subseteq C $, giving $ C_o\subseteq C\cup I' $ and $ o\in\w(C\cup I') $. Since also $ \w(I') \subseteq \w(C\cup I')$, we have $ \w'(C)\cup \w(I')\subseteq \w(C\cup I') $. So, $ \sigma(C\cup I')=\sigma'(C)\cup \mu(I')\subseteq \w'(C)\cup \w(I') \cup \{o^*\} \subseteq \w(C\cup I') \cup \{o^*\} $. 

		\textbf{The weak core.} To show that it is not consistent, consider a variant of Example \ref{Example:Kingdom} in which agent $ 1 $ owns an additional object $ c $ and all agents have the preferences $ a\succ b \succ c $. Then, the allocation $ \mu $ in which $ 1 $ receives $ a $ and other agents receive $ o^* $ belongs to the weak core. However, agent $ 1 $ is self-feasible under $ \mu $, and once she departs with object $ a $, objects $ b $ and $ c $ lose all of their owners and become publicly owned in the reduced economy. The two remaining agents can then both become strictly better off, with agent $ 2 $ receiving $ b $ and agent $ 3 $ receiving $ c $, so the restriction of $ \mu $ is strongly blocked and no longer belongs to the weak core.
	\end{proof}

		\subsection{Proofs of results in Section \ref{section:consistentEC}}\label{appendix:proofsEC}

	\begin{proof}[Proof of Lemma \ref{lemma:firstcondition}]
		Write $ \mu'=\mu_{I\backslash C^\diamond} $. We first prove that $ \Omega(C|\w,\mu)= \Omega(C_{\sigma=\mu}|C,\w,\mu) \cup \Omega(C_{\sigma>\mu}|\w',\mu') $.

	  Since $ C^\diamond $ is the set of agents given by \eqref{eq:cdiamond} such that $ \w(C^\diamond)= \Omega(C_{\sigma=\mu}|C,\w,\mu)$, all objects in $ \w(C^\diamond) $ are controlled by $C_{\sigma=\mu}$ in the original economy. So $ \w(C^\diamond)\subseteq \Omega(C_{\sigma=\mu}|\w,\mu) \subseteq \Omega(C|\w,\mu) $. Since $ \w' $ is the endowment function after the removal of $ C^\diamond $, every object controlled by $C_{\sigma>\mu} $ in the reduced economy must be controlled by $C_{\sigma>\mu} \cup C_{\sigma=\mu}$ in the original economy. So $ \Omega(C_{\sigma>\mu}|\w',\mu') \subseteq \Omega(C|\w,\mu) $. Therefore, we obtain $ \Omega(C_{\sigma=\mu}|C,\w,\mu) \cup \Omega(C_{\sigma>\mu}|\w',\mu') \subseteq  \Omega(C|\w,\mu)$. It remains to prove the other direction.

		Let $ \Omega(C|\w,\mu)=\w(\cup_{k=0}^\infty C^k) $, where $ C^0=C $ and $ C^k=C^{k-1}\cup \{i\in I\backslash C^{k-1}: \mu(i)\in \w(C^{k-1}) \} $ for every $ k\ge 1 $. We first check that $ \w(C)\subseteq \w(C_{\sigma=\mu})\cup \w'(C_{\sigma>\mu}) $. Let $ o\in\w(C) $. If $ C_o\subseteq C_{\sigma=\mu} $, then $ o\in\w(C_{\sigma=\mu}) $. Otherwise $ C_o \cap C_{\sigma>\mu}\neq \emptyset $, which is disjoint from $ C^\diamond $. Then $ o $ survives in $ \Gamma(\mu,I\backslash C^\diamond) $, and the owners of $ o $ in the reduced economy are therefore $ C_o\cap(I\backslash C^\diamond) $, a nonempty subset of $ C_{\sigma>\mu} $. So $ o\in\w'(C_{\sigma>\mu}) $. 

		For every $ k\ge 1 $, since $ \mu(C^\diamond)\subseteq\w(C^\diamond)\cup\{o^*\} $ and $ \w(C^\diamond)=\Omega(C_{\sigma=\mu}|C,\w,\mu) $,  we may restrict attention to $ \hat{C}^k=(C^k\backslash C^{k-1})\backslash C^\diamond $, and it suffices to prove that $ \mu(\hat{C}^k)\subseteq \Omega(C_{\sigma>\mu}|\w',\mu') $. We argue by induction on $ k $. Since $ C\cap C^\diamond=C_{\sigma=\mu} $,
		\[
		C^{k-1}\backslash C^\diamond=C_{\sigma>\mu}\cup\hat{C}^1\cup\cdots\cup\hat{C}^{k-1}\quad\text{for every }k\ge 1.
		\]
		Write $ D $ for the group reached by the recursion defining $ \Omega(C_{\sigma>\mu}|\w',\mu') $, so that this set equals $ \w'(D) $. Then $ C_{\sigma>\mu}\subseteq D $, and $ D $ is closed: any agent whose assignment is owned by $ D $ in $ \Gamma(\mu,I\backslash C^\diamond) $ belongs to $ D $. Let $ k\ge 1 $ and suppose that $ \mu(\hat{C}^\ell)\subseteq \w'(D) $ for every $ 1\le\ell<k $. Let $ i\in\hat{C}^k $. Then $ C_{\mu(i)}\subseteq C^{k-1} $, and $ C_{\mu(i)}\not\subseteq C^\diamond $. The owners of $ \mu(i) $ in $ \Gamma(\mu,I\backslash C^\diamond) $ are therefore $ C_{\mu(i)}\cap(I\backslash C^\diamond) $, a nonempty subset of $ C^{k-1}\backslash C^\diamond=C_{\sigma>\mu}\cup\hat{C}^1\cup\cdots\cup\hat{C}^{k-1} $. Every agent in that set either belongs to $ C_{\sigma>\mu}\subseteq D $ or, by the induction hypothesis, holds an object in $ \w'(D) $ and so belongs to $ D $ by closure. Hence $ D $ contains every owner of $ \mu(i) $, giving $ \mu(i)\in\w'(D)=\Omega(C_{\sigma>\mu}|\w',\mu') $.  Thus, we obtain $  \Omega(C|\w,\mu)\subseteq \Omega(C_{\sigma=\mu}|C,\w,\mu) \cup \Omega(C_{\sigma>\mu}|\w',\mu') $.

		Now, for every $j\in I\backslash C$, if $\mu(j)\in \Omega(C|\w,\mu)$ but $\mu(j)\notin \Omega(C_{\sigma=\mu}|C,\w,\mu)$, then $\mu(j)\in \Omega(C_{\sigma>\mu}|\w',\mu')$. Conversely, if $j\in I\backslash C$ and $ \mu(j)\in\Omega(C_{\sigma>\mu}|\w',\mu') $, then $\mu(j)\in \Omega(C|\w,\mu)$, but because $ \mu(j) $ remains after the removal of $C^\diamond$, $ j\notin C^\diamond $; since $ C^\diamond $ is closed under the recursion defining it, $ \mu(j)\notin\w(C^\diamond)=\Omega(C_{\sigma=\mu}|C,\w,\mu) $. 
\end{proof}

	\begin{proof}[Proof of Theorem \ref{theorem:C1core:char}]
		Recall that $ \mathcal{EC} $ and $ \mathcal{CEC} $ denote the exclusion core and the consistent exclusion core respectively. We show that $	\mathcal{CEC}=\Phi(\mathcal{EC})$.
		Lemma \ref{lemma:maxrefinement} then implies that $ \mathcal{CEC} $ is the largest consistent refinement of $ \mathcal{EC} $. Its nonemptiness follows from Proposition \ref{prop:YRMH-IGYT}.

		\textbf{First inclusion}: $ \mathcal{CEC}\subseteq\Phi(\mathcal{EC}) $. Fix an economy $ \Gamma $ and an allocation $ \mu\in\mathcal{CEC}(\Gamma) $. Every exclusion block is a consistent exclusion block, so $ \mu\in\mathcal{EC}(\Gamma) $. Let a nonempty $ I'\subsetneq I $ be self-feasible under $ \mu $, write $ \Gamma_1=\Gamma(\mu,I\backslash I') $, and let $ \w_1 $ denote its endowment function. We want to show that $ \mu_{I\backslash I'}\in\mathcal{EC}(\Gamma_1) $.  As $ I' $ is an arbitrary admissible departure, $ \mu\in\Phi(\mathcal{EC})(\Gamma) $.

		Suppose, toward a contradiction, that $ \mu_{I\backslash I'} $ is exclusion blocked in $ \Gamma_1 $ by a coalition $ \hat C $ via $ \hat\sigma $. Let $ C=\hat C\cup I' $ and let $ \sigma $ agree with $ \hat\sigma $ on $ I\backslash I' $ and with $ \mu $ on $ I' $. We then show that $ C $ consistently exclusion blocks $ \mu $ via $\sigma$ in $\Gamma$. Since $ C_{\sigma>\mu}=\hat C $, $ C_{\sigma=\mu}=I' $, and $I'$ is self-feasible under $ \mu $, conditions (1) and (2) of Definition \ref{definition:C1core} hold. It remains to verify  condition (3).

		For every $ J\subseteq I\backslash I' $, the reduction rule gives $ \w_1(J)\subseteq\w(J\cup I') $. Induction along the two control recursions therefore gives $\Omega(\hat C|\w_1,\mu_{I\backslash I'})\subseteq\Omega(C|\w,\mu)$. 

		Let $ C^\diamond $ be the set of agents given by \eqref{eq:cdiamond} such that $ \w(C^\diamond)= \Omega(I'|C,\w,\mu)$; it is self-feasible, and $ C^\diamond\cap C=I' $. Let $ D $ be the group reached by the recursion defining $ \Omega(\hat C|\w_1,\mu_{I\backslash I'}) $, that is, $ \w_1(D)=\Omega(\hat C|\w_1,\mu_{I\backslash I'})  $. We claim that $ D\cap C^\diamond=\emptyset $. That recursion starts from $ \hat C $ and runs inside $ \Gamma_1 $, so $ D\subseteq I\backslash I' $; and its starting set is already disjoint from $ C^\diamond $, because $ \hat C\cap C^\diamond\subseteq C\cap C^\diamond=I' $ while $ \hat C\cap I'=\emptyset $. Hence, if $ D \cap C^\diamond\neq \emptyset $, some agent of $ C^\diamond\backslash I' $ would have to be added at some step of the recursion; let $ i $ be the first one added. Adding $ i $ requires the agents reached before her to own $ \mu(i) $ in $ \Gamma_1 $. Let $ o=\mu(i) $; self-feasibility of $ C^\diamond $ gives $ C_o\subseteq C^\diamond $. If $ C_o\cap(I\backslash I')\neq\emptyset $, the owners of $ o $ in $ \Gamma_1 $ are $ C_o\cap(I\backslash I') $, a nonempty subset of $ C^\diamond\backslash I' $, none of whom has been reached before $ i $. If $ C_o\subseteq I' $, then $ o $ is publicly owned in $ \Gamma_1 $, so its owners there are all of $ I\backslash I' $, including $ i $ herself, who has not been reached. Either way the agents reached before $ i $ do not own $ o $, contradicting her addition.

		Now let $ j\in I\backslash C $ be harmed by $ \sigma $. The exclusion block in $ \Gamma_1 $ gives $ \mu(j)\in\Omega(\hat C|\w_1,\mu_{I\backslash I'}) $. Since we have shown that $\Omega(\hat C|\w_1,\mu_{I\backslash I'})\subseteq\Omega(C|\w,\mu)$, $ \mu(j) \in \Omega(C|\w,\mu) $. Closure of $ D $ gives $ j\in D $, hence $ j\notin C^\diamond $; since also $ j\notin C $, closure of $ C^\diamond $ gives $ \mu(j)\notin\w(C^\diamond)=\Omega(C_{\sigma=\mu}|C,\w,\mu) $. This is condition (3). So $ C $ consistently exclusion blocks $ \mu $ via $\sigma$, a contradiction.

		\textbf{Second inclusion}: $ \Phi(\mathcal{EC})\subseteq\mathcal{CEC} $. Fix an economy $\Gamma$ and let $ \mu\in\Phi(\mathcal{EC})(\Gamma) $. Suppose that $ \mu $ is consistently exclusion blocked by $ C $ via $ \sigma $. Without loss of generality, let $ C $ contain every agent who is strictly better off in $ \sigma $, so $ C_{\sigma>\mu}=I_{\sigma>\mu} $; this leaves $ C_{\sigma=\mu} $ unchanged, enlarges $ \Omega(C|\w,\mu) $, and shrinks $ \Omega(C_{\sigma=\mu}|C,\w,\mu) $, so the block survives. $ C_{\sigma=\mu}$ must be nonempty, since otherwise the block is an exclusion block in $ \Gamma $, contradicting $ \mu\in\mathcal{EC}(\Gamma) $.

		Let $ C^\diamond $ be the set of agents given by \eqref{eq:cdiamond} such that $ \w(C^\diamond)= \Omega(C_{\sigma=\mu}|C,\w,\mu)$; it is self-feasible. Every $ i\in C^\diamond\backslash C_{\sigma=\mu} $ lies outside $ C $ and holds an object in $ \w(C^\diamond) $, so condition (3) prevents her from being harmed; nor can she be strictly better off, since every such agent lies in $ C $. As preferences are strict, $ \sigma $ agrees with $ \mu $ on $ C^\diamond $. So, $ \sigma_{I\backslash C^\diamond} $ is an allocation in $ \Gamma(\mu,I\backslash C^\diamond) $, and every member of $ C_{\sigma>\mu} $ is strictly better off under it. By Lemma \ref{lemma:firstcondition}, every harmed outsider loses an object in $ \Omega(C_{\sigma>\mu}|\w',\mu_{I\backslash C^\diamond}) $, where $ \w' $ is the endowment function of $ \Gamma(\mu,I\backslash C^\diamond) $. Hence $ C_{\sigma>\mu} $ exclusion blocks $ \mu_{I\backslash C^\diamond} $ in $ \Gamma(\mu,I\backslash C^\diamond) $, contradicting $ \mu\in\Phi(\mathcal{EC})(\Gamma) $.
	\end{proof}

	\subsection{Proofs of results in Section \ref{section:YRMH}}\label{appendix:proofsYRMH}

	\begin{proof}[\textbf{Proof of Proposition \ref{prop:YRMH-IGYT}}]

		For any economy $ \Gamma=(I,O,\succ_I,\{C_o\}_{o\in O}) $ and any order of agents $ \rhd $, let $ \mu $ denote the outcome of YRMH-IGYT. We need to show that $\mu$ belongs to the consistent exclusion core of $\Gamma$. Suppose, toward a contradiction, that $ \mu $ is consistently exclusion blocked by a coalition $ C $ via another $ \sigma $; without loss of generality, let $ C $ contain all agents who are strictly better off in $ \sigma $. By condition (2) of Definition \ref{definition:C1core}, the unaffected group $ C_{\sigma=\mu} $ is self-feasible under $ \mu $, so $ \mu(C_{\sigma=\mu})\subseteq \w(C_{\sigma=\mu})\cup\{o^*\}$. Of condition (2), the proof below uses only the consequence that $\mu(C_{\sigma=\mu})\subseteq \w(C)\cup\{o^*\} $. This will allow us to use the same argument to prove that $\mu$ belongs to the refinement of the consistent exclusion core defined in Online Appendix \ref{appendix:refinedEC:definition}.

		Each agent removed in YRMH-IGYT obtains her favorite object among those remaining at that point, and $ o^* $ is always available. By condition (1) of Definition \ref{definition:C1core}, some agent is strictly better off in $ \sigma $; among such agents, let $ i^\star $ be one removed earliest, say in step $ t^\star $. Since $ \sigma(i^\star)\succ_{i^\star}\mu(i^\star) $, the object $ \sigma(i^\star) $ differs from $ o^* $ and is removed before step $ t^\star $, so $ \sigma(i^\star)=\mu(j) $ for some agent $ j $ removed before step $ t^\star $. Then $ \sigma(j)\neq\mu(j) $, and $ j $ is not strictly better off in $ \sigma $ by the choice of $ i^\star $, so $ j $ is worse off. Among the agents who are worse off in $ \sigma $, let $ i^\diamond $ be one removed earliest, say in step $ t^\diamond $, and let $ o^\diamond=\mu(i^\diamond) $. Since $ j $ is removed before step $ t^\star $, $ t^\diamond<t^\star $. Consequently, every agent of $ C $ removed in or before step $ t^\diamond $ is unaffected from $ \mu $ to $ \sigma $: by condition (1) she is not worse off, and she is removed before every agent who is strictly better off.

		Since $ i^\diamond $ is worse off in $ \sigma $, she is a harmed outsider, so condition (3) of Definition \ref{definition:C1core} requires $ o^\diamond\in\Omega(C|\w,\mu) $. Through three claims, we show that there exists a self-feasible $ C'\subseteq C_{\sigma=\mu} $ with $ o^\diamond\in\Omega(C'|C,\w,\mu) $. Since $ \Omega(C'|C,\w,\mu)\subseteq\Omega(C_{\sigma=\mu}|C,\w,\mu) $, this yields $ o^\diamond\in\Omega(C_{\sigma=\mu}|C,\w,\mu) $, a contradiction to condition (3) of Definition \ref{definition:C1core}.

		We say that an agent \emph{holds a right} over a remaining object $ o $ in step $ t $ if she belongs to $ C_o(t-1)\cup S_o(t-1) $, that is, if she is a remaining owner of $ o $ or has acquired shared ownership of it. By the rules of YRMH-IGYT, a remaining object points to an agent exactly when some agent holds a right over it, and it then points to such an agent; in particular, no agent holds a right over an object in the step in which the object is removed without being involved in a cycle. Rights are acquired only when a cycle is removed: each remaining agent holding a right over an object of the cycle acquires a right over every remaining object over which some agent of the cycle holds a right. An agent keeps her rights until she is removed.

	\begin{claim}\label{thm3:lemma1}
		In any step $ t\le t^\diamond $, (a) every agent holding a right over an object of $ \w(C) $ belongs to $ C $; and (b) if a cycle is removed and the cycle involves an agent from $ C $ or an object from $ \w(C) $, then in the cycle all agents belong to $ C_{\sigma=\mu} $ and all objects belong to $ \w(C) $.
	\end{claim}

	\begin{claimproof}
		We argue by induction on the step. In step $ 1 $ only owners hold rights, and every owner of an object of $ \w(C) $ belongs to $ C $, so (a) holds. Suppose that (a) holds in some step $ t\le t^\diamond $.

		We first derive (b) in step $ t $. In a cycle, each agent points to the object she obtains, and each object points to an agent holding a right over it. If an agent of the cycle belongs to $ C $, she belongs to $ C_{\sigma=\mu} $, having been removed in step $ t\le t^\diamond $, so the object she points to, her assignment, lies in $ \w(C)\cup\{o^*\} $, and in fact in $ \w(C) $, since $ o^* $ is never part of a cycle. If an object of the cycle belongs to $ \w(C) $, then by (a) the agent it points to belongs to $ C $. Applying the argument inductively to the remaining agents and objects in the cycle, we conclude that every agent in the cycle belongs to $ C $, and thus to $ C_{\sigma=\mu} $, and every object in the cycle belongs to $ \w(C) $.

		If $ t<t^\diamond $, we then derive (a) in step $ t+1 $. Rights change only when a cycle is removed. Suppose that an agent $ j' $ acquires a right over some $ b\in\w(C) $ when a cycle is removed in step $ t $. By the updating rule, some agent $ j $ of the cycle holds a right over $ b $, and $ j' $ holds a right over some object $ a $ of the cycle. By (a), $ j\in C $; by (b), $ a\in\w(C) $; and by (a) again, $ j'\in C $.
	\end{claimproof}

	\begin{claim}\label{thm3:lemma2}
		In any step $ t\le t^\diamond $, if an object $ o $ is removed without being involved in a cycle, then there exists a self-feasible $ C'\subseteq C_{\sigma=\mu} $ such that $ C_o\cap C\subseteq C' $.
	\end{claim}

	\begin{claimproof}
		Since $ o $ is removed without being involved in a cycle, no agent holds a right over $ o $ in step $ t $; in particular, every owner of $ o $ is removed before step $ t $. Let $ C' $ consist of the agents in $ C_o\cap C $, the owners of their assignments, the owners of those owners' assignments, and so on. Formally, let $ C'=\cup_{k=0}^\infty D^k $, where $ D^0=C_o\cap C $ and $ D^k=\cup_{o'\in\mu(D^{k-1})}C_{o'} $ for every $ k\ge 1 $. Each object in $ \mu(D^{k-1}) $ other than $ o^* $ is owned by $ D^k $, so $ \mu(C')\subseteq\w(C')\cup\{o^*\} $; that is, $ C' $ is self-feasible.

		We argue that no agent of $ C' $ is present in step $ t $. The idea is that the right over $ o $ held by the agents of $ D^0 $ passes along the chains of owners that define $ C' $ as these agents leave, so that if some agent of $ C' $ were present in step $ t $, some agent of $ C' $ would hold a right over $ o $, and $ o $ would still point to an agent in step $ t $. Suppose otherwise. Following a chain of owners from $ D^0 $ to a present agent and stopping at its first present agent, we obtain $ i_0,\ldots,i_m $ with $ i_0\in D^0 $ and $ i_r\in C_{\mu(i_{r-1})} $ for $ 1\le r\le m $, such that $ i_m $ is present in step $ t $ and $ i_0,\ldots,i_{m-1} $ are removed before it. Write $ o_0=o $ and $ o_r=\mu(i_{r-1}) $ for $ r\ge 1 $, so that $ i_r $ owns $ o_r $. For $ 0\le r\le m $ and a step $ s\le t $ in which $ i_r $ is present, let the \emph{index} of $ i_r $ be the smallest $ q $ such that $ i_q,\ldots,i_{r-1} $ are all removed before step $ s $; then $ o_q $ remains in step $ s $, as $ i_{q-1} $ is present or $ q=0 $. We show by induction on $ s $ that $ i_r $ holds a right over $ o_q $; since the index of $ i_m $ in step $ t $ is $ 0 $, she then holds a right over $ o $ in step $ t $, whereas no agent does. In step $ 1 $, $ q=r $ and $ i_r $ owns $ o_r $. Suppose that the statement holds in step $ s<t $, and let $ i_r $ remain in step $ s+1 $, with index $ q $ in step $ s $ and $ q' $ in step $ s+1 $. If $ q'=q $, she keeps her right. Otherwise $ q'<q $, and $ i_{q-1} $ is removed in step $ s $ with $ o_q $; as $ i_r $ holds a right over $ o_q $, this object points to an agent, so $ i_{q-1} $ is removed in a cycle, which contains every agent removed in step $ s $, and $ i_r $ acquires every right that the cycle's members hold over remaining objects. The first of $ i_{q'},\ldots,i_{q-1} $ present in step $ s $ has index $ q' $ there, so she holds a right over $ o_{q'} $, which remains in step $ s+1 $; and she is removed in step $ s $, in the cycle. Hence $ i_r $ holds a right over $ o_{q'} $ in step $ s+1 $.

		It remains to show that $ C'\subseteq C_{\sigma=\mu} $. The agents of $ D^0 $ belong to $ C $ and are removed before step $ t\le t^\diamond $, so they belong to $ C_{\sigma=\mu} $. If $ D^{k-1}\subseteq C_{\sigma=\mu} $ for some $ k\ge 1 $, then $ \mu(D^{k-1})\subseteq\w(C)\cup\{o^*\} $, so $ D^k\subseteq C $; since the agents of $ D^k $ are removed before step $ t $, $ D^k\subseteq C_{\sigma=\mu} $. By induction, $ C'\subseteq C_{\sigma=\mu} $.
	\end{claimproof}

		Recall that $ \Omega(C|\w,\mu)=\w(\cup_{k=0}^\infty C^k) $ where $ C^0=C $ and $ C^k=C^{k-1}\cup \{i\in I\backslash C^{k-1}: \mu(i)\in \w(C^{k-1}) \} $ for every $ k\ge 1 $.

	\begin{claim}\label{thm3:lemma3}
		In any step $ t\le t^\diamond $, if any $ i\in C^k\backslash C^{k-1} $ with $ k\ge 1 $ is removed, then $ i $ and $ \mu(i) $ are not involved in a cycle, and there exists a self-feasible $ C'\subseteq C_{\sigma=\mu} $ such that $ \mu(i)\in \Omega(C'|C,\w,\mu) $.
	\end{claim}

	\begin{claimproof}
		We argue by induction on $ k $. Fix $ k\ge 1 $, and suppose that the claim holds with $ \ell $ for every $ 1\le\ell<k $; for $ k=1 $ this hypothesis is vacuous. Let $ i\in C^k\backslash C^{k-1} $ be removed in step $ t\le t^\diamond $, and let $ o=\mu(i) $. Then $ o\in\w(C^{k-1}) $, so each owner of $ o $ belongs either to $ C $ or to $ C^{k-1}\backslash C $, and the induction hypothesis applies to owners of the latter kind.

		We first show that in every step up to $ t $, each agent holding a right over $ o $ belongs to $ C $ or is an owner of $ o $ in $ C^{k-1}\backslash C $. Owners of $ o $ satisfy this. A right over $ o $ is acquired only when a cycle containing an agent $ j $ who holds a right over $ o $ is removed, in some step $ s<t $; suppose that the property holds in step $ s $. An owner of $ o $ in $ C^{k-1}\backslash C $ is not removed in a cycle, by the induction hypothesis, so $ j\in C $. By Claim \ref{thm3:lemma1}(b), every object of the cycle belongs to $ \w(C) $, and the acquiring agent, who holds a right over one of them in step $ s $, belongs to $ C $ by Claim \ref{thm3:lemma1}(a).

		Now suppose that $ i $ and $ o $ are involved in a cycle in step $ t $, and let $ j $ be the agent that $ o $ points to. Then $ j $ holds a right over $ o $ in step $ t $. If $ j\in C $, Claim \ref{thm3:lemma1}(b) places every agent of the cycle, including $ i $, in $ C $, contradicting $ i\notin C^{k-1}\supseteq C $. Otherwise $ j $ is an owner of $ o $ in $ C^{k-1}\backslash C $, and the induction hypothesis says that $ j $ is not removed in a cycle, again a contradiction. So $ i $ and $ o $ are not involved in a cycle.

		It remains to find $ C' $. Since $ o $ is removed in step $ t $ without being involved in a cycle, Claim \ref{thm3:lemma2} gives a self-feasible $ C'_0\subseteq C_{\sigma=\mu} $ with $ C_o\cap C\subseteq C'_0 $, and every owner of $ o $ is removed before step $ t $. For each $ j\in C_o\backslash C $, the induction hypothesis therefore gives a self-feasible $ C'_j\subseteq C_{\sigma=\mu} $ with $ \mu(j)\in\Omega(C'_j|C,\w,\mu) $. Let $ C'=C'_0\cup(\cup_{j\in C_o\backslash C}C'_j) $. It is a subcoalition of $ C_{\sigma=\mu} $, and it is self-feasible because self-feasible coalitions are closed under union. Every $ j\in C_o\cap C $ belongs to $ C' $, and every $ j\in C_o\backslash C $ lies outside $ C $ and satisfies $ \mu(j)\in\Omega(C'_j|C,\w,\mu)\subseteq\Omega(C'|C,\w,\mu) $, so it is added by the recursion defining $ \Omega(C'|C,\w,\mu) $. Hence that recursion reaches all of $ C_o $, giving $ o\in\Omega(C'|C,\w,\mu) $.
	\end{claimproof}

		Since $ o^\diamond\in \Omega(C|\w,\mu)$, there exists $ k\ge 1 $ such that $ o^\diamond\in \w(C^{k-1}) $ and $ i^\diamond\in C^k\backslash C^{k-1} $. Because $ i^\diamond $ is removed with $ o^\diamond $ in step $ t^\diamond $, Claim \ref{thm3:lemma3} yields a self-feasible $ C'\subseteq C_{\sigma=\mu} $ with $ o^\diamond \in \Omega(C'|C,\w,\mu)\subseteq\Omega(C_{\sigma=\mu}|C,\w,\mu) $, contradicting condition (3) of Definition \ref{definition:C1core}. Hence $ \mu $ belongs to the consistent exclusion core.
	\end{proof}

\subsection{Proofs of results in Section \ref{section:incentives}}\label{appendix:incentives}

\begin{proof}[Proof of Theorem \ref{theorem:general:strategyproof}]
	Suppose, toward a contradiction, that a strategy-proof mechanism $ \varphi $ selects from such a solution $ f $. Consider an economy with five agents $ I=\{1,2,3,4,5\} $ and three objects $ O=\{a,b,c\} $, where $C_{a}=\{1\}$, $C_{b}=\{2,3\}$, and $C_{c}=\{4,5\}$.\footnote{I thank Zhibin Tan and Ertan Zhuang for drawing my attention to a version of this example, which they use to establish a different result \citep{tan2026freeriding}.} Consider the following two preference profiles, in which the objects omitted from a column are unacceptable to that agent:
	\[
	\begin{array}{ccccc}
		 \succ_1 & \succ_2 & \succ_3 & \succ_4 & \succ_5 \\ \hline
		 o^* & a & a & a & a\\
		  & b & b & c & c\\
	    & o^* & o^* & o^* & o^*
	\end{array}
	\qquad
	\begin{array}{ccccc}
		 \succ_1 & \succ'_2 & \succ_3 & \succ_4 & \succ_5 \\ \hline
		 o^* & c & a & a & a\\
		  & a & b & c & c\\
	    & o^* & o^* & o^* & o^*
	\end{array}
	\]

\textbf{The economy $ \Gamma $.} Denote the economy with the preference profile $\succ_I$ by $\Gamma$. Because $f$ is Pareto efficient, $ \varphi (\Gamma)$ must be a Pareto efficient allocation. So in $\varphi(\Gamma)$, object $ a $ must be assigned to one of agents $ 2,3,4,5 $, all of whom most prefer it. Let $ r $ denote its recipient. If $ r\in\{2,3\} $, the argument below applies with $ r $ in the role of agent $ 2 $ and the other member of that pair in the role of agent $ 3 $. If $ r\in\{4,5\} $, apply the same argument after interchanging the roles of the pairs $ \{2,3\} $ and $ \{4,5\} $ and of their co-owned objects $ b $ and $ c $. These are separate cases determined by the selected allocation and do not require the mechanism to be anonymous. We present the case with $ r=2 $ to simplify notation.

\textbf{The economy $ \Gamma' $.} Now suppose that agent $2$ changes her preferences to $\succ'_2$. We denote the economy with the preference profile $(\succ'_2,\succ_{-2})$ by $ \Gamma' $. We claim that every allocation $\mu$ in $ f(\Gamma') $ assigns $ o^* $ to agent $ 2 $.

	First, Pareto efficiency precludes assigning an unacceptable object to any agent. It also precludes assigning $ a $ to agent $ 2 $: agent $ 2 $ prefers $ c $ to $ a $, so $ c $ cannot be left unassigned, and if its holder is agent $ 4 $ or agent $ 5 $, exchanging $ a $ and $ c $ then strictly improves both.

	Second, we show that $ \mu(2)\neq c $. Let $ \mu\in f(\Gamma') $ have $ \mu(2)=c $. Pareto efficiency gives $ \mu(1)=o^* $ and assigns $ a $ to one of agents $ 3,4,5 $. In each of the two cases below, one owner of $ c $ belongs to a self-feasible coalition $ I' $ while the other owner of $ c $ receives $ o^* $. If $ \mu(3)=a $, then agents $ 4 $ and $ 5 $ both receive $ o^* $, and $ I'=\{4\} $ is therefore self-feasible. If instead $ a $ is assigned to some $ m\in\{4,5\} $, then the other owner $ \ell $ of $ c $ receives $ o^* $, since both objects she finds acceptable are held by agents $ 2 $ and $ m $; $ I'=\{1,m\} $ is therefore self-feasible because agent $ 1 $ holds $ o^* $ and agent $ m $ holds $ a $, which agent $ 1 $ privately owns. In either case, after $ I' $ departs with its assignments, object $ c $ remains and is privately owned by the owner who stays, who still receives $ o^* $ and prefers $ c $. Consistency requires the restriction of $ \mu $ to belong to $ f(\Gamma'(\mu,I\backslash I')) $, whereas individual rationality excludes it, a contradiction.

	Thus every allocation in $ f(\Gamma') $ assigns $ o^* $ to agent $ 2 $; so $ \varphi(\Gamma')(2)=o^* $. Suppose $ \succ'_2 $ is agent $ 2 $'s true preference. By reporting $ \succ_2 $ instead, she turns $ \Gamma' $ into $ \Gamma $ and receives $ a $. Since $ a\succ'_2 o^* $, this is a profitable manipulation, contradicting strategy-proofness of $\varphi$.
\end{proof}

\subsection{Proofs of results in Section \ref{section:special}}\label{appendix:special}

As noted in Section \ref{section:special}, the economies can be classified into three classes: public ownership, private ownership, and mixed ownership. We prove Proposition \ref{prop:specialsetting} for each of them.

\begin{proof}[Proof of Proposition \ref{prop:specialsetting} in public-ownership economies]

 Consider any public-ownership economy. In each step of YRMH-IGYT, every remaining object points to the $ \rhd $-highest remaining agent. That agent therefore receives her most preferred remaining object and is removed; the procedure then repeats with the next agent. YRMH-IGYT is thus serial dictatorship with order $ \rhd $, which is the TTC algorithm in which every object ranks agents by $ \rhd $; it is therefore group strategy-proof by Claim \ref{claim:ttcgsp} below.  
 Since serial dictatorship generates every Pareto efficient allocation as its order varies \citep{abdulkadirouglu1998random}, YRMH-IGYT can find all Pareto efficient allocations.  Since both the exclusion core and the consistent exclusion core lie between the set of YRMH-IGYT outcomes and the set of Pareto efficient allocations, these sets must coincide.
\end{proof}

\begin{proof}[Proof of Proposition \ref{prop:specialsetting} in private-ownership economies]
In a private-ownership economy no agent ever acquires shared ownership in YRMH-IGYT: every object in a removed cycle points to its unique owner, so the rule for updating ownership transfers nothing. Each object therefore points to its own owner whenever she remains, and to no agent otherwise. To prove the first statement, we show that YRMH-IGYT can find every allocation in the exclusion core.  Let $ \mu $ be any such allocation. We construct the following algorithm to find an order $ \rhd $ of agents and show that $ \mu $ is the outcome of YRMH-IGYT with $ \rhd $. 

		\textbf{An algorithm to construct $\rhd$ from $\mu$.} Start with all agents and all objects. In each step $ t\ge 1 $, given the remaining agents and objects, conduct two operations.

		\textit{Operation A}: Let remaining agents point to their favorite remaining objects, let remaining objects point to their owners if their owners remain and otherwise to no agent, and let $ o^* $ point to every agent who points to it; since $ o^* $ has unlimited copies, each such agent forms a cycle with her own copy. Distinct cycles then share no agent and no real object; remove them and repeat until no cycles are generated. Claim \ref{claim:operationA}(a) proves that the object pointed to by every agent in every cycle is her assignment under $ \mu $. As cycles are removed, place their agents in $ \rhd $ below all agents placed earlier, one cycle after another; the agents of a cycle, and cycles removed together, may be ranked arbitrarily.  Go to Operation B if any agents remain.

		\textit{Operation B}: After Operation A, the pointing graph has no cycle. Because every agent points to one object and every object points to at most one agent, the agents and objects form disjoint trees; the root of each tree is an object that points to no agent, and every remaining agent is linked to one root through a unique directed path.\footnote{Formally, an agent $ i $ is linked to an object $ o_\ell $ via a directed path if there exist distinct agents $ \{i_1,\ldots, i_k\} $ and distinct objects $ \{o_1,\ldots, o_k\}  $ such that $ i \rightarrow o_1\rightarrow i_1 \rightarrow o_2\rightarrow i_2\rightarrow \cdots \rightarrow i_{k-1}\rightarrow o_k \rightarrow i_k \rightarrow o_\ell. $} For every root $ o_\ell $, denote the set of agents who point to $ o_\ell $ by $ I_\ell $. Claim \ref{claim:operationA}(b) proves that there exist a root $ o_\ell $ and an agent $ i\in I_\ell $ with $ \mu(i)=o_\ell $. Denote the set of such $ i $ by $ U_t $, remove them with their assignments under $ \mu $, and place them in $ \rhd $ below all agents placed earlier, ranked among themselves arbitrarily. Move to step $ t+1 $ if any agents remain; otherwise stop. 

		At least one agent is removed in each step, so the algorithm stops in finitely many steps. Thus $ \rhd $ ranks the agents in the order in which the algorithm removes them, and by Claim \ref{claim:operationA} each is removed with her assignment under $ \mu $. We show that YRMH-IGYT with $ \rhd $ removes the same agents with the same objects, in the same order. Suppose that it has done so up to some point. If the algorithm next removes a cycle, the highest-ranked remaining agent belongs to it. The objects remaining in YRMH-IGYT are those that remain when the algorithm forms this cycle, except for the objects of other cycles that the algorithm forms at the same time and that come earlier in $ \rhd $; so every agent of the cycle has the same favorite remaining object as in the algorithm, namely the object she points to in the cycle. The highest-ranked agent therefore points to her object in the cycle, whose owner is the next agent of the cycle; that agent is promoted and points to the next object of the cycle, and so on until the cycle closes and is removed, leaving no promoted agent behind. An agent who points to $ o^* $ simply obtains it. If the algorithm next removes an agent $ i\in U_t $, she is the highest-ranked remaining agent. The objects remaining in YRMH-IGYT are those that remain in Operation B of step $ t $, except for the roots assigned to agents of $ U_t $ ranked above her; so she still points to $ \mu(i) $, which points to no agent because its owner has been removed, and she obtains it. Hence the outcome of YRMH-IGYT with $ \rhd $ is $ \mu $.

		\begin{claim}\label{claim:operationA}
			In each step $ t $ of the above algorithm: (a) In Operation A, the object pointed to by every agent in every cycle is her assignment under $ \mu $; (b) In Operation B, there exists at least one root $ o_\ell $ and an agent $ i\in I_\ell $ such that $ \mu(i)=o_\ell $.
		\end{claim}

		\begin{claimproof}[Proof of Claim \ref{claim:operationA}]

		We prove the claim by induction. Suppose that for all agents removed before step $ t $, they are removed with their assignments under $ \mu $. The objects removed before step $ t $ are then exactly the assignments under $ \mu $ of those agents, so $ \mu(i) $ is still remaining for every remaining agent $ i $. We then consider step $ t $. 

			(a)		Let $ (C_1,C_2,\ldots, C_K) $ denote the order of cycles removed in Operation A; each $ C_k $ denotes the set of agents involved in a cycle. If multiple cycles are removed simultaneously, their relative ranking in the order is arbitrary. 

			We argue by induction on the order of cycles. Consider any $ k\ge 1 $, and suppose that for every $ \ell<k $, all agents in $ C_\ell $ point to their assignments under $ \mu $; for $ k=1 $ this hypothesis is vacuous. The cycles removed before $ C_k $ then take exactly their assignments under $ \mu $ with them, so $ \mu(i) $ is still remaining for every $ i\in C_k $. Suppose, toward a contradiction, that some agent in $ C_k $ points to an object different from her assignment under $ \mu $, and denote by $ C'_k $ the set of such agents. We show that $ C'_k $ can exclusion block $ \mu $ via another $ \sigma $ in which, for all $ i\in C_k $, $ \sigma(i) $ is the object pointed to by $ i $ in the cycle, for all $ j\in I\backslash C_k $ with $ \mu(j)\in \sigma(C_k) $, $ \sigma(j)=o^* $, and for all other $ j $, $ \sigma(j)=\mu(j) $. First, since each agent in $ C'_k $ points to her most preferred remaining object, which differs from her own assignment, she is strictly better off in $ \sigma $. Second, for any $ j\in I\backslash C'_k $ with $ \mu(j)\succ_j \sigma(j) $, we have $ \mu(j)\in \sigma(C_k) $ and $ \mu(j)\neq o^* $. Since $ C_k $ forms a cycle, every object in $ \sigma(C_k) $ other than $ o^* $ points to its owner in $ C_k $, so $ \sigma(C_k)\subseteq\w(C_k)\cup\{o^*\} $. Moreover $ \w(C_k)\subseteq\Omega(C'_k|\w,\mu) $: an agent of $ C_k\backslash C'_k $ points to her own assignment, which is owned by the next agent around the cycle, so the recursion defining $ \Omega $ absorbs her once it has absorbed that agent; walking backwards around the cycle from an agent of $ C'_k $ therefore absorbs all of $ C_k $. So, $ \mu(j)\succ_j \sigma(j) $ implies $ \mu(j)\in \Omega(C'_k|\w,\mu) $, a contradiction.

		(b) Since $ \mu $ is Pareto efficient, every root $ o_\ell $ with $I_\ell\neq \emptyset$ in Operation B must be assigned to an agent under $ \mu $. Suppose, toward a contradiction, that every such root $ o_\ell$ is assigned to some $ i_\ell \notin I_\ell $ under $ \mu $. We then show that $ \mu $ is exclusion blocked. There are two cases.

		\textit{Case 1:} There exists such a root $ o_\ell $ for which $i_\ell $ is linked to $ o_\ell $ but does not directly point to $ o_\ell $.  Suppose that $ i_\ell $ is linked to $ o_\ell $ through the following path:
		\[
		i_\ell \rightarrow o_1\rightarrow i_1 \rightarrow o_2\rightarrow i_2\rightarrow \cdots \rightarrow i_{k-1}\rightarrow o_k \rightarrow i_k \rightarrow o_\ell.
		\] Let $ C=\{i_1,i_2,\ldots,i_k,i_\ell\} $, the set of agents on the path. Let $ C' $ consist of the agents in $ C $ who point to objects different from their assignments under $ \mu $. Since both $ i_\ell$ and $i_k$ belong to $ C' $, $ C' $ is nonempty. We then prove that $ C' $ can exclusion block $ \mu $ via another $ \sigma $ in which, for all $ i\in C $, $ \sigma(i) $ is the object pointed to by $ i $, for all $ j\notin C $ with $ \mu(j)\in \sigma(C) $, $ \sigma(j)=o^* $, and for all other $ j $, $ \sigma(j)=\mu(j) $. First, all agents in $ C' $ are strictly better off in $ \sigma $. Second, for every $ j\in I\backslash C' $ with $ \mu(j)\succ_j \sigma(j)$, we have $ \mu(j)\in \sigma(C) $. Every object in $ \sigma(C) $ other than $ o_\ell $ points to its owner in $ C $, so $ \sigma(C)\backslash\{o_\ell\}\subseteq \w(C) $; walking backwards along the path from $ i_k\in C' $ gives $ \w(C)\subseteq \Omega(C'|\w,\mu) $; and because $ o_\ell=\mu(i_\ell) $ with $ i_\ell\in C' $, no agent outside $ C' $ holds $ o_\ell $ under $ \mu $. Therefore, $ \mu(j)\succ_j \sigma(j)$ implies $ \mu(j)\in  \Omega(C'|\w,\mu)$.

		\textit{Case 2:} For every such root $ o_\ell $, $ i_\ell $ is not linked to $ o_\ell $. Then, every $ i_\ell $ must be linked to some $ o_k\neq o_\ell $. 	
		Let $ R $ be a nonempty set of such roots, of minimum cardinality, such that for every $ o_{y}\in R $, $ i_{y} $ is linked to some $o_z\in R$ with $ o_{z}\neq o_y$. 
		As in Case 1, every $ i_{y} $ is linked to $ o_{z}$ through a path. Denote by $ C $ the set of all agents in those paths; see Figure \ref{figure:proofillustration} for an illustration. Among these agents, denote by $ C' $ the set of agents who point to objects different from their assignments under $ \mu $. $ C' $ is nonempty because, for every $ o_y $ in the subset,  every $ i_{y} $ belongs to $ C' $, and every agent who directly points to $ o_{y} $ also belongs to $ C' $. We then prove that $ C' $ can exclusion block $ \mu $ via another $ \sigma $ in which, for every $ i\in C $, $ \sigma(i) $ is the object pointed to by $ i $, for every $ j\notin C $ with $ \mu(j)\in \sigma(C) $, $ \sigma(j)=o^* $, and for every other $ j $,  $ \sigma(j)=\mu(j) $. First, all agents in $ C' $ are strictly better off in $ \sigma $. Second, for every $ j\in I\backslash C' $ with $ \mu(j)\succ_j \sigma(j)$, $ \mu(j)\in \sigma(C) $. As in Case 1, every object in $ \sigma(C) $ that is not a root points to its owner in $ C $, so all such objects lie in $ \w(C) $; walking backwards along each of the disjoint paths from its agent in $ C' $ gives $ \w(C)\subseteq \Omega(C'|\w,\mu) $; and each root $ o_y $ equals $ \mu(i_y) $ with $ i_y\in C' $, so no agent outside $ C' $ holds a root under $ \mu $.  Therefore, $ \mu(j)\succ_j \sigma(j)$ implies $ \mu(j)\in  \Omega(C'|\w,\mu)$.
		\begin{figure}[!h]
			\small
			\[
			i_y \rightarrow o_1\rightarrow i_1 \rightarrow o_2\rightarrow i_2\rightarrow \cdots \rightarrow i_{k-1}\rightarrow o_k \rightarrow i_k \rightarrow o_z.
			\] 
			\[
			i_z \rightarrow o'_1\rightarrow i'_1 \rightarrow o'_2\rightarrow i'_2\rightarrow \cdots \rightarrow i'_{k-1}\rightarrow o'_k \rightarrow i'_k \rightarrow o_y.
			\] 
			\caption{\small Case 2 when the selected subset of roots is $ \{o_y,o_z\} $. Agent $ i_y $, assigned $ o_y $ under $ \mu $, is linked to $ o_z $, and agent $ i_z $, assigned $ o_z $, is linked to $ o_y $. The set $ C $ consists of the agents on the two paths, and $ C' $ of those among them who point to objects other than their assignments under $ \mu $. In particular, $ \{i_y,i_z,i_k,i'_k\}\subseteq C' $: agents $ i_y $ and $ i_z $ are assigned roots but point elsewhere, while $ i_k $ and $ i'_k $ point to roots assigned to others. Every non-root object on the two paths is controlled by $ C' $, and the two roots are held under $ \mu $ by $ i_y $ and $ i_z $, so no agent outside $ C' $ holds one.}\label{figure:proofillustration}
		\end{figure}
	\end{claimproof}

	 It remains to prove that YRMH-IGYT with any $\rhd$ is group strategy-proof. In any economy, for each object $ o $ privately owned by agent $ i_o $, define a fixed priority order $ \rhd_o $ that ranks $ i_o $ first and ranks all other agents according to $ \rhd $. The outcome of YRMH-IGYT then coincides with that of the generalized TTC for these object priorities. Apart from finding cycles one at a time, the two procedures differ only in that an object whose owner has been removed points to the $ \rhd $-highest remaining agent in the generalized TTC but to no agent in YRMH-IGYT; so a chain of requests that closes into a cycle in the former is resolved one agent at a time in the latter, with the same assignments. By Claim \ref{claim:ttcgsp}, the generalized TTC is group strategy-proof. So, YRMH-IGYT is group strategy-proof.

	\begin{claim}\label{claim:ttcgsp}
		The TTC algorithm with fixed strict priorities over agents, one for each object, is group strategy-proof.
	\end{claim}

	\begin{claimproof}
		In each step, every remaining agent points to her favorite remaining object or to $ o^* $, every remaining object points to its highest-priority remaining agent, and an agent pointing to $ o^* $ forms a cycle on her own. A cycle remains a cycle after any disjoint cycle is removed, since its agents' favorite objects and its objects' highest-priority agents remain; hence the outcome does not depend on the order in which cycles are removed. Fix an agent $ i $ and the reports of the others, and first remove the cycles that avoid $ i $, one at a time, until none remains. This phase does not depend on $ i $'s report. Let $ M_i $ be the set of real objects then remaining. Every path of pointers from an object in $ M_i $ leads to $ i $, since otherwise it would end in a cycle avoiding her; so $ i $ closes a cycle with whichever object of $ M_i $ she points to, or takes $ o^* $. Her assignment is therefore her reported favorite element of $ M_i\cup\{o^*\} $, a set that does not depend on her report, and the algorithm is strategy-proof \citep{abdulkadirouglu2003school}. Moreover, if two reports of $ i $ give her the same assignment, she closes the same cycle, or leaves alone, after the same first phase, so the remaining problem and hence the whole allocation are the same; that is, the algorithm is nonbossy. Strategy-proofness and nonbossiness together imply group strategy-proofness \citep{papai2000strategyproof}; the argument requires only that preferences be strict.
	\end{claimproof}
	\end{proof}

	To prove the proposition in mixed-ownership economies, we first prove Lemma \ref{lemma:consistent}.

\begin{proof}[\textbf{Proof of Lemma \ref{lemma:consistent}}]

	   We first prove the lemma for the exclusion core. For every mixed-ownership economy $ \Gamma $ and every allocation $ \mu $ in the exclusion core of $ \Gamma^P $, we prove that $ \mu $ is the extension of $ \mu_I $ and that $ \mu_I $ is in the exclusion core of $ \Gamma $. Because $ \mu $ is Pareto efficient in $ \Gamma^P $, $ \mu $ must assign $ o^* $ to $ i^P $, so it is the extension of $ \mu_I $, and $ \mu_I $ must be Pareto efficient in $ \Gamma $. Suppose, toward a contradiction, that $ \mu_I $ is exclusion blocked by a coalition $ C $ via another $ \sigma $ in $ \Gamma $. Because $ \mu_I $ is Pareto efficient, $ C\subsetneq I $. For every $ j\in I\backslash C $ with $\mu(j)\succ_j \sigma(j)$, $\mu(j)\in  \Omega(C|\w,\mu_I)$, and $ \mu(j) $ cannot be a public endowment in $ \Gamma $. Let $ \w^P $ be the endowment function in $ \Gamma^P $. Since $ \w $ and $ \w^P $ agree on every proper subset of $ I $, $ \Omega(C|\w^P,\mu) $ and $ \Omega(C|\w,\mu_I) $ contain the same privately owned objects. Therefore, for every $ j\in I\backslash C $ with $\mu(j)\succ_j \sigma(j)$, $\mu(j)\in \Omega(C|\w^P,\mu) $. Then in $ \Gamma^P $, $ \mu $ is exclusion blocked by $ C $ via the extension $ \sigma^P $ of $ \sigma $, a contradiction. 

		Conversely, consider any allocation $ \mu $ in the exclusion core of $ \Gamma $. We prove that its extension $ \mu^P $ is in the exclusion core of $ \Gamma^P $. Since $ \mu $ is Pareto efficient in $ \Gamma $, $ \mu^P $ is Pareto efficient in $ \Gamma^P $. Suppose, toward a contradiction, that $ \mu^P $ is exclusion blocked by a coalition $ C $ via another $ \sigma $ in $ \Gamma^P $. Then $ i^P\notin C $, since $ i^P $ cannot be made strictly better off, and $ C\subsetneq I $, since $ \mu^P $ is Pareto efficient. Since $ \w^P(S)\subseteq \w(S) $ for every $ S\subseteq I $, we have $ \Omega(C|\w^P,\mu^P)\subseteq \Omega(C|\w,\mu) $. But this means that in $ \Gamma $, $ \mu $ is exclusion blocked by $ C $ via $ \sigma_I $.

		We then prove the lemma for YRMH-IGYT. The following claim proves a stronger result: the algorithm produces the same outcome order by order in the two economies.

		\begin{claim}\label{claim:yrmhinvariance}
			For every mixed-ownership economy $ \Gamma $, every initial order $ \rhd $ of $ I $, and every initial order $ \rhd^P $ in $ \Gamma^P $ that ranks the agents in $ I $ identically to $ \rhd $, the outcome of YRMH-IGYT in $ \Gamma $ under $ \rhd $ agrees on $ I $ with its outcome in $ \Gamma^P $ under $ \rhd^P $.
		\end{claim}

			\begin{claimproof}[\textbf{Proof of Claim \ref{claim:yrmhinvariance}}]
		Fix $ \Gamma $, $ \rhd $, and $ \rhd^P $ as in the claim. When the object pointed to by the highest-ranked agent points to some agent and no cycle forms, YRMH-IGYT moves that agent to the top of the current order; we say that she is \emph{promoted}.

		\emph{$ i^P $ may be assumed $ \rhd^P $-first.} Since $ \Gamma^P $ is a private-ownership economy, no agent ever acquires shared ownership in it, as observed in the proof of Proposition \ref{prop:specialsetting} in private-ownership economies. Whenever $ i^P $ becomes the highest-ranked agent she points to $ o^* $ and is removed, taking away no real object and leaving the relative order of the agents in $ I $ unchanged; until then she is pointed to only by the objects in $ O^P $, and an agent who points to such an object merely promotes $ i^P $ and is highest-ranked again once $ i^P $ has gone, with the same favorite remaining object. The outcome of the algorithm in $ \Gamma^P $ therefore does not depend on where $ i^P $ stands in $ \rhd^P $, and we may assume that $i^P$ is ranked highest and removed in step $ 1 $.

		\emph{In $ \Gamma $, each object is held by its owner alone, by everyone, or by no one.} Call an object \emph{free} if it is publicly owned or its owner has been removed. We claim that after each step $ t $ of YRMH-IGYT in $ \Gamma $, the set $ C_o(t)\cup S_o(t) $ of agents holding a right over a remaining object $ o $ consists only of its owner if $ o $ is not free, and is $ I(t) $ or $ \emptyset $ if $ o $ is free. This holds before step $ 1 $. Removing agents restricts these sets to $ I(t) $, which turns $ I(t-1) $ into $ I(t) $ and leaves unheld an object whose owner has left; so the claim can fail only through sharing. The updating rule shares only when a removed cycle contains an object held by some remaining agent. By the claim for step $ t-1 $, an object in the cycle that is not free is held only by its owner, who is in the cycle, and a free object held by no one points to no agent and so lies in no cycle. Sharing therefore requires a free object in the cycle held by all of $ I(t-1) $, and it then gives every agent of $ I(t) $ a right over each remaining object $ o' $ over which a member of the cycle held one. Such an $ o' $ is free after step $ t $: otherwise its only holder, its owner, would have been in the cycle and removed. So $ o' $ is held by all of $ I(t) $.

		Consequently, an object that is not free points to its owner in both economies. In either economy, write $ h $ for the $ \rhd $-highest remaining agent; this is well defined in $ \Gamma^P $ because $ i^P $ leaves in step $ 1 $ and $ \rhd^P $ ranks the agents in $ I $ as $ \rhd $ does. A free object points to no agent in $ \Gamma^P $, which is private-ownership, whereas in $ \Gamma $ it points to no agent if no one holds it, and otherwise to $ h $; this pointing is maintained while $ h $ remains, since she stays $ \rhd $-highest as others leave.

		\emph{Promoted agents form a chain.} In either economy, at any point in the algorithm, the agents promoted so far who have not yet been removed form, together with $ h $, a chain
		\[
		h=i_1 \rightarrow a_1\rightarrow i_2 \rightarrow a_2\rightarrow \cdots \rightarrow a_{m-1} \rightarrow i_m,
		\]
		in which each $ i_r $ with $ r<m $ points to her favorite remaining object $ a_r $, which points to its owner $ i_{r+1} $, who was promoted when $ i_r $ pointed to it. Here $ i_2,\ldots,i_m $ are the promoted agents, in the order of their promotion, and $ h $ is not one of them. The current order ranks $ i_m $ first, then $ i_{m-1} $, and so on down to $ h=i_1 $, followed by the agents outside the chain in their order under $ \rhd $. This follows by induction on the steps. The highest-ranked agent $ i_m $ points to an object. If the object points to an agent outside the chain, that agent is promoted. If it points to an agent in the chain, a cycle forms and removes a terminal segment of the chain, after which the agent just below that segment is highest-ranked. If it is $ o^* $ or points to no agent, $ i_m $ obtains it and is removed alone. Once no agent of the chain remains, the highest-ranked agent is the new $ h $, who starts a new chain. Since a free object points to no agent or to $ h $, which lies in the chain, no link $ a_r $ is free; and $ a_r $ leaves no earlier than $ i_r $, so it remains her favorite remaining object.

		\emph{The two procedures make the same assignments.} The procedures may take different numbers of steps, so we compare them block by block. Suppose that at some point both have made the same assignments to the agents in $ I $ and face the same remaining agents, objects, and current order, and hence the same chain; by the normalization above, this holds before step $ 1 $ in $ \Gamma $ and after step $ 1 $ in $ \Gamma^P $. Let $ o $ be the favorite remaining object of $ i_m $. If $ o $ is not free, it points to its owner in both economies, so the next step promotes the same agent or removes the same cycle in both. If $ o $ is $ o^* $, or is free and held by no one in $ \Gamma $ (so that it points to no agent in either economy), then $ i_m $ obtains $ o $ and is removed in both. Otherwise $ o $ is free and held by every remaining agent in $ \Gamma $, so it points to $ h=i_1 $ there and to no agent in $ \Gamma^P $. In $ \Gamma $ the whole chain then forms a cycle, in which $ i_r $ obtains $ a_r $ for $ r<m $ and $ i_m $ obtains $ o $. In $ \Gamma^P $, $ i_m $ obtains $ o $ and is removed; then $ i_{m-1} $ is highest-ranked, and $ a_{m-1} $, whose owner has just left, points to no agent and is still her favorite, so she obtains it and is removed; and so on down to $ i_1 $. The links are available in turn because they have distinct owners and differ from the free object $ o $. Either way, the two procedures again make the same assignments and face the same remaining agents, objects, and current order. By induction, the two outcomes agree on $ I $.
	\end{claimproof}

	Since $ i^P $ ranks $ o^* $ first, she obtains $ o^* $ under every initial order, so by Claim \ref{claim:yrmhinvariance} the YRMH-IGYT outcomes in $ \Gamma^P $ are exactly the extensions of those in $ \Gamma $. These two results, together with Proposition \ref{prop:specialsetting} in private-ownership economies, imply that the exclusion core and the set of YRMH-IGYT outcomes coincide in mixed-ownership economies. Since the consistent exclusion core lies between them, it coincides with them in both private-ownership and mixed-ownership economies, so the lemma holds for it as well.	
	\end{proof}  

	\begin{proof}[\textbf{Proof of Proposition \ref{prop:specialsetting} in mixed-ownership economies}]
		The first statement is shown in the last paragraph of the proof of Lemma \ref{lemma:consistent}: in every mixed-ownership economy, the exclusion core, the consistent exclusion core, and the set of YRMH-IGYT outcomes coincide.

		By Claim \ref{claim:yrmhinvariance}, the outcome of YRMH-IGYT in any mixed-ownership economy $ \Gamma $ under any $\rhd$ agrees on $ I $ with its outcome in $ \Gamma^P $ under any $\rhd^P$ that ranks the agents in $I$ identically to $\rhd$. A profitable joint manipulation by a coalition in any such $ \Gamma $ would therefore induce the same profitable manipulation in the private-ownership economy $ \Gamma^P $, with the artificial agent's report held fixed, contradicting the group strategy-proofness of YRMH-IGYT there. So, YRMH-IGYT is group strategy-proof in mixed-ownership economies.
	\end{proof}

\setlength{\bibsep}{0pt plus 0.5ex}

\bibliography{reference} 

\clearpage

\begin{center}
	\large Online Appendix to \\
	 ``Consistency and the exclusion core under co-ownership''
\end{center}

\setcounter{page}{1}
\renewcommand{\thepage}{OA-\arabic{page}}

\counterwithin{theorem}{section}
\counterwithin{proposition}{section}
\counterwithin{lemma}{section}
\counterwithin{corollary}{section}
\counterwithin{definition}{section}
\counterwithin{example}{section}
\makeatletter
\let\c@proposition\c@theorem
\let\c@lemma\c@theorem
\let\c@corollary\c@theorem
\let\c@definition\c@theorem
\let\c@example\c@theorem
\makeatother

\section{The exchange-rights concepts of \cite{sun2020core}}\label{appendix:rectified}

This appendix supplements Section \ref{section:Sunetal}. Online Appendix \ref{appendix:rectifiedcore} states \citeapos{sun2020core} rectified core formally and proves Proposition \ref{prop:rectifiedcore:consistency}; Online Appendix \ref{appendix:comparisons} gives two economies showing that the consistent exclusion core has no set-inclusion relationship with either of their concepts.

\subsection{Consistency of \citeapos{sun2020core} rectified core}\label{appendix:rectifiedcore}

\citeapos{sun2020core} rectified core sharpens their effective core within the exchange-rights view, and goes beyond what consistency alone justifies: by Proposition \ref{prop:rectifiedcore:consistency}, it is consistent, yet it lies strictly inside the largest consistent refinement of the effective core.

\begin{definition}\label{definition:rectifiedcore}
	In an economy, an allocation $ \mu $ is \emph{rectification blocked} by a coalition $ C $ via another allocation $ \sigma $ if (1) $ \forall i\in C $, $ \sigma(i) \succsim_i \mu(i) $ and $ \exists j\in C $, $ \sigma(j)\succ_j\mu(j) $; (2) $ \sigma(C)\subseteq \w(C)\cup \{o^*\}$; 
	\begin{enumerate}
		\item[(3')] for every self-feasible $ C'\subseteq C_{\sigma=\mu} $ and every $ o\in \w(C')\backslash \sigma(C') $, if $ o\in \mu(I\backslash C) $, then $ o\notin \sigma(C) $.
	\end{enumerate}
	The \textbf{rectified core} consists of allocations that are not rectification blocked.
\end{definition}

Compared with the effective core, the rectified core permits assigning an object $ o $ owned but not used by a self-feasible $C'\subseteq C_{\sigma=\mu} $ to $ C\backslash C' $, provided $ o $ is not assigned to any agent outside the coalition under $ \mu $. Thus the rectified core is a superset of the strong core and a subset of the effective core.

We restate Proposition \ref{prop:rectifiedcore:consistency} of Section \ref{section:Sunetal} and prove it here.

\medskip
\noindent\textbf{Proposition \ref{prop:rectifiedcore:consistency}.} \emph{The rectified core of \cite{sun2020core} is consistent, but it is not the largest consistent refinement of the effective core.}
\medskip

\begin{proof}[Proof of Proposition \ref{prop:rectifiedcore:consistency}]
	We prove consistency here; Example \ref{Example:notmax} illustrates why the rectified core is not the largest consistent refinement of the effective core.

	Consider any economy $ \Gamma=(I,O,\succ_I,\{C_o\}_{o\in O}) $. Let $ \mu $ be any allocation in the rectified core. For any self-feasible $ I'\subsetneq I $ under $ \mu $, we show that $ \mu_{I\backslash I'} $ belongs to the rectified core of $ \Gamma(\mu,I\backslash I') $. Let $ \w' $ denote the endowment function in $ \Gamma(\mu,I\backslash I') $.

	Suppose, toward a contradiction, that $ \mu_{I\backslash I'} $ is rectification blocked by a coalition $ C $ via another $ \sigma' $ in $ \Gamma(\mu,I\backslash I') $. Since the rectified core is Pareto efficient, $ \mu $ is Pareto efficient in $ \Gamma $, and $ \mu_{I\backslash I'} $ is Pareto efficient in $ \Gamma(\mu,I\backslash I') $. So $ C\subsetneq I\backslash I' $. Since $ \sigma'(C) \subseteq \w'(C) \cup \{o^*\}$, for every $ i\in C $, $ \sigma'(i)$ is not publicly owned in the reduced economy. So, $ \sigma'(C) \cap [\w(I')\backslash \mu(I')]=\emptyset $. 
	Below we prove that in $ \Gamma $, $ \mu $ is rectification blocked by $ C^*= C\cup I' $ via another $ \sigma $, where $ \sigma_{I\backslash I'}=\sigma' $ and $ \sigma_{I'}=\mu_{I'} $. We verify the conditions of Definition \ref{definition:rectifiedcore} in turn.

	\textit{Condition (1)}: The members of $ I' $ are unaffected, and those of $ C $ fare as under $ \sigma' $.

	\textit{Condition (2)}: Since $ \w'(C)\subseteq \w(C\cup I') $ and $ \w(I') \subseteq \w(C\cup I')$, we have $ \w'(C)\cup \w(I')\subseteq \w(C\cup I') $. So, $ \sigma(C\cup I')=\sigma'(C)\cup \mu(I')\subseteq \w'(C)\cup \w(I') \cup \{o^*\} \subseteq \w(C\cup I') \cup \{o^*\} $.

	\textit{Condition (3$'$)}: Let $ C'\subseteq C^*_{\sigma=\mu} $ be self-feasible, and let $ o\in \w(C')\backslash \sigma(C') $ with $ \mu^{-1}(o)\in I\backslash C^* $. Since $ \sigma(I')=\mu(I') $ and $ \mu^{-1}(o)\notin I' $, $ o\notin\sigma(I') $, and it remains to show that $ o\notin\sigma'(C) $. If $ C_o\subseteq I' $, then $ o\in\w(I')\backslash\mu(I') $, which is disjoint from $ \sigma'(C) $. Otherwise, $ o $ is not publicly owned in the reduced economy, and its owners there are $ C_o\cap(I\backslash I')\subseteq C'\cap C $, because $ C_o\subseteq C'\subseteq C\cup I' $; hence $ o\in\w'(C'\cap C)\backslash\sigma'(C'\cap C) $, and $ o $ is assigned under $ \mu_{I\backslash I'} $ to an agent outside $ C $. It therefore suffices to show that $ C'\cap C $, a subcoalition of $ C_{\sigma'=\mu} $, is self-feasible in the reduced economy, for condition (3$'$) of the block of $ \mu_{I\backslash I'} $ by $ C $ then gives $ o\notin\sigma'(C) $. Take any $ o'\in\sigma(C'\cap C) $ with $ o'\neq o^* $. Since $ C' $ is self-feasible and $ \sigma $ agrees with $ \mu $ on $ C' $, $ o'\in\w(C') $. Moreover, $ o'\in\sigma'(C) $, and $ o'\notin\mu(I') $, because $ o' $ is assigned under $ \mu $ to an agent of $ C $. As $ \sigma'(C) $ is disjoint from $ \w(I')\backslash\mu(I') $, $ o'\notin\w(I') $, so the owners of $ o' $ in the reduced economy are $ C_{o'}\cap(I\backslash I')\subseteq C'\cap C $.

	Therefore $ C^* $ rectification blocks $ \mu $ via $ \sigma $ in $ \Gamma $, contradicting the assumption that $ \mu $ belongs to the rectified core of $ \Gamma $.
\end{proof}

The following example proves the second claim of Proposition \ref{prop:rectifiedcore:consistency}.

\begin{example}[The rectified core is not the largest consistent refinement of the effective core]\label{Example:notmax}
	Consider the following economy:
	\begin{center}
		\begin{tabular}{rcc}
			& $ a $ & $ b $  \\ \hline
			$ C_o $: & $ 1 $ & $ 2 $ \\ \hline
			$ \mu $: & $ 3 $ & $ 4 $ \\
			$ \sigma $: & $ 4 $ & $ 1 $
		\end{tabular}
		\quad \quad
		\begin{tabular}{cccc}
			$ \succ_1 $	& $ \succ_2 $ & $ \succ_3 $ & $ \succ_4 $  \\ \hline
			$ b $ & $ o^* $ & $ a $ & $ a $\\
			$ o^* $ & $ a $ & $ b $ & $ b $\\
			$ a $ & $ b $ & $ o^* $ & $ o^* $
		\end{tabular}
	\end{center}

	The allocation $ \mu $ belongs to the largest consistent refinement of the effective core but not to the rectified core.

	To show that $ \mu $ belongs to the largest consistent refinement of the effective core, by Lemma \ref{lemma:maxrefinement}, it suffices to show, for $ R=I $ and for $ R=I\backslash S $ with every nonempty self-feasible $ S\subsetneq I $, that $ \mu_R $ belongs to the effective core of the economy with agents $ R $, which is $ \Gamma $ if $ R=I $ and $ \Gamma(\mu,R) $ otherwise. Suppose that some $ C\subseteq R $ effectively blocks $ \mu_R $ there via some $ \sigma $. Only agents $ 1 $ and $ 4 $ can be made strictly better off, agent $ 1 $ only by receiving $ b $, which agent $ 4 $ holds, and agent $ 4 $ only by receiving $ a $, which agent $ 3 $ holds. So whenever some agent gains, agent $ 3 $ or agent $ 4 $ is harmed; hence $ C\neq R $, and $ C $ cannot use an object that is publicly owned there. If agent $ 1 $ gains, then $ b $ is still owned by agent $ 2 $, who belongs to $ C $ and cannot gain; holding $ o^* $, she forms a self-feasible unaffected subcoalition whose redundant object $ b $ goes to agent $ 1 $, which condition (3) of Definition \ref{definition:effectivecore} forbids because $ C\neq R $. So agent $ 4 $ gains, and the same argument applies to agent $ 1 $, the owner of $ a $, who then belongs to $ C $ without gaining.

	However, $ \mu $ does not belong to the rectified core: the coalition $ \{1,2,4\} $ rectification blocks it via $ \sigma $. Its only unaffected member, agent $ 2 $, is self-feasible and owns the redundant object $ b $, but $ b $ is assigned under $ \mu $ to agent $ 4 $, a member of the coalition, so condition (3$'$) permits its use. 
\end{example}

\subsection{Comparison with the consistent exclusion core}\label{appendix:comparisons}

Example \ref{Example:HET} below and Example \ref{Example:combinedfeature} in Section \ref{section:YRMH} show that the consistent exclusion core does not have a set-inclusion relationship with \citeapos{sun2020core} two concepts: in the former, the consistent exclusion core is strictly smaller; in the latter, an allocation of the consistent exclusion core lies outside both.

\begin{example}[The consistent exclusion core may be strictly smaller than Sun et al.'s concepts]\label{Example:HET}
	Consider the following house allocation with existing tenants economy.
	\begin{center}
		\begin{tabular}{rcccc}
			& $ a $ & $ b $ & $ c $ & $ d $ \\ \hline
			$ C_o $: & $ 1 $ & $ 2 $ & $ 3 $ & $ 1,2,3,4 $\\ \hline
			$ \mu $: & $ 1 $ & $ 3 $ & $ 2 $ & $ 4 $\\
			$ \sigma_1 $: & $ 4 $ & $ 1 $ & $ 3 $ & $ 2 $\\
			$ \sigma_2 $: & $ 1 $ & $ 3 $ & $ 4 $ & $ 2 $
		\end{tabular}
		\quad \quad
		\begin{tabular}{cccc}
			$ \succ_1 $	& $ \succ_2 $ & $ \succ_3 $ & $ \succ_4 $ \\ \hline
			$ b $ & $ d $ & $ b $ & $ a $\\
			$ a $ & $ c$ & $ c $ & $ d $\\
			$ c $& $ b $ & $ a $ & $ c $ \\
			$ d $& $ a $ & $ d $ & $ b $
		\end{tabular}
	\end{center}

	The reader may check that both the exclusion core and the consistent exclusion core are $ \{\sigma_1,\sigma_2\} $, whereas both the effective core and the rectified core are $ \{\mu,\sigma_1,\sigma_2\} $. In fact, all three allocations belong to the strong core, so the rectified core, which contains the strong core and is contained in the effective core, coincides with both. The allocation $\mu$ is exclusion blocked by the coalition $ \{1,2,4\} $ via $ \sigma_1 $. The coalition has the right to exclude $ 3 $ from $ b $, because $ b $ is owned by $ 2 $. Once this exclusion takes place, assigning the public object $ d $ to $ 2 $ harms no outsider.
\end{example}

In the economy of Example \ref{Example:combinedfeature}, the reader may check that both the exclusion core and the consistent exclusion core are $ \{\mu,\sigma,\delta\} $, whereas both the effective core and the rectified core are $ \{\mu,\sigma\} $. In fact, both $\mu$ and $\sigma$ belong to the strong core. The allocation $\delta$ is effectively blocked by $ \{1,2\} $ via $ \mu $, since agent $ 2 $ strictly gains $ b $ while agent $ 1 $ keeps $ a $ and remains unaffected, and $ \{1,2\} $ owns both objects. Because agent $ 1 $ alone is not self-feasible---her assignment $ a $ is co-owned with agent $ 2 $---no self-feasible subcoalition of the unaffected agent shields $ b $ from reassignment. 	In contrast, $ \delta $ is not exclusion blocked by $ \{1,2\} $, because $ 1 $ cannot be made strictly better off; $ \delta $ is not consistently exclusion blocked by $ \{1,2\} $, because the unaffected agent $ 1 $ is not self-feasible.

\section{The autonomous exclusion core}\label{appendix:refinedEC:definition}

The autonomous exclusion core sharpens the consistent exclusion core within the exclusion-rights view, and likewise goes beyond what consistency alone justifies. This appendix motivates it, states it formally, and establishes its properties.

Consistency alone does not rule out every implausible allocation. In Example \ref{Example:combinedfeature} of Section \ref{section:YRMH}, the consistent exclusion core retains the allocation $ \delta $, in which agent $ 1 $ keeps $ a $ and agent $ 3 $ holds $ b $. The coalition $ \{1,2\} $ owns both objects, yet it cannot block: the unaffected agent $ 1 $ is not self-feasible, so condition (2) of Definition \ref{definition:C1core} keeps her out. By Theorem \ref{theorem:C1core:char}, consistency alone does not remove $ \delta $, so ruling it out requires a refinement of the consistent exclusion core.

There can be more than one such refinement, and this appendix proposes the one we find most natural. Consistency alone requires that self-feasible groups of unaffected agents be permitted to join a blocking coalition. Our refinement further permits the unaffected group to fail self-feasibility, provided that the coalition owns every real object assigned to that group. We call such a block \emph{autonomous}. The definition below therefore modifies Definition \ref{definition:C1core} in two places. Condition (2) no longer requires the unaffected group $ C_{\sigma=\mu} $ to be self-feasible, but only that the coalition own what that group holds. Condition (3) is unchanged in substance, except that it now applies to the largest self-feasible subcoalition of $ C_{\sigma=\mu} $,\footnote{A largest such subcoalition exists because self-feasible coalitions are closed under union: if $ \mu(S_1)\subseteq \w(S_1)\cup\{o^*\} $ and $ \mu(S_2)\subseteq \w(S_2)\cup\{o^*\} $, then $ \mu(S_1\cup S_2)\subseteq \w(S_1)\cup \w(S_2)\cup\{o^*\}\subseteq \w(S_1\cup S_2)\cup\{o^*\} $. The empty set is self-feasible, so the union of all self-feasible subcoalitions of $ C_{\sigma=\mu} $ is the largest one. Throughout we adopt the convention $ C_{o^*}=\emptyset $: the null object has no owners.} which we denote by $ C^\circ_{\sigma=\mu} $,  rather than to $ C_{\sigma=\mu} $ itself.

\begin{definition}\label{Definition:refined}
	In an economy, an allocation $ \mu $ is \emph{autonomously exclusion blocked} by a coalition $ C $ via another allocation $ \sigma $ if
	\begin{enumerate}
		\item $ \forall i\in C $, $ \sigma(i) \succsim_i \mu(i) $ and $ \exists j\in C $, $ \sigma(j)\succ_j\mu(j) $;

		\item $ \mu(C_{\sigma=\mu})\subseteq \w(C)\cup \{o^*\} $;

		\item $ \forall j\in I\backslash C $, $\mu(j)\succ_j \sigma(j)\implies \mu(j)\in  \Omega(C|\w,\mu)$ and $ \mu(j)\notin  \Omega(C^\circ_{\sigma=\mu}|C,\w,\mu) $.

	\end{enumerate}
	The \textbf{autonomous exclusion core} consists of allocations that are not autonomously exclusion blocked.
\end{definition}

Since $ \mu(C^\circ_{\sigma=\mu})\subseteq \w(C^\circ_{\sigma=\mu})\cup \{o^*\} \subseteq \w(C)\cup \{o^*\}$, condition (2) in fact imposes a restriction only on $ C_{\sigma=\mu}\backslash C^\circ_{\sigma=\mu} $, and condition (3) asks of $ C^\circ_{\sigma=\mu} $ exactly what condition (3) of Definition \ref{definition:C1core} asks of $ C_{\sigma=\mu} $. In Example \ref{Example:combinedfeature}, the autonomous exclusion core rules $ \delta $ out, since the coalition $ \{1,2\} $ blocks $ \delta $ via $ \mu $---agent $ 2 $ strictly gains, agent $ 1 $ is unaffected, and agent $ 1 $'s unchanged assignment $ a $ is owned by the coalition because agent $ 2 $ co-owns it.

Lemma \ref{lemma:secondcondition} formalizes the economic implication of  condition (2): every $i\in C_{\sigma=\mu} \backslash C^\circ_{\sigma=\mu} $ directly or indirectly draws on the ownership of $ C_{\sigma>\mu}  $ to remain unaffected. To state it, for each $ i\in C_{\sigma=\mu} \backslash C^\circ_{\sigma=\mu} $, let $ C_i^1=C_{\mu(i)} $ and $ C_i^k= \cup_{j\in C_i^{k-1}} C_{\mu(j)}$ for $ k \ge 2$. We say that such an $ i $ is \textbf{strict-gainer dependent} in $ C $ from $ \mu $ to $ \sigma $ if there exists $ k\in\mathbb{N} $ such that $ C_i^\ell\subseteq C_{\sigma=\mu} $ for all $ 1\le \ell<k $, $ C_i^k\subseteq C $, and $ C_i^k\cap C_{\sigma>\mu}\neq\emptyset $; that is, either all owners of $ \mu(i) $ lie in $ C$ with one owner in $ C_{\sigma>\mu} $, or all owners of $ \mu(i) $ lie in $ C_{\sigma=\mu} $ but some such owner's assignment has an owner in $ C_{\sigma>\mu} $, and so on. So the chain of owners starting from $i$'s assignment stays inside the unaffected group until it reaches a strict gainer, without ever leaving the coalition.

\begin{lemma}\label{lemma:secondcondition}
	Suppose that $ \sigma(h)\succsim_h\mu(h) $ for every $ h\in C $. Then the following are equivalent:
	\begin{enumerate}
		\item $ \mu(C_{\sigma=\mu})\subseteq \w(C)\cup\{o^*\} $;
		\item Every $ i\in C_{\sigma=\mu} \backslash C^\circ_{\sigma=\mu} $ is strict-gainer dependent.
	\end{enumerate}
\end{lemma}

Every consistent exclusion block is an autonomous exclusion block, so the autonomous exclusion core is contained in the consistent exclusion core. It is therefore also Pareto efficient, since the exclusion core is. The autonomous exclusion core is nonempty: the closing step of the proof of Proposition \ref{prop:YRMH-IGYT} in Appendix \ref{appendix:proofsYRMH} in fact shows that \emph{every outcome of YRMH-IGYT belongs to the autonomous exclusion core}. It is also consistent. Theorem \ref{thm:refined EC} summarizes its properties.  

\begin{theorem}\label{thm:refined EC}
	The autonomous exclusion core contains every outcome of YRMH-IGYT, and hence is nonempty; it is consistent and contained in the consistent exclusion core.
\end{theorem}

Proposition \ref{prop:subset:rectified core} places the autonomous exclusion core inside \citeapos{sun2020core} rectified core: whenever an allocation is rectification blocked by a coalition, it is autonomously exclusion blocked by the same coalition.

\begin{proposition}\label{prop:subset:rectified core}
	The autonomous exclusion core is a subset of the rectified core and of the effective core.
\end{proposition}

Example \ref{Example:HET} in Online Appendix \ref{appendix:rectified} illustrates that the containment can be strict: the allocation $ \mu $ belongs to the rectified core, yet it does not belong to the exclusion core and hence not to the autonomous exclusion core. 

\subsection{Proofs}\label{appendix:autonomous:proof}

	\begin{proof}[\textbf{Proof of Lemma \ref{lemma:secondcondition}}]
		First suppose that $ \mu(C_{\sigma=\mu})\subseteq \w(C)\cup\{o^*\} $. Fix $ i\in C_{\sigma=\mu} \backslash C^\circ_{\sigma=\mu} $, let $ C_i^0=\{i\} $, and define $ C_i^1,C_i^2,\ldots $ as in the text. Suppose that $ C_i^\ell\subseteq C_{\sigma=\mu} $ for all $ \ell<k $, as holds for $ k=1 $. Then every object other than $ o^* $ assigned to an agent of $ C_i^{k-1} $ is owned by $ C $, so $ C_i^k\subseteq C $; since every member of $ C $ weakly improves, either $ C_i^k\cap C_{\sigma>\mu}\neq\emptyset $, and $ i $ is strict-gainer dependent, or $ C_i^k\subseteq C_{\sigma=\mu} $. If the latter holds for every $ k $, then $ \cup_{k=0}^\infty C_i^k $ is a subcoalition of $ C_{\sigma=\mu} $ that contains $ i $ and is self-feasible, since every object other than $ o^* $ assigned to an agent of $ C_i^k $ is owned by $ C_i^{k+1} $; this contradicts the maximality of $ C^\circ_{\sigma=\mu} $.

		Conversely, suppose that every agent in $ C_{\sigma=\mu}\backslash C^\circ_{\sigma=\mu} $ is strict-gainer dependent, and let $ h\in C_{\sigma=\mu} $. If $ h\in C^\circ_{\sigma=\mu} $, then $ C_{\mu(h)}\subseteq C^\circ_{\sigma=\mu}\subseteq C $ unless $ \mu(h)=o^* $. Otherwise, $ C_{\mu(h)}=C_h^1\subseteq C $, because the chain either stops at $ k=1 $ or has $ C_h^1\subseteq C_{\sigma=\mu} $. Either way, $ \mu(h)\in \w(C)\cup\{o^*\} $.
	\end{proof}

\begin{proof}[Proof of Theorem \ref{thm:refined EC}]
	\textbf{Nonemptiness.} The proof of Proposition \ref{prop:YRMH-IGYT} in Appendix \ref{appendix:proofsYRMH} invokes condition (2) of Definition \ref{definition:C1core} only through the fact that $ \mu(C_{\sigma=\mu})\subseteq\w(C)\cup\{o^*\} $, and the coalition $ C' $ it produces is a self-feasible subcoalition of $ C_{\sigma=\mu} $, hence contained in $ C^\circ_{\sigma=\mu} $. Therefore, the proof finds a harmed outsider $ i^\diamond $ with $\mu(i^\diamond)=o^\diamond$ such that $ o^\diamond\in\Omega(C|\w,\mu) $ and $ o^\diamond\in\Omega(C'|C,\w,\mu) \subseteq\Omega(C^\circ_{\sigma=\mu}|C,\w,\mu) $, a contradiction to condition (3) of Definition \ref{Definition:refined}. So, every outcome of YRMH-IGYT belongs to the autonomous exclusion core.

	\textbf{Consistency.} Consider any economy $ \Gamma=(I,O,\succ_I,\{C_o\}_{o\in O}) $. Let $ \mu $ be an allocation in the autonomous exclusion core. For every self-feasible $ I'\subsetneq I $ under $ \mu $, we show that $ \mu_{I\backslash I'} $ belongs to the autonomous exclusion core of  $ \Gamma(\mu,I\backslash I') $. Let $ \w' $ denote the endowment function in $ \Gamma(\mu,I\backslash I') $.

	Suppose, toward  a contradiction, that in $ \Gamma(\mu,I\backslash I') $, $ \mu_{I\backslash I'} $ is autonomously exclusion blocked by a coalition $ C $ via another $ \sigma' $. The autonomous exclusion core is Pareto efficient since it is a refinement of the exclusion core. Since $ \mu $ is Pareto efficient in $ \Gamma $, $ \mu_{I\backslash I'} $ is Pareto efficient in $ \Gamma(\mu,I\backslash I') $. So, $ C\subsetneq I\backslash I' $. Below, we prove that in $ \Gamma $, $ \mu $ is autonomously exclusion blocked by $ C^*= C\cup I' $ via another $ \sigma $ where $ \sigma_{I\backslash I'}=\sigma' $ and $ \sigma_{I'}=\mu_{I'} $. We verify the three conditions of Definition \ref{Definition:refined} in turn.

	\textit{Condition (1)}: The members of $ I' $ are unaffected, and those of $ C $ fare as under $ \sigma' $.

	\textit{Condition (2)}: Since $ \sigma $ agrees with $ \mu $ on $ C^*_{\sigma=\mu}=C_{\sigma'=\mu} \cup I'$, we have $ \mu(C^*_{\sigma=\mu})= \mu(C_{\sigma'=\mu})\cup \mu(I')$. Now $ \mu(C_{\sigma'=\mu})\subseteq \w'(C) \cup \{o^*\}$ by condition (2) of the block in the reduced economy, and $ \w'(C)\subseteq \w(C^*) $, since $ C'_o\subseteq C $ forces $ C_o\subseteq C\cup I' $. Also $ \mu(I')\subseteq \w(I') \cup \{o^*\}\subseteq \w(C^*)\cup \{o^*\}$ because $ I' $ is self-feasible. Hence 
	$\mu(C^*_{\sigma=\mu}) \subseteq \w(C^*)\cup \{o^*\}$. 

	\textit{Condition (3)}: $ \forall j\in I\backslash C^* $, if $\mu(j)\succ_j \sigma(j)$, then $ j \in I\backslash I' $ and $ \sigma(j)=\sigma'(j) $. Since $ C $ autonomously exclusion blocks $ \mu_{I\backslash I'} $ via $ \sigma' $ in $ \Gamma(\mu,I\backslash I') $, $ \mu(j)\in \Omega(C|\w',\mu_{I\backslash I'}) $ and $ \mu(j)\notin\Omega(C^\circ_{\sigma'=\mu}|C,\w',\mu_{I\backslash I'}) $. Here $ \Omega(C|\w',\mu_{I\backslash I'})=\w'(\cup_{k=0}^\infty \tilde{C}^k) $, where $ \tilde{C}^0=C $ and $ \tilde{C}^k=\tilde{C}^{k-1}\cup \{i\in (I\backslash I')\backslash \tilde{C}^{k-1}: \mu(i)\in \w'(\tilde{C}^{k-1}) \} $ for every $ k\ge 1 $. One can verify that $ \Omega(C|\w',\mu_{I\backslash I'})\subseteq \Omega(C^*|\w,\mu) $,\footnote{Let $ \Omega(C^*|\w,\mu)=\w(\cup_{k=0}^\infty B^k) $, where $ B^0=C^* $ and $ B^k=B^{k-1}\cup \{i\in I\backslash B^{k-1}: \mu(i)\in \w(B^{k-1}) \} $ for every $ k\ge 1 $. Since $ \w'(S)\subseteq\w(S\cup I') $ for every $ S\subseteq I\backslash I' $, induction on $ k $ gives $ \tilde{C}^k\cup I'\subseteq B^k $, and hence $ \w'(\tilde{C}^k)\subseteq\w(B^k) $, for every $ k\ge 0 $.} so $ \mu(j)\in  \Omega(C^*|\w,\mu) $.

	It remains to prove that $ \mu(j)\notin \Omega(\overline{C}|C^*,\w,\mu) $, where $ \overline{C} $ denotes the largest self-feasible subcoalition of $ C^*_{\sigma=\mu} $. By Definition \ref{defn:ownexclusionright}, $ \Omega(\overline{C}|C^*,\w,\mu)=\w(\cup_{k=0}^\infty C^k) $, where $ C^0=\overline{C} $ and $ C^k=C^{k-1}\cup \{i\in I\backslash (C^*\cup C^{k-1}): \mu(i)\in \w(C^{k-1}) \} $ for every $ k\ge 1 $. Only agents outside $ C^* $ are ever added, so $ (\cup_{k=0}^\infty C^k)\cap C^*=\overline{C} $. Also, $ I'\subseteq\overline{C} $, since $ I' $ is self-feasible.

	We first show that $ \overline{C}\cap C $, which may be empty, is a self-feasible subcoalition of $ C_{\sigma'=\mu} $ in $ \Gamma(\mu,I\backslash I') $, and hence is contained in $ C^\circ_{\sigma'=\mu} $. Take any $ o\in\sigma(\overline{C}\cap C) $ with $ o\neq o^* $. Since $ \overline{C} $ is self-feasible and $ \sigma $ agrees with $ \mu $ on it, $ C_o\subseteq\overline{C} $. By condition (2) of the block in the reduced economy, $ o\in\w'(C) $; since $ C\subsetneq I\backslash I' $, $ o $ is not publicly owned there, so its owners there are $ C_o\cap(I\backslash I')\subseteq\overline{C}\cap C $, as $ \overline{C}\subseteq C\cup I' $.

	As $ \Omega(\cdot|C,\w',\mu_{I\backslash I'}) $ is monotone in its first argument, it therefore suffices to show that $ \mu(j)\in\Omega(\overline{C}|C^*,\w,\mu) $ would imply $ \mu(j)\in \Omega(\overline{C}\cap C|C,\w',\mu_{I\backslash I'}) $, which lies in $ \Omega(C^\circ_{\sigma'=\mu}|C,\w',\mu_{I\backslash I'}) $, contradicting the block in the reduced economy. We prove, by induction on $ k\ge 0 $, that every $ i\in I\backslash C^* $ with $ \mu(i)\in\Omega(\overline{C}|C^*,\w,\mu) $, for whom $ k $ is the least index with $ \mu(i)\in\w'(\tilde{C}^k) $, satisfies $ \mu(i)\in \Omega(\overline{C}\cap C|C,\w',\mu_{I\backslash I'}) $; agent $ j $ is one of them. Let $ o=\mu(i) $. Agent $ i $ does not belong to $ \tilde{C}^k $: she lies outside $ C $, and entering at step $ m\le k $ would require $ o\in\w'(\tilde{C}^{m-1}) $, contrary to the choice of $ k $. So $ \tilde{C}^k\neq I\backslash I' $, and $ o\in\w'(\tilde{C}^k) $ is not publicly owned in the reduced economy; its owners there are $ C_o\cap(I\backslash I')\subseteq\tilde{C}^k $. Let $ h $ be any of them. If $ h\in\overline{C} $, then $ h\in\overline{C}\cap C $, as $ h\notin I' $. Otherwise, since $ C_o\subseteq\cup_{m\ge 0}C^m $ and this set meets $ C^* $ only in $ \overline{C} $, agent $ h $ lies in $ I\backslash C^* $ and is added by the recursion defining $ \Omega(\overline{C}|C^*,\w,\mu) $, so $ \mu(h)\in\Omega(\overline{C}|C^*,\w,\mu) $; and as $ h\in\tilde{C}^k\backslash C $, she enters the recursion defining $ \Omega(C|\w',\mu_{I\backslash I'}) $ at some step $ m\le k $, so $ \mu(h)\in\w'(\tilde{C}^{m-1}) $. By the induction hypothesis, $ \mu(h)\in \Omega(\overline{C}\cap C|C,\w',\mu_{I\backslash I'}) $, and the recursion defining this set, which adds agents outside $ C $, adds $ h $. Every owner of $ o $ in the reduced economy is thus reached by that recursion, so $ o\in \Omega(\overline{C}\cap C|C,\w',\mu_{I\backslash I'}) $.

	Hence no coalition autonomously exclusion blocks $ \mu_{I\backslash I'} $ in $ \Gamma(\mu,I\backslash I') $, and the autonomous exclusion core is consistent.
\end{proof}

\begin{proof}[Proof of Proposition \ref{prop:subset:rectified core}]
	We show that in any economy, if an allocation $ \mu $ does not belong to the rectified core, then it does not belong to the autonomous exclusion core either. The proposition then follows, because the rectified core is contained in the effective core. 

	Suppose that $ \mu $ is rectification blocked by a coalition $ C $ via another $ \sigma $. Define $ I_1=\{i\in I\backslash C:\mu(i)\in \sigma(C)\} $. Let $ \sigma' $ be an allocation such that, $ \forall i\in I\backslash (I_1 \cup C) $, $ \sigma'(i)=\mu(i) $, $ \forall i\in C $, $ \sigma'(i)=\sigma(i) $, and $ \forall i\in I_1 $, $ \sigma'(i)=o^* $. We then show that $\mu$ is autonomously exclusion blocked by $C$ via $\sigma'$.

	\textit{Condition (1)}: For every $ i\in C $, $ \sigma'(i) \succsim_i \mu(i) $, and $ \sigma'(j)\succ_j \mu(j) $ for some $ j\in C $.

	\textit{Condition (2)}: If $ C_{\sigma'=\mu}\neq \emptyset $, then $ \sigma'(C_{\sigma'=\mu})=\sigma(C_{\sigma=\mu})\subseteq \w(C)\cup \{o^*\} $.

	\textit{Condition (3)}: For any $ j\in I\backslash C $, if $ \mu(j)\succ_j \sigma'(j) $, then $ j\in I_1 $, so $ \sigma'(j)=o^* $ and $ \mu(j) $ is a real object with $ \mu(j)\in \sigma(C) $. Condition (2) of a rectification block gives $ \sigma(C)\subseteq \w(C)\cup\{o^*\} $, so $ C_{\mu(j)}\subseteq C $ and hence $ \mu(j)\in \w(C)\subseteq \Omega(C|\w,\mu) $, which is the first requirement of condition (3) of Definition \ref{Definition:refined}.

	It remains to show that $ \mu(j)\notin \Omega(C^\circ_{\sigma'=\mu}|C,\w,\mu) $. Write $ \Omega(C^\circ_{\sigma'=\mu}|C,\w,\mu)=\w(D) $, where $ D=\cup_{k=0}^\infty C^k $ with $ C^0=C^\circ_{\sigma'=\mu} $ and $ C^k $ as in Definition \ref{defn:ownexclusionright}. Every agent added at a step $ k\ge 1 $ lies outside $ C $, so $D\cap C=C^\circ_{\sigma'=\mu}$. Suppose that $ \mu(j)\in \w(D) $, that is, $ C_{\mu(j)}\subseteq D $. Together with $ C_{\mu(j)}\subseteq C $ established above, this yields $ C_{\mu(j)}\subseteq D\cap C=C^\circ_{\sigma'=\mu} $, so $ \mu(j)\in \w(C^\circ_{\sigma'=\mu}) $. Economically, an exchange-rights block can reallocate only objects the coalition owns outright, so the chains through outsiders that enlarge $ \Omega $ beyond $ \w $ never come into play, and indirect control collapses onto direct control.

	Since $ C^\circ_{\sigma'=\mu}\subseteq C $ and $ j\in I\backslash C $, we have $ j\notin C^\circ_{\sigma'=\mu} $, and $ \sigma' $ agrees with $ \mu $ on $ C^\circ_{\sigma'=\mu} $; hence $ \mu(j)\notin \sigma'(C^\circ_{\sigma'=\mu})=\mu(C^\circ_{\sigma'=\mu}) $. Therefore $ \mu(j)\in \w(C^\circ_{\sigma'=\mu})\backslash \sigma(C^\circ_{\sigma'=\mu}) $ with $ \mu^{-1}(\mu(j))=j\in I\backslash C $, and $ C^\circ_{\sigma'=\mu} $ is a self-feasible subcoalition of $ C_{\sigma=\mu} $. Condition (3') of a rectification block then gives $ \mu(j)\notin \sigma(C) $, contradicting $ j\in I_1 $. So $ \mu(j)\notin \Omega(C^\circ_{\sigma'=\mu}|C,\w,\mu) $.
\end{proof}

\section{Further results}\label{appendix:further}

This appendix collects supporting results on four topics: independence of the requirements in Theorem~\ref{theorem:general:strategyproof}, a stronger form of consistency, BK's own refinements of the exclusion core, and \citeapos{tan2026freeriding} bilateral core.

\subsection{Independence of the requirements in Theorem~\ref{theorem:general:strategyproof}}\label{appendix:independence}

Remark \ref{remark:independence} asserts that the three requirements (individual rationality, Pareto efficiency, consistency) on the solution are independent. That consistency cannot be dropped follows from BK's results on the exclusion core. Their generalized TTC algorithm, run with any fixed extension of the priorities, is strategy-proof and selects from their weak exclusion core, which under the embedding of the co-ownership model is contained in the exclusion core (Online Appendix~\ref{appendix:strongEC}); the exclusion core is not consistent by Corollary~\ref{prop:exclusioncore:consistency}. We verify the other two claims here.

	\textbf{Individual rationality}: Under serial dictatorship, agents move in a strict order $\rhd$, each taking her favorite object among those still available, the null object included; ownership plays no role, which is why the solution may deny an agent an object she privately owns. It is clearly \textit{Pareto efficient} and \textit{strategy-proof}. It is \textit{consistent} because a self-feasible departing group leaves with exactly the objects its members take, and the relative order of the remaining agents is unchanged: by induction along the order, when a remaining agent's turn comes in the reduced economy, her original assignment is still available, and every object available to her is also available in the original economy, so she again takes her original assignment.

	\textbf{Pareto efficiency}: Fix a strict order $ \rhd $ over the grand set of agents, and let each object rank its owners above the others, ordering the agents within each group by $ \rhd $. Let $ S^\rhd $ be the correspondence of stable matchings for these priorities, so $ S^\rhd $ is nonempty. The agent-proposing deferred acceptance mechanism produces a stable matching, and it is \textit{strategy-proof} because the priorities do not depend on preference reports \citep{gale1962college,roth1982economics}. Every stable matching is \textit{individually rational} in the sense of Section \ref{section:model}: stability already grants each agent an assignment weakly preferred to $ o^* $, and an agent who prefers an object she privately owns to her assignment would form a blocking pair with it, since she has its highest priority. The correspondence is \textit{consistent} because, after a self-feasible departure, every remaining object's priority order is its original order restricted to the remaining agents; this holds even when all of its owners depart, since the remaining agents are then all nonowners and become all owners, retaining their order under $ \rhd $. Any blocking pair in the reduced economy would therefore already block the original matching. \emph{Pareto efficiency}, however, may fail. Take $ I=\{1,2,3\} $, $ O=\{a,b\} $, $ C_a=\{1\} $, $ C_b=\{3\} $, and $ 1\rhd2\rhd3 $, so that object $ a $ ranks $ 1\rhd_a 2\rhd_a 3 $ and object $ b $ ranks $ 3\rhd_b 1\rhd_b 2 $, with preferences $ b\succ_1 a\succ_1 o^* $, $ a\succ_2 o^*\succ_2 b $, and $ a\succ_3 b\succ_3 o^* $. Deferred acceptance yields $ (a,o^*,b) $, listing assignments in agent order, which is Pareto dominated by $ (b,o^*,a) $.

\subsection{Stronger consistency}\label{appendix:extraexample}

The following example shows that no nonempty, Pareto efficient, and individually rational solution satisfies the stronger consistency in Remark~\ref{remark:strongerconsistency}, which permits any group of agents to depart with its assignments.

\begin{example}[Stronger consistency is too demanding]\label{Example:stronger:consistency}
	Consider an economy consisting of three agents $ \{1,2,3\} $ and two objects $ \{a,b\} $.
	\begin{center}
		\begin{tabular}{rcc}
			& $ a $ & $ b $  \\ \hline
			$ C_o $: & $ {1,2} $ & $ 3 $ \\ \hline
			$ \mu $: & $ 3 $ & $ 2 $ \\
			$ \sigma $: & $ 3 $ & $ 1 $
		\end{tabular}
		\quad \quad
		\begin{tabular}{ccc}
			$ \succ_1 $	& $ \succ_2 $ & $ \succ_3 $  \\ \hline
			$ b $ & $ b $ & $ a $ \\
			$ a $ & $ a $ & $ b $ \\
			&
		\end{tabular}
	\end{center}

Since agents $ 1 $ and $ 2 $ have preferences different from agent $ 3 $'s, $ \mu $ and $ \sigma $ are intuitive: one of the two owners of object $ a $ receives $ 3 $'s private endowment and $ 3 $ receives $ a $. Indeed, they are the Pareto efficient and individually rational allocations. To see this, individual rationality requires only that agent $ 3 $, who privately owns $ b $, receive $ a $ or $ b $. If she receives $ b $, then whoever holds $ a $ would rather trade with her, and if $ a $ is unassigned it can be given away for free, so Pareto efficiency fails in either case. Hence agent $ 3 $ receives $ a $, and Pareto efficiency then gives $ b $ to agent $ 1 $ or agent $ 2 $.

However, if we consider stronger consistency by permitting any group of agents to leave with their assignments, then, under $ \mu $, removing $ 2 $ with object $ b $ leaves a reduced economy in which $ 1 $ privately owns object $ a $ and $ 3 $ owns nothing, so $ 1 $ alone can strongly block $ \mu_{\{1,3\}} $ by claiming object $ a $. Similarly, under $ \sigma $, removing $ 1 $ with object $ b $ allows $ 2 $ alone to strongly block $ \sigma_{\{2,3\}} $ in the reduced economy. A nonempty, Pareto efficient, and individually rational solution must select $ \mu $ or $ \sigma $ in this economy, and stronger consistency would then carry it to an allocation that individual rationality forbids in the reduced economy. So no solution with these three properties satisfies stronger consistency. The problem here is that, although $ 3 $ privately owns $ b $ in the original economy, removing $ b $ with $ 1 $ or with $ 2 $ strips $ 3 $ of that ownership in the reduced economy, which therefore cannot be viewed as a reduced version of the original.
\end{example}

\subsection{BK's other solution concepts}\label{appendix:strongEC}

BK extend the co-ownership model to a setting in which agents' rights over objects are described by a priority structure $ (\rhd_o)_{o\in O} $; here $ i\rhd_o j $ means that $ i $ has a higher-level right over $ o $ than $ j $, and each $ \rhd_o $ is a strict partial order (irreflexive and transitive, possibly incomplete). Write $ i\unrhd_o j $ if $ i\rhd_o j $ or $ i=j $. For a coalition $ C $, write $ C\unrhd_o j $ if $ i\unrhd_o j $ for some $ i\in C $, and $ C\rhd_o j $ if $ C\unrhd_o j $ and $ j\notin C $. If $ o $ is unassigned under an allocation $ \mu $, write $ \mu^{-1}(o)=\emptyset $; by BK's convention, every agent has priority over $ \emptyset $. From the priorities, BK derive three endowment systems. For every allocation $ \mu $, coalition $ C $, and object $ o $:
\begin{itemize}
	\item $ o\in\w_\mu(C) $ if and only if $ C\unrhd_o\mu^{-1}(o) $;
	\item $ o\in\w^*_\mu(C) $ if and only if $ C\unrhd_o\mu^{-1}(o) $ and, in case $ C\rhd_o\mu^{-1}(o) $, every agent $ k $ with $ C\ntrianglerighteq_o k $ satisfies $ k\rhd_{o'}\mu^{-1}(o) $ for every object $ o'\neq o $;
	\item $ o\in\w^{**}(C) $ if and only if $ C $ contains every $ \rhd_o $-maximal agent, that is, every agent $ i $ such that no agent $ j $ satisfies $ j\rhd_o i $.
\end{itemize}
The \textbf{strong}, \textbf{weak}, and \textbf{unconditional} exclusion cores consist of the allocations $ \mu $ that are not exclusion blocked in the sense of Definition~\ref{Definition:exclusion:core} when $ \w $ is replaced by $ \w_\mu $, $ \w^*_\mu $, and $ \w^{**} $, respectively, in the definition of $ \Omega(C|\w,\mu) $.\footnote{BK call $ \w_\mu $ the weak and $ \w^*_\mu $ the strong conditional endowment system: the weak system imposes fewer conditions, grants coalitions more exclusion rights, and therefore yields the smaller, strong exclusion core.} Since $ \w^{**}(C)\subseteq\w^*_\mu(C)\subseteq\w_\mu(C) $ for every $ C $, the three cores satisfy $ \text{strong}\subseteq\text{weak}\subseteq\text{unconditional} $. 

The co-ownership model embeds into the priority model by setting $ i\rhd_o j $ if and only if $ i\in C_o $ and $ j\notin C_o $. Under this embedding, the $ \rhd_o $-maximal agents are exactly the owners of $ o $, so $ \w^{**}=\w $; the unconditional exclusion core therefore coincides with the exclusion core, and the strong and weak exclusion cores can be viewed as refinements of the exclusion core.

\paragraph{The strong exclusion core.} It may be empty, since the embedded priorities may fail BK's acyclicity condition. See the following example.

\begin{example}[BK's strong exclusion core may be empty]\label{Example:strongexclusioncore:empty}
		Consider the following economy.	
		\begin{center}
			\begin{tabular}{ccc}
				& $ a $ & $ b $ \\ \hline
				$ C_o $: & $ 1$ & $ 2,3 $\\ \hline
				$ \mu_1 $: & $ 2 $ & $ 1 $\\
				$ \mu_2 $: & $ 3 $ & $ 1 $\\
				$ \mu_3 $: & $ 2 $ & $ 3 $\\
				$ \mu_4 $: & $ 3 $ & $ 2 $
			\end{tabular}
			\quad \quad
			\begin{tabular}{ccc}
				$ \succ_1 $	& $ \succ_2 $ & $ \succ_3 $  \\ \hline
				$ b $ & $ a $ & $ a $  \\
				$ a $ & $ b $ & $ b $  \\
				& & \\
				& &\\
				& &
			\end{tabular}
		\end{center}

		Under BK's embedding, the priority structure satisfies $1\rhd_a 2$, $1\rhd_a 3$, $2\rhd_b 1$, and $3\rhd_b 1$, with no other priority comparisons. The endowment system $ \w_\mu $ of the strong exclusion core then takes a simple form.\footnote{That is, for any coalition $ C $ and object $ o $, we have $ o\in \w_\mu(C) $ if and only if $ o $ is unassigned and $ C\neq \emptyset $, or $ \mu^{-1}(o)\in C $, or $ C\cap C_o\neq \emptyset $ and $ \mu^{-1}(o)\notin C_o $.} 		
		We show that the strong exclusion core in the corresponding priority structure is empty. 

		Any Pareto inefficient allocation is exclusion blocked by the set of agents who strictly gain in a Pareto improvement, so only the Pareto efficient allocations remain to be examined, and these are exactly $ \mu_1,\mu_2,\mu_3 $, and $ \mu_4 $. In each of them exactly one agent holds no object. Take that agent as the blocking coalition, blocking via the allocation that gives $ a $ to agent $ 1 $ and $ b $ to whichever of agents $ 2 $ and $ 3 $ does not hold $ a $.

		Under $ \mu_1 $ and $ \mu_2 $ the blocking agent is $ 3 $ and $ 2 $ respectively. She controls $ b $, since she co-owns it while agent $ 1 $, who holds it, does not; the chain then reaches agent $ 1 $, and the enlarged coalition controls $ a $, since agent $ 1 $ owns it while its holder does not. So $ \Omega(\{3\}|\w_{\mu_1},\mu_1)=\{a,b\} $, and likewise for $ \mu_2 $, which places both harmed outsiders' objects within the coalition's control. Under $ \mu_3 $ and $ \mu_4 $ the blocking agent is $ 1 $, who controls $ a $ outright, since she owns it while its holder does not; the only harmed outsider loses $ a $. Every Pareto efficient allocation is therefore exclusion blocked, and the strong exclusion core is empty.
	\end{example}

\paragraph{The weak exclusion core.} It is always nonempty, but it is not consistent: it coincides with the exclusion core in every housing market and in every augmented housing market, so it is a singleton in every housing market and contains every Pareto efficient allocation in every augmented housing market. 

Specifically, in a housing market, the weak exclusion core is nonempty and contained in the unconditional exclusion core, which is the exclusion core and contains only the TTC allocation. In an augmented housing market, let $ \mu $ be Pareto efficient, so that every object is assigned, and let $ C $ be a coalition without $ i^* $; since $ i^* $ can never gain strictly, only such coalitions can block. Consider any object $ o $ held by some $ j\notin C $. If $ j $ is the original owner of $ o $, no agent has priority over her at $ o $, so $ C\unrhd_o j $ fails. Otherwise, $ C\unrhd_o j $ requires the original owner of $ o $ to belong to $ C $; the second requirement in the definition of $ \w^*_\mu $ then fails for $ k=i^* $, since $ i^* $ has no priority over $ j $ at the object that $ j $ owns. Hence $ \w^*_\mu(C) $ consists of the objects held by members of $ C $, the recursion defining $ \Omega(C|\w^*_\mu,\mu) $ adds no agent, and a block by $ C $ could harm no one, which the Pareto efficiency of $ \mu $ rules out. The weak exclusion core thus contains every Pareto efficient allocation.

\paragraph{The direct exclusion core.} BK also define the \emph{direct exclusion core}, which replaces condition (2) of Definition~\ref{Definition:exclusion:core} by the requirement that every agent made worse off lose an object the coalition directly controls: $ \mu(j)\in\w(C) $ in place of $ \mu(j)\in\Omega(C|\w,\mu) $. It therefore contains the exclusion core, and is likewise nonempty and Pareto efficient. It is not consistent either. Being Pareto efficient and containing the exclusion core, it equals the set of Pareto efficient allocations in every augmented housing market by Proposition~\ref{prop:fragility}; yet in some housing markets it is strictly smaller than that set.\footnote{Let two agents privately own $ a $ and $ b $ respectively and let both prefer $ a $ to $ b $. Both allocations that assign the two objects are Pareto efficient, but the one giving $ b $ to the owner of $ a $ is directly exclusion blocked by that owner alone, who reclaims $ a $ from its holder. The direct exclusion core is therefore a singleton.} Since every housing market is the reduced economy obtained from its augmentation by removing $ i^* $, consistency fails.

\subsection{\citeapos{tan2026freeriding}  bilateral core}\label{appendix:bilateral}

\cite{tan2026freeriding} propose the \textbf{bilateral core}, a refinement of \citeapos{sun2020core} rectified core built on a veto-type interpretation of exclusion rights. This appendix shows that it is not consistent and not comparable to the exclusion core or to the consistent exclusion core, yet it contains the autonomous exclusion core of Online Appendix \ref{appendix:refinedEC:definition}.

For an allocation $ \mu $ and an object $ o $, let $ C_{o|\mu} $ denote the set of \textit{extended co-owners} of $ o $ under $ \mu $: begin with $ C_o $ and repeatedly add the owners of the assignments of the agents already reached.\footnote{Formally, $ C_{o|\mu}=\cup_{k=0}^\infty C^k_{o|\mu} $, where $ C^0_{o|\mu}=C_o $ and $ C^k_{o|\mu}=C^{k-1}_{o|\mu}\cup (\cup_{i\in C^{k-1}_{o|\mu}}C_{\mu(i)}) $ for every $ k\ge 1 $. The convention $ C_{o^*}=\emptyset $ gives $ C_{o^*|\mu}=\emptyset $.} This construction runs in the opposite direction to the one behind $ \Omega(C|\w,\mu) $ in Section \ref{section:standardcore}, collecting agents from an object rather than objects from a coalition. \cite{tan2026freeriding} show that an agent holds a \textit{weak exclusion right} over $ o $ under $ \mu $---that is, belongs to every self-feasible coalition whose endowment contains $ o $---exactly when she is an extended co-owner of $ o $ under $ \mu $. A coalition holds the right as soon as one of its members does, so it is veto-type: where $ \w $ asks that a coalition contain all co-owners of an object, the weak exclusion right asks only that it contain one extended co-owner. A similar idea underlies BK's strong exclusion core (Online Appendix~\ref{appendix:strongEC}), whose conditional endowment $ \w_\mu $ places an object in a coalition's endowment as soon as one of its members has a higher-level right over it than its occupant.

Write $ I_{\sigma>\mu}=\{i\in I:\sigma(i)\succ_i\mu(i)\} $ and $ I_{\mu>\sigma}=\{i\in I:\mu(i)\succ_i\sigma(i)\} $. A nonempty coalition $ C $ \textit{bilaterally blocks} $ \mu $ via another allocation $ \sigma $ if (1) $ C\subseteq I_{\sigma>\mu} $; (2) $ C\cap C_{\mu(j)|\mu}\neq\emptyset $ for every $ j\in I_{\mu>\sigma} $; (3) $ I_{\mu>\sigma}\cap C_{\sigma(i)|\sigma}=\emptyset $ for every $ i\in C $; and (4) for every $ i\in I_{\sigma>\mu}\backslash C $ with $ I_{\mu>\sigma}\cap C_{\sigma(i)|\sigma}\neq\emptyset $, $ \sigma(j)\succ_j\sigma(i) $ for every $ j\in I_{\mu>\sigma} $. The \textbf{bilateral core} consists of the allocations that cannot be bilaterally blocked.

Conditions (2) and (3) are the bilateral requirement after which the concept is named: the coalition holds a weak exclusion right over every object a harmed agent gives up under $ \mu $, while the harmed agents hold none over the objects the coalition receives under $ \sigma $. Condition (4) requires that, whenever some harmed agent holds a weak exclusion right over an object that goes to a gainer outside the coalition, every harmed agent prefer her new assignment to that object, so that no harmed agent has reason to withhold it.

The following example shows that the bilateral core is not consistent.

\begin{example}[The bilateral core is not consistent]\label{Example:bilateral}
	Consider the following economy with $ I=\{1,\ldots,7\} $ and $ O=\{a,b,c,d\} $. All objects omitted from a preference column are ranked below $o^*$, in arbitrary strict order.
	\begin{center}
		\begin{tabular}[t]{rcccc}
			& $ a $ & $ b $ & $ c $ & $ d $ \\ \hline
			$ C_o $: & $ 6 $ & $ 1 $ & $ 3,4 $ & $ 7 $ \\ \hline
			$ \mu $: & $ 5 $ & $ 4 $ & $ 3 $ & $ 2 $ \\
			$ \sigma $: & $ 4 $ & $ 3 $ & $ 2 $ & $ 1 $
		\end{tabular}
		\quad \quad
		\begin{tabular}[t]{ccccccc}
			$ \succ_1 $ & $ \succ_2 $ & $ \succ_3 $ & $ \succ_4 $ & $ \succ_5 $ & $ \succ_6 $ & $ \succ_7 $ \\ \hline
			$ d $ & $ c $ & $ b $ & $ b $ & $ a $ & $ o^* $ & $ o^* $\\
			$ o^* $ & $ d $ & $ c $ & $ a $ & $ o^* $ & & \\
			& $ o^* $ & $ o^* $ & $ c $ & & & \\
			& & & $ o^* $ & & &
		\end{tabular}
	\end{center}

	The allocation $ \mu $ belongs to the bilateral core. Only agents $ 1 $, $ 2 $, and $ 3 $ can be made strictly better off, so by condition (1) any blocking coalition $ C $ is a nonempty subset of $ \{1,2,3\} $. Since $ C_{a|\mu}=\{6\} $, $ C_{d|\mu}=\{7\} $, and $ C_{o^*|\mu}=\emptyset $, condition (2) protects agents $ 2 $ and $ 5 $, who hold $ d $ and $ a $, together with every agent who holds $ o^* $: none of them can be harmed.

	Suppose that $ C $ bilaterally blocks $ \mu $ via some $ \tau $. We first show that $ 1\in C $. Agent $ 2 $ gains only by receiving $ c $, and whenever $ \tau(2)=c $ agent $ 3 $ cannot be harmed: she co-owns $ c $, so harming her would violate condition (3) if $ 2\in C $, while if $ 2\notin C $ condition (4) would require $ \tau(3)\succ_3 c $ and so leave her unharmed. As $ c $ is then taken, agent $ 3 $ must receive $ b $. Agent $ 3 $ likewise gains only by receiving $ b $. So if $ 2\in C $ or $ 3\in C $, then $ \tau(3)=b $, agent $ 4 $ loses her most preferred object, and condition (2) with $ C_{b|\mu}=\{1\} $ forces $ 1\in C $.

	Now $ 1\in C $ gives $ \tau(1)=d $. Agent $ 2 $ cannot be harmed, so $ \tau(2)=c $, and then, as above, $ \tau(3)=b $. Agent $ 5 $ cannot be harmed either, so she keeps $ a $, and agent $ 4 $ is left with $ o^* $. Since agent $ 4 $ co-owns $ c $ and is harmed, condition (3) places agent $ 2 $ outside $ C $, and condition (4) then requires $ o^*=\tau(4)\succ_4\tau(2)=c $, contradicting $ c\succ_4 o^* $.

	We show that the restriction of $\mu$ is bilaterally blocked once agent $ 6 $, who is self-feasible under $ \mu $, departs. After she departs, object $ a $ loses its only owner and becomes publicly owned, so $ C'_a=I\backslash\{6\} $ while the other owner sets are unchanged; no real object leaves. A prime marks extended co-ownership in the reduced economy. There the restriction of $ \mu $ is bilaterally blocked by $ C=\{1,3\} $ via $ \sigma_{I\backslash\{6\}} $, written $ \sigma $ below, for which $ I_{\sigma>\mu}=\{1,2,3\} $ and $ I_{\mu>\sigma}=\{4,5\} $. Condition (1) is immediate. For condition (2), the harmed agents $ 4 $ and $ 5 $ hold $ b $ and $ a $ under $ \mu $, and $ C'_{b|\mu}=\{1\} $ and $ C'_{a|\mu}=I\backslash\{6\} $, both of which meet $ C $. For condition (3), $ C'_{d|\sigma}=\{7\} $ and $ C'_{b|\sigma}=\{1,7\} $, and neither contains $ 4 $ or $ 5 $. For condition (4), the only gainer outside $ C $ is agent $ 2 $, whose assignment $ c $ is co-owned by the harmed agent $ 4 $; the condition requires $ \sigma(4)\succ_4 c $ and $ \sigma(5)\succ_5 c $, and indeed $ a\succ_4 c $ and $ o^*\succ_5 c $. So the bilateral core is not consistent.

	The failure is driven by the departure itself. Agents $ 6 $ and $ 7 $ hold the null object and serve only as owners of $ a $ and $ d $, and that ownership keeps $ C_{a|\mu} $ and $ C_{d|\mu} $ beyond the reach of the only agents who could gain. Once agent $ 6 $ leaves with nothing, $ a $ becomes publicly owned, every remaining agent becomes an extended co-owner of it, and the coalition $ \{1,3\} $ acquires against agent $ 5 $ a claim it did not previously have.
\end{example}

The bilateral core is not comparable to the exclusion core or to the consistent exclusion core. On the one hand, take $ \mu $ in the original economy of Example~\ref{Example:bilateral}, which belongs to the bilateral core. The coalition $ \{1,2,3\} $ blocks it via the allocation $ \tau $ in which agents $ 1 $, $ 2 $, and $ 3 $ receive $ d $, $ c $, and $ b $, agent $ 4 $ receives the null object, and all other assignments are unchanged; Example~\ref{Example:bilateral} shows that $ \tau $ is the only candidate for a bilateral block of $ \mu $ and then rules it out. All three coalition members strictly improve, and the only harmed outsider, agent $ 4 $, loses $ b $, which agent $ 1 $ privately owns. Since no member of the coalition is unaffected, this is an exclusion block in the sense of Definition~\ref{Definition:exclusion:core}, so $ \mu $ lies outside the exclusion core, and hence outside the consistent exclusion core. On the other hand, the allocation $ \delta $ of Example~\ref{Example:combinedfeature} belongs to the consistent exclusion core, and hence to the exclusion core, but not to the rectified core (Online Appendix~\ref{appendix:comparisons}), which contains the bilateral core.

Despite this incomparability, the autonomous exclusion core of Online Appendix~\ref{appendix:refinedEC:definition} refines the bilateral core. The proof converts any bilateral block into an autonomous exclusion block by retaining the assignments of all strict gainers and restoring the original assignments of the other agents wherever feasibility allows.

\begin{proposition}\label{prop:autonomous:bilateral}
	In every economy, the bilateral core contains the autonomous exclusion core.
\end{proposition}

\begin{proof}
	Suppose that an allocation $ \mu $ is bilaterally blocked by a coalition $ C_0 $ via another $ \sigma $. Write $ G=I_{\sigma>\mu} $ and $ H=I_{\mu>\sigma} $, and let $ \overline{C}=\{i\in G:C_{\sigma(i)|\sigma}\cap H=\emptyset\} $. By conditions (1) and (3) of bilateral blocking, $ C_0\subseteq\overline{C}\subseteq G $; so $ \overline{C} $ is nonempty and, by condition (2), meets $ C_{\mu(j)|\mu} $ for every $ j\in H $. Moreover, $ \mu(H)\cap\sigma(G\backslash\overline{C})=\emptyset $. Otherwise, $ \mu(h)=\sigma(g) $ for some $ h\in H $ and $ g\in G\backslash\overline{C} $. As $ g\notin C_0 $ and $ C_{\sigma(g)|\sigma}\cap H\neq\emptyset $, condition (4) applies to $ g $ and requires $ \sigma(h)\succ_h\sigma(g)=\mu(h) $, contradicting $ h\in H $.

	\textit{The blocking coalition.} Define $ T=\cup_{i\in\overline{C}}C_{\sigma(i)|\sigma} $ and $ C=G\cup T $. We will use $ C $ as the autonomous blocking coalition. By construction $ T\cap H=\emptyset $, and hence $ C\cap H=\emptyset $. The recursion defining extended co-owners also gives $ \sigma(T)\subseteq\w(T)\cup\{o^*\} $ and $ \sigma(\overline{C})\subseteq\w(C)\cup\{o^*\} $. For the first inclusion, whenever $ t\in T $, all owners of $ \sigma(t) $ also belong to $ T $. For the second, all owners of $ \sigma(i) $ belong to $ C_{\sigma(i)|\sigma}\subseteq T\subseteq C $ for every $ i\in\overline{C} $. Every agent in $ C\backslash G $ belongs to neither $ G $ nor $ H $, so her assignment is unchanged under $ \sigma $; with the first inclusion above this gives $ \mu(C\backslash G)\subseteq\w(C)\cup\{o^*\} $.

	\textit{The blocking allocation.} Let $ D=\{j\in I\backslash G:\mu(j)\in\sigma(G)\backslash\{o^*\}\} $, and define an allocation by
	\[
	\tau(i)=
	\begin{cases}
		\sigma(i),& i\in G,\\
		o^*,& i\in D,\\
		\mu(i),& i\in I\backslash(G\cup D).
	\end{cases}
	\]
	This allocation is feasible: the real objects assigned to $ G $ are distinct because $ \sigma $ is an allocation, those retained from $ \mu $ are distinct because $ \mu $ is, and no agent outside $ G $ retains a real object of $ \sigma(G) $, since every such agent belongs to $ D $. Moreover, $ D\subseteq H $. To see this, take $ j\in D $. Her assignment $ \mu(j) $ is a real object held under $ \sigma $ by a member of $ G $, and $ j\notin G $, so $ \sigma(j)\neq\mu(j) $; strict preferences then imply $ j\in H $. Since $ C\cap H=\emptyset $, we have $ C\cap D=\emptyset $, so every member of $ C\backslash G $ keeps her original assignment under $ \tau $. Thus every member of $ C $ is weakly better off and every member of $ G $ strictly better off; as $ \overline{C}\subseteq G $ is nonempty, condition (1) of Definition~\ref{Definition:refined} holds. Also, $ C_{\tau=\mu}=C\backslash G $, so the inclusion $ \mu(C\backslash G)\subseteq\w(C)\cup\{o^*\} $ established above is condition (2).

	\textit{Condition (3).} Consider any agent $ j\in I\backslash C $ harmed under $ \tau $. Every agent outside $ G\cup D $ keeps her assignment and every agent in $ G $ improves, so $ j\in D\subseteq H $. Her original object is assigned under $ \sigma $ to some member of $ G $, who belongs to $ \overline{C} $ because $ \mu(H)\cap\sigma(G\backslash\overline{C})=\emptyset $. Since $ \sigma(\overline{C})\subseteq\w(C)\cup\{o^*\} $, we obtain $ \mu(j)\in\sigma(\overline{C})\backslash\{o^*\}\subseteq\w(C)\subseteq\Omega(C|\w,\mu) $, which is the first requirement of condition (3) of Definition~\ref{Definition:refined}.

	For the remaining requirement, let $ N^\circ=C^\circ_{\tau=\mu} $ be the largest self-feasible subcoalition of $ C\backslash G $. Suppose, to the contrary, that $ \mu(j)\in\Omega(N^\circ|C,\w,\mu) $. Let $ F $ be the final agent set in the recursion defining this restricted exclusion right. By Definition~\ref{defn:ownexclusionright}, that recursion starts from $ N^\circ $ and adds only agents outside $ C $, so $ F\cap C=N^\circ $ and $ \Omega(N^\circ|C,\w,\mu)=\w(F) $. Together with $ \mu(j)\in\w(C) $, this gives $ C_{\mu(j)}\subseteq F\cap C=N^\circ $. Since $ N^\circ $ is self-feasible under $ \mu $, following the owners of its members' assignments never leaves $ N^\circ $, so $ C_{\mu(j)|\mu}\subseteq N^\circ $. But $ N^\circ\subseteq C\backslash G $ is disjoint from $ \overline{C}\subseteq G $, so $ \overline{C}\cap C_{\mu(j)|\mu}=\emptyset $. This contradicts the fact that $ \overline{C} $ meets $ C_{\mu(j)|\mu} $, as $ j\in H $. The remaining requirement follows.

	We have shown that $ C $ autonomously exclusion blocks $ \mu $ via $ \tau $. Thus every allocation outside the bilateral core is outside the autonomous exclusion core.
\end{proof}

The inclusion can be strict: the allocation $ \mu $ of Example~\ref{Example:bilateral} belongs to the bilateral core but, as shown above, not to the exclusion core, and hence not to the autonomous exclusion core.

\end{document}